\documentclass[11pt]{article}
\usepackage[dvips]{graphicx}
\usepackage{graphics}
\usepackage{graphicx}
\usepackage{amssymb}
\usepackage{amsmath}
\usepackage{amssymb}
\usepackage{overcite}
\usepackage{epsf}
\input{epsf}
\numberwithin{equation}{section}
\begin{document}
\begin{titlepage}
\title{\bf Bogoliubov's Vision: Quasiaverages and Broken Symmetry  to Quantum 
Protectorate and Emergence\thanks{International Journal of Modern Physics B (IJMPB), Volume: 24, Issue: 8
(2010)  p.835-935.}}  
\author{A. L. Kuzemsky 
\\
{\it Bogoliubov Laboratory of Theoretical Physics,} \\
{\it  Joint Institute for Nuclear Research,}\\
{\it 141980 Dubna, Moscow Region, Russia.}\\
{\it E-mail:  kuzemsky@theor.jinr.ru} \\
{\it http://theor.jinr.ru/\symbol{126}kuzemsky}}
\date{}
\maketitle
\begin{abstract}
In the present interdisciplinary review we focus on the
applications of the  symmetry principles to quantum and statistical physics in connection with some other branches
of science.
The profound and innovative idea of
quasiaverages formulated by  N.~N.~Bogoliubov,  gives the so-called macro-objectivation of the degeneracy in domain 
of quantum statistical mechanics, quantum field theory and in the quantum physics in general. 
We discuss the complementary unifying ideas of modern physics, namely: spontaneous symmetry breaking, 
quantum protectorate and emergence.
The interrelation of the concepts of symmetry breaking, quasiaverages and quantum protectorate 
was analyzed in the context of quantum theory and statistical physics.
The chief purposes of this paper were to demonstrate the connection and interrelation of these conceptual advances
of the many-body physics and to try to show explicitly that
those concepts, though different in details, have a certain common features.
Several problems  in the field  of    
statistical physics of  complex materials and systems (e.g. the chirality of molecules) and the foundations of the
microscopic theory of magnetism and superconductivity were discussed in relation to these ideas. 
\vspace{1cm}

%

%
%
\end{abstract}
\end{titlepage}
\newpage
\tableofcontents
\newpage
%
%
%
%
%
\section{Introduction}
%
%
There have been many interesting and important developments in statistical physics during the last decades. 
It is well known that symmetry principles play a crucial role in physics~\cite{weyl,wign,yang,yang91,yang03,bog66,bas90,rose}. 
The theory of symmetry is
a basic tool for understanding and formulating the fundamental notions of physics~\cite{ugo,baro}.
Symmetry considerations show that symmetry arguments are very powerful tool for bringing order into the
very complicated picture of the real world~\cite{ham62,symm79,group86,group94}.
As was rightly noticed by R. L. Mills,   "symmetry is a driving force in the shaping of physical theory"~\cite{rmi89}.
 According to D. Gross 
"the primary lesson of physics of this century is that the secret of nature is symmetry"~\cite{gross}.
Every symmetry leads to a conservation law~\cite{wign54,wign68,ryd06};  the well known examples are the conservation of
energy, momentum and electrical charge. A variety of other conservation laws can be deduced from symmetry
or invariance properties of the corresponding Lagrangian or  Hamiltonian of the system.  According to Noether
theorem, every continuous symmetry transformation under which the Lagrangian of a given system remains invariant
implies the existence of a conserved function~\cite{rose,group86}.\\
Many fundamental laws of physics in addition to their detailed features possess various symmetry properties. These
symmetry properties lead to  certain constraints and regularities on the possible properties of matter. 
Thus the  principles of symmetries  belong to the \textbf{underlying principles} of physics. 
Moreover, the idea of symmetry is a useful and workable tool for many areas of the quantum field theory, statistical 
physics and  condensed matter physics~\cite{group94,bir64,sene,jos91,chat08}. 
However, it is worth to stress the fact that all symmetry principles have an empirical basis.\\
The invariance
principles of nonrelativistic quantum mechanics~\cite{wign54,wign68,slg,col88,grei,itz} include those associated with space translations, space inversions, 
space rotations, Galilean transformations, and time reversal. In relation to these transformations the important
problem was to give a presentation in terms of the properties of the dynamical equations under appropriate 
coordinate transformations and to establish the relationship to certain contact transformations.\\
The developments in many-body theory and quantum field theory, in the theory of phase transitions,
and in the general theory of symmetry provided a new perspective. As it was emphasized by Callen~\cite{cal74,hbcal},
it appeared that symmetry considerations lie ubiquitously at the very roots of thermodynamic theory, so universally
and so fundamentally that they suggest a new conceptual basis. The interpretation which was 
proposed by Callen~\cite{cal74,hbcal}, suggests that thermodynamics is the study of those properties of macroscopic 
matter that follow from the symmetry properties of physical laws, mediated through the statistics of large systems.\\
In the many body problem and statistical mechanics one studies systems with infinitely many degrees of freedom.
Since actual systems are finite but large, it means that one studies a model which not only mathematically simpler than 
the actual system, but also allows a more precise formulation of phenomena such as phase transitions, transport processes, 
which are typical for macroscopic systems.  States not invariant under symmetries of the Hamiltonian are
of importance in many fields of physics~\cite{orig91,nnb85,pei91,pei92}.
In principle, it is necessary to clarify and generalize the notion of state of a system~\cite{robi67,hug68}, depending on the
algebra of observables $\mathcal{U}$. In the case of truly finite system the normal states are the most general states. However
all states in statistical mechanics are of the more general states~\cite{hug68}.
From this point of view a study of the automorphisms of $\mathcal{U}$ is of significance for a classification of
states~\cite{hug68}.
In  other words, the transformation $\Psi(\eta) \rightarrow \Psi(\eta) \exp (i \alpha)$ for all $\eta$ leaves
the commutation relations invariant. Gauge transformation define a one-parameter group of automorphisms. In most cases the 
three group of transformations, namely translation in space, evolution in time and gauge transformation, commute with
each other. Due to the quasi-local character of the observables one can prove that~\cite{hug68}
$$ \lim_{|x| \rightarrow \infty} ||[A_{\textbf{x}}, B]|| = 0.$$
It is possible to say therefore that the algebra $\mathcal{U}$ of observables is asymptotically abelian for space translation.
A state which is invariant with respect to translations in space and time we can call respectively homogeneous and 
stationary. If a state is invariant for gauge transformation we say that the state has a fixed particle number.\\
In physics, spontaneous symmetry breaking occurs when a system that is symmetric with respect to some symmetry 
group goes into a vacuum state that is not symmetric. When that happens, the system no longer appears to behave in 
a symmetric manner. It is a phenomenon that naturally occurs in many situations.
The symmetry group can be discrete, such as the space group of a crystal, or continuous (e.g., a Lie group), 
such as the rotational symmetry of space~\cite{ham62,symm79,group86,group94}. However if the system contains only a single spatial dimension then only 
discrete symmetries may be broken in a vacuum state of the full quantum theory, although a classical solution 
may break a continuous symmetry.
The problem of a great importance is to understand the domain of validity of the broken symmetry concept~\cite{pei91,pei92}.
It is of significance to understand is it valid only at low energies(temperatures) or it is universally applicable~\cite{moha93}.\\ 
Symmetries and  breaking of symmetries play an important role in  statistical 
physics~\cite{bb,drue69,lanf69,for75,cico81,asto,birt,fang08}, classical mechanics~\cite{sati80,cico81,asto,birt,fang08},
condensed matter physics~\cite{mich80,pwa84,scha08} and particle physics~\cite{namb,slg,col88,ait,huang07,lsb92,bs99,fstr05}.
Symmetry is a crucial concept in the theories that describe the subatomic world~\cite{orig91,nnb85} because it has an intimate connection 
with the laws of conservation. For example, the fact that physics is  invariant everywhere in 
the universe means that linear momentum is conserved. Some symmetries, such as rotational invariance, are perfect. 
Others, such as parity, are broken by small amounts, and the corresponding conservation law therefore only holds 
approximately.\\
 In  particle physics the natural question sounds  as what is it that determines the mass of a given particle 
and how is this mass related to the mass of other particles~\cite{wilc}. The partial answer to this question has been given within the frame work 
of a broken symmetry concept. For example, in order to
describe properly the $SU(2) \times U(1)$ theory in terms of electroweak interactions, it is necessary to
deduce how massive gauge quanta can emerge from a gauge-invariant theory. To resolve this problem, the
 idea of spontaneous symmetry breaking was used~\cite{namb,fstr05,scha08}. From the other hand, the application 
of the Ward identities reflecting the $U(1)_{em} \times SU(2)_{spin}$--gauge invariance of non-relativistic quantum 
mechanics~\cite{grei} leads to a variety of generalized quantized Hall effects~\cite{stud91,girv}.\\
 The mechanism of spontaneous symmetry breaking is usually understood as the mechanism
responsible for the occurrence of asymmetric states in quantum systems in the thermodynamic limit and is used
in various fields of quantum physics. However, broken symmetry concept can be used as well in classical 
physics~\cite{jona86}. It was
shown at Ref.~\cite{scha96} that starting from a standard description of an ideal, isentropic fluid, it was possible to
derive the effective theory governing a gapless non-relativistic mode -- the sound mode. The theory, which was dictated
by the requirement of Galilean invariance, entails the entire set of hydrodynamic equations. The gaplessness of the
sound mode was explained by identifying it as the Goldstone mode associated with the spontaneous breakdown of the
Galilean invariance. Thus the presence of sound waves in an isentropic fluid was explained as an \emph{emergent property}.\\
It is appropriate to note here that the emergent properties of matter were analyzed and discussed
by R. Laughlin and D. Pines~\cite{pnas,pines} from a general point of view (see also Ref.~\cite{wen05}). 
They introduced a unifying idea of \emph{quantum protectorate}.
This concept  belongs also to the underlying principles  of physics.   
The idea of quantum protectorate reveals the essential
difference in the behavior of the complex many-body systems at the low-energy
and high-energy scales.  The existence of two
scales, low-energy and high-energy, in the description
of physical phenomena is used in physics, explicitly or implicitly. 
 It is worth noting that standard thermodynamics and statistical mechanics are intended to describe
the properties of many-particle system at low energies, like the temperature and pressure of the gas.
For example, it was known for many years that a system in the low-energy limit can be
characterized by a small set of "collective" (or hydrodynamic)
variables and equations of motion corresponding
to these variables. Going beyond the framework of the low-energy region would require the consideration
of high-energy excitations.\\ 
It should be stressed that symmetry implies degeneracy. The greater the symmetry, the greater the degeneracy.
The study of the degeneracy of the energy levels plays a very important role in quantum physics. There is an additional aspect
of the degeneracy problem in quantum mechanics when  a system possess more subtle symmetries. This is the case when
degeneracy of the levels arises from the invariance of the Hamiltonian $H$ under groups involving simultaneous
transformation of coordinates and momenta that contain as subgroups the usual geometrical groups based on point
transformations of the coordinates. For these groups the free part of $H$ is not invariant, so that the symmetry is established only for
interacting systems. For this reason they are usually called dynamical groups. Particular case  is the hydrogen
atom~\cite{band,band2,gun67}, whose so-called \emph{accidental degeneracy} of the levels of given principal quantum number is due to the symmetry
of $H$ under the four-dimensional rotation group $O(4)$.\\
It is of importance to emphasize that when spontaneous symmetry breaking takes place, the ground state of the system
is degenerate.
Substantial progress in the understanding of the spontaneously broken symmetry concept is connected with
Bogoliubov's fundamental   ideas about quasiaverages~\cite{nnb61,bb,bs75,petr95}. 
Studies of degenerated systems led Bogoliubov in 1960-61 to the formulation of \textbf{the method of quasiaverages}.
This method has proved to be a universal tool for systems whose ground states become unstable under small
perturbations.
Thus the role of symmetry (and the breaking of symmetries) in combination 
with the degeneracy of the system was reanalyzed and essentially clarified by N. N. Bogoliubov in 1960-1961. He 
invented and formulated
a powerful innovative idea of \emph{quasiaverages} in statistical mechanics~\cite{nnb61,bb,bs75,petr95}.
The very elegant work of N. N. Bogoliubov on \emph{quasiaverages}~\cite{nnb61} has been
of great importance for a deeper understanding of phase  transitions, superfluidity and superconductivity, 
magnetism and other fields of equilibrium and nonequilibrium
statistical mechanics~\cite{nnb61,bb,bs75,petr95,zub71,kuz07}. The  Bogoliubov's idea  of \emph{quasiaverages} is an essential
conceptual advance of modern physics. \\
According to F. Wilczek~\cite{wil05},  "the primary goal of fundamental physics is to discover profound concepts that 
illuminate our understanding of nature".
The chief purposes of this paper are to demonstrate the connection and interrelation of three conceptual advances 
( or "profound concepts")
of the many-body physics, namely the broken symmetry, quasiaverages and quantum protectorate, 
and to try to show explicitly that
those concepts, though different in details, have a certain common features.
%
%
%
%
%
\section{Gauge Invariance}
%
%
An important class of symmetries  is the so-called dynamical symmetry. The symmetry of electromagnetic equation under
gauge transformation can be considered as a prototype of the class of dynamical symmetries~\cite{huang07}. The conserved quantity corresponding to gauge 
symmetry is the electric charge. 
A gauge transformation is a unitary transformation $U$ which produces a local phase change
\begin{equation} \label{}
U \phi(x) \rightarrow  e^{i \Lambda(x)}    \phi(x),
\end{equation}
where $\phi(x)$ is the classical local field describing a charged particle at point $x$. The phase factor $e^{i \Lambda(x)}$
is the representation of the one-dimensional unitary group $U(1)$.\\
F. Wilczek pointed out that ''gauge theories lie at the heart of modern formulation of the fundamental laws
of physics. The special characteristic of these theories is their extraordinary degree of symmetry, known as
gauge symmetry or gauge invariance''~\cite{wil89}.\\ 
The usual gauge transformation has the form
\begin{equation} \label{}
A_{\mu} \rightarrow  A_{\mu}' - \left (\partial/\partial x^{\mu} \right ) \lambda,
\end{equation}
where $\lambda$ is an arbitrary differentiable function from space-time to the real numbers, $\mu$ being 1, 2, 3, or 4.
If every component of $A$ is changed in this fashion,  the $\vec{E}$ and $\vec{B}$ vectors, which by Maxwell equations
characterize the electromagnetic field, are left unaltered, so therefore the field described by $A$ is equally
well characterized by $A'$.\\
Few conceptual advances in theoretical physics have been as exciting and influential as gauge invariance~\cite{or79,ok01}.
Historically, the definition of gauge invariance was originally introduced in the Maxwell theory of
electromagnetic field~\cite{huang07,yang06,brown91}. 
The introduction of potentials is a common procedure in dealing with problems in electrodynamics.
In this way Maxwell equations were rewritten in forms which are rather simple and more appropriate for analysis.
In this theory, common choices of gauge are $\vec{\nabla}\cdot \vec{A} = 0$, called the
Coulomb gauge.  There are many other gauges.  In general, it is necessary to select  the scalar gauge function $\chi(x,t)$ whose spatial and temporal
derivatives transform one set of electromagnetic potential into another equivalent set. A violation of gauge invariance
means that there are some parts of the potentials that do not cancel. For example, Yang and Kobe~\cite{yang86}
have used  the gauge dependence of the conventional interaction Hamiltonian to show that the conventional
interpretation of the quantum mechanical probabilities violates causality in those gauges with advanced potentials or 
faster-than-$c$ retarded potentials~\cite{yang76,yang76a}. Significance of electromagnetic potentials in the quantum theory
was demonstrated by Aharonov and Bohm~\cite{bohm59} in 1959 (see also Ref.~\cite{yang83}).\\
The gauge principle implies an invariance under internal symmetries performed independently
at different points of space and time~\cite{stro66}. The known example of gauge invariance is a change in phase of the
Schr\"{o}dinger wave function for an electron
\begin{equation} \label{}
\Psi(x,t) \rightarrow  e^{i q \varphi(x,t)/\hbar}    \Psi(x,t)
\end{equation}
In general, in quantum mechanics the wave function is complex, with a phase factor $\varphi(x,t)$. The phase
change varies from point to point in space and time. It is well known~\cite{stud91,girv}   that such phase changes
form a $U(1)$ group at each point of space and time, called the gauge group. The constant $q$ in the phase change is the
electric charge of the electron. It should be emphasized that not all theories of the gauge type can be 
internally consistent when quantum mechanics is fully taken into account.\\
Thus the gauge principle, which might also be described as a principle of local symmetry, is a statement about the
invariance properties of physical laws. It requires that every continuous symmetry be a local symmetry.
The concepts of local and global symmetry  are highly non-trivial. The operation of global symmetry  acts simultaneously 
on all variables of a system whereas the operation of local symmetry  acts independently on each variable.
Two known examples of phenomena that indeed
associated with local symmetries are electromagnetism (where we have a local $U(1)$ invariance), and gravity
(where the group of Lorenz transformations is replaced by general, local coordinate transformations).
According to D. Gross~\cite{gross}, "there is an essential difference between gauge invariance and global symmetry
such as translation or rotational invariance. Global symmetries are symmetries of the laws of nature... 
we search now for a synthesis of these two forms of symmetry [local and global],
a unified theory that contains both as a consequence of a greater and deeper symmetry,
of which these are the low energy remnants...".\\
There is the  general Elitzur's theorem~\cite{eli75}, which states that a spontaneous  breaking of local symmetry
for symmetrical gauge theory without gauge fixing is impossible. In other words,
local symmetry can never be broken and a non gauge invariant quantity never acquires nonzero vacuum expectation value.
This theorem was analyzed and refined in many papers~\cite{raif91,split03}. K. Splittorff~\cite{split03}
analyzed the impossibility of spontaneously breaking local symmetries and the sign problem.
Elitzur's theorem stating the impossibility of spontaneous breaking of local symmetries in a gauge theory was reexamined. 
The existing proofs of this theorem rely on gauge invariance as well as positivity of the weight in the Euclidean partition 
function. Splittorff examined the 
validity of Elitzur's theorem in gauge theories for which the Euclidean measure of the partition function is not 
positive definite. He found that Elitzur's theorem does not follow from gauge invariance alone. A  general criterion under 
which spontaneous 
breaking of local symmetries in a gauge theory is excluded was formulated.\\ 
Quantum field theory and the principle of gauge symmetry provide a theoretical framework for constructing 
effective models of systems consisting   of many particles~\cite{matsu95}  and condensed 
matter physics problems~\cite{klein}.
It was shown also recently~\cite{wang08} that the gauge symmetry principle inherent in Maxwell's electromagnetic theory
can be used in the efforts to reformulate general relativity into a gauge field theory.
The gauge symmetry principle has been  applied  in various forms to quantize gravity. \\
Popular unified theories of weak and electromagnetic interactions  are based on the notion of a spontaneously broken 
gauge symmetry. The hope has also been expressed by several authors that suitable generalizations of such theories 
may account for strong interactions as well. 
It was conjectured that the spontaneous breakdown of gauge symmetries may 
have a cosmological origin. As a consequence it was proposed that at some early stage of development of an expanding universe, 
a phase transition takes place. Before the phase transition, weak and electromagnetic interactions 
(and perhaps strong interactions too) were of comparable strengths. The presently observed differences 
in the strengths of the various interactions develop only after the phase transition takes place.\\
To summarize, the following 
  sentence of D. Gross  is appropriate for the case: "the most advanced form of 
symmetries we have understood are local symmetries --
general coordinate invariance and gauge symmetry. In contrast we do not believe that 
global symmetries are fundamental. Most global symmetries are approximate and even those
that, so far, have shown no sign of been broken, like baryon number and perhaps $C P T$, are
likely to be broken. They seem to be simply accidental features of low energy physics. Gauge
symmetry, however is never really broken -- it is only hidden by the asymmetric macroscopic
state we live in. At high temperature or pressure gauge symmetry will always be restored"~\cite{gross}. 
%
%
%
%
\section{Spontaneous Symmetry Breaking}
%
%
As it was mentioned earlier, a symmetry can be exact or approximate~\cite{orig91,pei91,pei92}. 
Symmetries inherent in the physical laws may be dynamically and spontaneously broken, i.e., they may not
manifest themselves in the actual phenomena. It can be as well broken by certain reasons.
C. N. Yang~\cite{yangcp83} pointed, that
non-Abelian gauge field become very useful in the second half of the twentieth century in the unified theory of electromagnetic and 
weak interactions, combined with symmetry breaking. Within the literature the 
term \emph{broken symmetry} is used both very often and with different meanings. There are two terms, the 
spontaneous breakdown of symmetries and dynamical symmetry breaking~\cite{dynam82}, which sometimes have been used as opposed but such a distinction is
irrelevant. According to Y. Nambu~\cite{namb}, the two terms may be used interchangeably. As it was mentioned previously,
a symmetry implies degeneracy. In general there are a multiplets of equivalent states related to each other by congruence
operations. They can be distinguished only relative to a weakly coupled external environment which breaks the
symmetry. Local gauged symmetries, however, cannot be broken this way because such an extended environment is not
allowed (a superselection rule), so all states are singlets, i.e., the multiplicities are not observable except possibly for
their global part. In other words, since a symmetry implies degeneracy of energy eigenstates, each multiplet of states forms
a representation of a symmetry group $G$. Each member of a multiplet is labeled by a set of quantum numbers for which one
may use the generators and Casimir invariants of the chain of subgroups, or  else some observables which form a
representation of $G$. It is a dynamical question whether or not the ground state, or the most stable state, is a singlet,
a most symmetrical one~\cite{namb}.\\
Peierls~\cite{pei91,pei92} gives a general definition of the notion of the spontaneous breakdown of symmetries which is
suited equally well for the physics of particles and condensed matter physics.   According to Peierls~\cite{pei91,pei92}, the term \emph{broken symmetries}
relates to situations in which symmetries which we expect to hold are valid only approximately or fail completely 
in certain situations.\\
The intriguing mechanism of spontaneous symmetry breaking is a unifying concept that lie at the basis of most 
of the recent developments in theoretical physics, from statistical mechanics to many-body theory and to elementary 
particles theory. 
It is known that when the Hamiltonian of a system is invariant under a symmetry operation, but the
ground state is not, the symmetry of the system can be spontaneously broken~\cite{group86}. 
Symmetry breaking is termed \emph{spontaneous} when there is no explicit term in a Lagrangian which
manifestly breaks the symmetry~\cite{bak62,bern74,grib}.\\
The existence of degeneracy in the energy states of a quantal system is related to the invariance 
or symmetry properties of the system. By applying the 
symmetry operation to the ground state, one can transform it to a different but equivalent ground state. Thus the
ground state is degenerate, and in the case of a continuous symmetry, infinitely degenerate. The real, or relevant,
ground state of the system can  only be one of these degenerate states. A system may exhibit the full symmetry of its
Lagrangian, but it is characteristic of infinitely large systems that they also may condense into states 
of lower symmetry. According to Anderson~\cite{and72}, this leads to an essential difference 
between infinite systems and finite systems. For infinitely extended systems a symmetric Hamiltonian can account 
for non symmetric behaviors, giving rise to non symmetric realizations of a physical system. \\
In terms of group theory~\cite{group86,ghab87,chan70}, it can be formulated that if for a specific problem in physics, we
can write down a basic set of equations which are invariant under a certain symmetry group $G$, then we would
expect that solutions of these equations would reflect the full symmetry of the basic set of equations. If for some
reason this is not the case, i.e., if there exists a solution which reflects some asymmetries with respect to the group
$G$, then we say that a spontaneous symmetry breaking has occurred. 
Conventionally one may describes a breakdown of symmetry by introducing a noninvariant term into the Lagrangian.
Another way of treating of this problem is to consider noninvariance under a group of transformations. It is known
from nonrelativistic many-body theory, that solutions of the field equations exist  that have less symmetry than that
displayed by the Lagrangian.\\
The breaking of the symmetry establishes a multiplicity of ''vacuums'' or ground states, related by the
transformations of the (broken) symmetry group~\cite{group86,ghab87,chan70}. 
What is important, it is that the broken symmetry state is distinguished by the appearance of a 
\emph{macroscopic order parameter}. The various values of the macroscopic order parameter are in a certain 
correspondence with the several ground states. Thus the problem arises how to establish the relevant ground state.
According to Coleman arguments~\cite{col88}, this ground state should exhibit the maximal lowering of the symmetry of
all its associated macrostates.\\
It is worth mentioning that the idea of spontaneously broken symmetries was invented and 
elaborated by  N. N. Bogoliubov~\cite{bb,nnb58,bts58,nnb60}, 
P. W. Anderson~\cite{pwa84,ander58,ander75}, Y. Nambu~\cite{namb07,namb09},  G. Jona-Lasinio and others.
This idea was applied to the elementary particle physics by Nambu in his 
 1960 article~\cite{namb60} (see also Ref.~\cite{namDok}).  Nambu was guided in his work by an analogy with the theory of
superconductivity~\cite{nnb58,bts58,nnb60}, to which Nambu himself had made important contribution~\cite{ynamb60}. 
According to Nambu~\cite{ynamb60,namb84},
the situation  in the elementary particle physics  may be understood better by making  an analogy to the theory of superconductivity originated by
Bogoliubov~\cite{nnb58} and  Bardeen, Cooper and Schrieffer~\cite{bcs57}. There gauge invariance, 
the energy gap, and the collective excitations were logically related to each other. This analogy was the leading idea
which stimulated him greatly.
A model with a broken gauge symmetry has been discussed by Nambu and Jona-Lasinio~\cite{namb61}. This model starts 
with a zero-mass baryon and a massless pseudoscalar meson, accompanied by a broken-gauge symmetry. 
The authors considered a theory with a Lagrangian possessing $\gamma_{5}$ invariance and found that, although the
basic Lagrangian contains no mass term, since such terms violate $\gamma_{5}$ invariance, a solution exists that admits
fermions of finite mass.\\ 
The appearance of spontaneously broken symmetries and its bearing on the physical mass spectrum were analyzed in 
variety of papers~\cite{wilc,fwil05,wein04,hoof07,mass07}.
Kunihiro and  Hatsuda~\cite{kuni84}  elaborated
a self-consistent mean-field approach to the dynamical symmetry breaking by considering
the effective potential of the Nambu and Jona-Lasinio model. In their study
the dynamical symmetry breaking phenomena in the Nambu and Jona-Lasinio model were reexamined in the 
framework of a self-consistent mean-field (SCMF) theory. They formulated the SCMF theory in a lucid manner 
based on a successful decomposition on the Lagrangian into semiclassical and residual interaction parts by imposing a 
condition that "the dangerous term" in Bogoliubov's sense~\cite{nnb58} should vanish. It was shown that the difference of the energy 
density between the super and normal phases, the correct expression of which the original authors failed to give, can 
be readily  obtained by applying the SCMF theory. Furthermore, it was shown that the expression thus 
obtained is identical to that 
of the effective potential  given by the path-integral method with an auxiliary field up to the one loop 
order in the loop expansion, then one finds a new and simple way to get the effective potential. Some numerical results of 
the effective potential and the dynamically generated mass of fermion were also obtained.\\ 
The concept of spontaneous symmetry breaking is delicate. It is worth to emphasize that it can never take place
when the normalized ground state $|\Phi_{0} \rangle$ of the many-particle Hamiltonian (possibly interacting) is
non-degenerate, i.e., unique up to a phase factor. Indeed, the transformation law of the ground
state $|\Phi_{0}\rangle $ under any symmetry of the Hamiltonian must then be multiplication by a phase factor.
Correspondingly, the ground state $|\Phi_{0}\rangle $ must transforms according to the trivial representation of
the symmetry group, i.e., $|\Phi_{0} \rangle$ transforms as a singlet. In this case there is no room for the phenomenon
of spontaneous symmetry breaking by which the ground state transforms non-trivially
under some symmetry group of the Hamiltonian. Now, the Perron-Frobenius theorem for finite dimensional
matrices with positive entries or its extension   to single-particle Hamiltonians of the
form 
$H = - \Delta/2m + U(r)$
guarantees that the ground state is non-degenerate for non-interacting $N$-body Hamiltonians
defined on the Hilbert space $\bigotimes^{N}_{symm}\mathcal{H}^{(1)}$.
 Although there is no rigorous proof that the
same theorem holds for interacting $N$-body Hamiltonians, it is believed that the ground state of
interacting Hamiltonians defined on  
$\bigotimes^{N}_{symm}\mathcal{H}^{(1)}$ is also unique. It is believed also
that spontaneous symmetry breaking is always ruled out for interacting Hamiltonian defined on
the Hilbert space $\bigotimes^{N}_{symm}\mathcal{H}^{(1)}$.\\
Explicit symmetry breaking indicates a situation where the dynamical equations are not manifestly invariant under 
the symmetry group considered. This means, in the Lagrangian (Hamiltonian) formulation, that the Lagrangian (Hamiltonian) 
of the system contains one or more terms explicitly breaking the symmetry. Such terms, in general,  can have different 
origins. Sometimes
symmetry-breaking terms may be introduced into the theory by hand on the basis of theoretical or experimental results, 
as in the case of the quantum field theory of the weak interactions.  This theory was  constructed in a way that 
manifestly violates mirror symmetry or parity. The underlying result in this case is parity non-conservation in the 
case of the weak interaction, as it was formulated  by T. D. Lee and C. N. Yang.
It  may be of interest to  remind in this context  the general principle, formulated 
by C. N. Yang~\cite{yangcp83}:
''\textbf{symmetry dictates interaction}''.\\
C. N. Yang~\cite{yangcp83} noted also that, "the lesson we have learned from it that  keeps as much symmetry  as possible.
Symmetry is good for renormalizability \ldots The concept of broken symmetry does not really break the symmetry, it is
only breaks the symmetry phenomenologically. So the broken symmetric non-Abelian gauge field theory keeps formalistically
the symmetry. That is  reason why it is renormalizable. And that produced unification of electromagnetic and 
weak  interactions".\\
In fact,
the symmetry-breaking terms may appear because of non-renormalizable effects. 
One can think of current renormalizable field theories as effective field theories, which may be a sort of 
low-energy approximations to a more general theory. 
 The effects of non-renormalizable interactions  
are,  as a rule,  not big and can therefore be ignored at the low-energy regime. 
In this sense   the coarse-grained description thus obtained may possess 
 more symmetries than the anticipated general  theory. That is, the effective Lagrangian obeys symmetries that 
 are not symmetries of the underlying theory. Weinberg has called them the "accidental" symmetries.
They  may then be violated by the non-renormalizable terms arising from higher mass 
scales and suppressed in the effective Lagrangian.\\
R. Brout and F. Englert has reviewed~\cite{brout07} the concept of spontaneous broken symmetry
in the presence of global symmetries both in matter and particle 
physics. This concept was then taken over to confront local symmetries in relativistic field theory. Emphasis was placed 
on the basic concepts where, in the former case, the vacuum of spontaneous broken symmetry was degenerate whereas that 
of local (or gauge) symmetry was gauge invariant.\\ 
The notion of broken symmetry permits one to look more deeply at many complicated problems~\cite{pei91,pei92,koss,zur08}, such as
scale invariance~\cite{drago87}, stochastic interpretation of quantum mechanics~\cite{java},
quantum measurement problem~\cite{mori06} and many-body nuclear physics~\cite{bely06}.
The problem of a great importance is to understand the domain of validity of the broken symmetry concept.
Is it valid only at low energies (temperatures) or it is universally applicable.\\
In spite of the fact that the term \emph{spontaneous symmetry breaking} was coined in elementary particle physics to describe the
situation that the vacuum state had less symmetry than the group invariance of the equations, this notion is of use
in classical mechanics where it  arose in bifurcation theory~\cite{sati80,cico81,asto,birt,fang08}. The physical systems on the
brink of instability are described by the new solutions which appear often possess a lower isotropy symmetry group.
The governing equations themselves continue to be invariant under the full transformation group and that is the reason
why the symmetry breaking is spontaneous.\\ 
These results are of value for the nonequilibrium systems~\cite{prig80,prig82}.
Results in nonequilibrium thermodynamics have shown that bifurcations
require two conditions. First, systems have to be far from equilibrium.
We have to deal with open systems exchanging energy, matter and information
with the surrounding world. Secondly, we need non-linearity. This
leads to a multiplicity of solutions. The choice of the
branch of the solution in the non-linear problem depends on probabilistic
elements. Bifurcations provide a mechanism for the appearance of novelties
in the physical world. In general, however, there are successions of
bifurcations, introducing a kind of memory aspect. It is now generally well
understood that all structures around us are the specific outcomes of such
type of processes. The simplest example is the behavior of chemical reactions
in far-from-equilibrium systems. These conditions may lead to oscillating
reactions, to so-called Turing patterns, or to chaos in which initially
close trajectories deviate exponentially over time. The main point is that, for
given boundary conditions (that is, for a given environment), allowing us to
change of perspective is mainly due to our progress in dynamical systems
and spectral theory of operators. \\ 
J. van Wezel, J. Zaanen and J. van den Brink~\cite{zaa05} studied
an intrinsic limit to quantum coherence due to spontaneous symmetry breaking.
They investigated the influence of spontaneous symmetry breaking on the decoherence of a many-particle
quantum system. This decoherence process was analyzed in an exactly solvable model system that is known
to be representative of symmetry broken macroscopic systems in equilibrium. It was shown that spontaneous
symmetry breaking imposes a fundamental limit to the time that a system can stay quantum coherent. This
universal time scale is $t_{spon} \sim  2 \pi N \hbar/(k_{B}T)$, given in terms of the number of microscopic degrees of
freedom $N$, temperature $T$, and the constants of Planck $(\hbar)$ and Boltzmann $(k_{B})$.
According to their viewpoint, the relation between quantum physics at
microscopic scales and the classical behavior of macroscopic
bodies need a thorough study.
This subject has revived in recent years
both due to experimental progress, making it possible to
study this problem empirically, and because of its possible
implications for the use of quantum physics as a computational
resource. This ''micro-macro'' connection actually
has two sides. Under equilibrium conditions it is
well understood in terms of the mechanism of spontaneous
symmetry breaking. But in the dynamical realms its precise
nature is still far from clear. The question is ''Can spontaneous
symmetry breaking play a role in a dynamical reduction
of quantum physics to classical behavior?'' This is
a highly nontrivial question as spontaneous symmetry
breaking is intrinsically associated with the difficult problem
of many-particle quantum physics. Authors analyzed a
tractable model system, which is known to be representative
of macroscopic systems in equilibrium, to find the
surprising outcome that spontaneous symmetry breaking
imposes a fundamental limit to the time that a system can
stay quantum coherent.\\
In the next work~\cite{zaa06} J. van Wezel, J. Zaanen and J. van den Brink
studied a relation between decoherence and spontaneous symmetry breaking in many-particle qubits.
They used the fact  that  spontaneous symmetry breaking can lead to
decoherence on a certain time scale 
and that there is a limit to quantum coherence in many-particle spin qubits due to
spontaneous symmetry breaking. These results were derived for the Lieb-Mattis spin model. Authors shown
that the underlying mechanism of decoherence in systems with spontaneous symmetry breaking is in fact more
general. J. van Wezel, J. Zaanen and J. van den Brink presented here a generic route to finding the decoherence time 
associated with spontaneous symmetry
breaking in many-particle qubits, and subsequently  applied this approach to two model systems, indicating
how the continuous symmetries in these models are spontaneously broken. They discussed the relation of
this symmetry breaking to the thin spectrum.\\ 
The number of works on broken symmetry within the axiomatic frame is large;
this topic was reviewed by Reeh~\cite{reeh} and many others.
%
%
%
\section{Goldstone Theorem}
%
%
The Goldstone theorem~\cite{gold61} is remarkable in so far it connects the phenomenon of spontaneous breakdown of an 
internal symmetry with a property of the mass spectrum. In addition the Goldstone theorem states that breaking of 
global continuous symmetry implies the existence of massless, spin-zero bosons. The
presence of massless particles accompanying broken gauge symmetries seems to be quite general~\cite{blud63}.
The \emph{Goldstone theorem} states that, if system described by a Lagrangian which
has a continuous symmetry (and only short-ranged interactions)  has a broken symmetry state then the system
support a branch of small amplitude excitations with a dispersion relation $\varepsilon(k)$ that vanishes 
at $k \rightarrow 0$. Thus the Goldstone theorem ensures 
the existence of massless excitations if a continuous symmetry is spontaneously broken.\\ 
A more precisely, the Goldstone theorem examines a generic continuous symmetry which is spontaneously broken, i.e., its currents are conserved, 
but the ground state (vacuum) is not invariant under the action of the corresponding charges. Then, necessarily, new 
massless (or light, if the symmetry is not exact) scalar particles appear in the spectrum of possible excitations. There 
is one scalar particle - called a Goldstone boson  (or Nambu-Goldstone boson).
In particle and condensed matter physics, Goldstone bosons  are bosons that 
appear in models exhibiting spontaneous breakdown of continuous symmetries~\cite{burg1,burg2}.
Such a particle can be ascribed for each generator of the symmetry that is broken, i.e., that 
does not preserve the ground state. The Nambu-Goldstone mode is a long-wavelength fluctuation of the corresponding order 
parameter.\\ 
In other words,   zero-mass excitations always appear
when a gauge symmetry is broken~\cite{blud63,salam,weinb94,gura64,kib67}. 
Some (incomplete) proofs of the initial Goldstone "conjecture"  on the
massless particles required by symmetry breaking were worked out by
Goldstone, Salam and Weinberg~\cite{salam}. As S. Weinberg~\cite{weinb94} formulated it later, ''as everyone knows now, 
broken global symmetries in general don't look at all like approximate ordinary symmetries, but show up instead as low
energy theorems for the interactions of these massless Goldstone bosons''. 
 These spinless bosons correspond to the spontaneously broken internal symmetry generators, and are characterized by the 
quantum numbers of these. They transform nonlinearly (shift) under the action of these generators, and can thus be excited 
out of the asymmetric vacuum by these generators. Thus, they can be thought of as the excitations of the field in the broken 
symmetry directions in group space   and are massless if the spontaneously broken symmetry is not also broken explicitly. 
In the case of approximate symmetry,  i.e., if it is explicitly broken as well as spontaneously broken, then the
Nambu-Goldstone bosons are not massless, though they typically remain relatively light~\cite{kovn91}.\\
In paper~\cite{gilb64} a clear statement and proof of Goldstone theorem was carried out. It was shown that any 
solution of a Lorenz-invariant
theory (and of some other theories also) that violates an internal symmetry of the theory will contain
a massless scalar excitation  i.e., particle (see also Refs.~\cite{str65,fu65,gura09}).\\
The Goldstone theorem has applications in many-body nonrelativistic quantum theory~\cite{lang65,lange,wagn66}. In that case it states
that if symmetry is spontaneously broken, there are excitations (Goldstone excitations) whose frequency vanishes
$(\varepsilon(k) \rightarrow 0)$ in the long-wavelength limit $(k \rightarrow 0)$. In these cases we similarly
have that the ground state is degenerate. Examples are the isotropic ferromagnet in which the Goldstone excitations
are spin waves, a Bose gas in which the breaking of the phase symmetry $\psi \rightarrow \exp (i \alpha) \psi$
and of the Galilean invariance implies the existence of phonons as Goldstone excitations, and a crystal where
breaking of translational invariance also produces phonons. 
Goldstone theorem was applied also to a number of nonrelativistic many-body systems~\cite{lange,wagn66} and the question has arisen
as to whether such systems as a superconducting electron gas and an electron plasma which have an energy gap
in their spectrum (analog of a nonzero mass for a particle) are not a violation of the Goldstone theorem.
An inspection the situation in which the system is coupled by long-ranged interactions, as modelled by an
electromagnetic field leads to a better understanding of the limitations of Goldstone theorem. As first pointed
out by Anderson~\cite{pwa58,ander63}, the long-ranged interactions alter the excitation spectrum of the symmetry broken state by
removing the Goldstone modes and generating a branch of massive excitations (see also Refs.~\cite{weinb86,weinb08}).\\
It is worth to note that S. Coleman~\cite{scol73} proved that in two dimensions the Goldstone phenomenon can not occur.
This is related with the fact that 
in four dimensions, it is possible for a scalar field to have a vacuum expectation
value that would be forbidden if the vacuum were invariant under some continuous
transformation group, even though this group is a symmetry group in the sense that the
associated local currents are conserved. This is the Goldstone phenomenon, and Goldstone's
theorem states that this phenomenon is always accompanied by the appearance of
massless scalar bosons.  In two dimensions
Goldstone's theorem does not end with two
alternatives (either manifest symmetry or Goldstone bosons) but with only one (manifest
symmetry).\\
There are many extensions and generalizations of the Goldstone theorem~\cite{kast66,buch92}.
L. O'Raifeartaigh~\cite{raif91}  has
shown that the Goldstone theorem is actually a special case of the Noether theorem in the presence of spontaneous symmetry 
breakdown, and is thus immediately valid for quantized as well as classical fields. The situation when gauge fields are 
introduced was discussed as well. Emphasis being placed on some points that are not often discussed in the literature 
such as the compatibility of the Higgs mechanism and the Elitzur theorem~\cite{eli75} and the extent to which the vacuum configuration 
is determined by the choice of gauge.
A. Okopinska~\cite{okop96}  have shown that the Goldstone theorem is fulfilled in the $O(N)$  symmetric scalar quantum field theory
with  $\lambda \Phi^{4}$ interaction in the Gaussian approximation for arbitrary $N$.
Chodos and Gallatin~\cite{chod01} pointed
out that standard discussions  of Goldstone's theorem were based on a symmetry of the action assume constant fields and global 
transformations, i.e., transformations which are independent of space-time coordinates. By allowing for arbitrary 
field distributions in a general representation of the symmetry they derived a generalization of the standard Goldstone's 
theorem. When applied to gauge bosons coupled to scalars with a spontaneously broken symmetry the generalized theorem 
automatically imposes the Higgs mechanism, i.e., if the expectation value of the scalar field is nonzero then the gauge 
bosons must be massive. The other aspect of the Higgs mechanism, the disappearance of the "would be" Goldstone boson, 
follows directly from the generalized symmetry condition itself. They also used the generalized Goldstone's theorem to 
analyze the case of a system in which scale and conformal symmetries were both spontaneously broken.
The consistency between the Goldstone theorem and the Higgs mechanism was established in a manifestly covariant way by
N. Nakanishi~\cite{naka75}.  
%
%
%
\section{Higgs Phenomenon}
%
The most characteristic feature of spontaneously broken gauge theories is the Higgs mechanism~\cite{ph1,ph2,ph3,ph07}. It 
is that mechanism through which the Goldstone fields disappear and gauge fields acquire masses~\cite{mass07,bern74,kle64,lee73}. 
When spontaneous symmetry breaking takes place in theories with local symmetries, then the zero-mass 
Goldstone bosons combine with the vector gauge bosons to form massive vector particles. Thus in a situation
of spontaneous broken  local symmetry, the gauge boson gets its mass from the interaction of gauge bosons
with the spin-zero bosons.\\
The mechanism proposed by Higgs for the
elimination, by symmetry breakdown, of zero-mass quanta of gauge fields have led to a substantial progress in the
unified theory of particles and interactions. The Higgs mechanism could explain, in principle, the fundamental particle
masses in terms of the energy interaction between particles and the Higgs field.\\
P. W. Anderson~\cite{pwa84,ander58,ander75,pwa58,ander63} first pointed out that 
several cases in nonrelativistic condensed matter physics may be interpreted as due to \emph{massive photons}.
It was  Y. Nambu~\cite{namb09} who pointed clearly that the idea of a spontaneously broken symmetry being the way 
in which the mass of particles could be generated. He used an analogy of a theory of elementary particles
with the Bogoliubov-BCS theory of superconductivity. Nambu  showed how fermion masses would be generated in a way that was analogous to 
the formation of the energy gap in a superconductor. In 1963, P. W. Anderson~\cite{ander63} shown that the 
equivalent of a Goldstone boson in a superconductor could become massive due to its electromagnetic interactions.
Higgs was able to show that the introduction of a subtle form of symmetry known as gauge 
invariance invalidated some of the assumptions made by Goldstone, Salam and Weinberg in their paper~\cite{salam}.
Higgs formulated a theory in which 
there was one massive spin-one particle - the sort of particle that can carry a force - and one left-over massive 
particle that did not have any spin. 
Thus he invented a new type of particle, which was called later by the Higgs boson. 
The so-called Higgs mechanism is the mechanism of generating vector boson masses; it was big breakthrough in the field
of particle physics.\\
According to F. Wilczek~\cite{wil05}
''BCS theory traces superconductivity to the existence of a special sort of long-range
correlation among electrons. This effect is purely quantum-mechanical. A classical phenomenon that is only very roughly
analogous, but much simpler to visualize, is the occurrence of ferromagnetism
owing to long-range correlations among electron
spins (that is, their mutual alignment in a single direction). The
sort of correlations responsible for superconductivity are of a
much less familiar sort, as they involve not the spins of the electrons,
but rather the phases of their quantum-mechanical wavefunctions \ldots
 But as it is the leading idea
guiding our construction of the Higgs system, I think it is appropriate
to sketch an intermediate picture that is more accurate than
the magnet analogy and suggestive of the generalization required
in the Higgs system. Superconductivity
occurs when the phases of the Cooper pairs all align in the
same direction\ldots
Of course, gauge transformations
that act differently at different space-time points will spoil
this alignment. Thus, although the basic equations of electrodynamics
are unchanged by gauge transformations, the state of a
superconductor does change. To describe this situation, we say that in
a superconductor gauge symmetry is spontaneously broken.
The phase alignment of the Cooper pairs gives them a form of rigidity.
Electromagnetic fields, which would tend to disturb this alignment, are
rejected. This is the microscopic explanation of the Meissner effect, or
in other words, the mass of photons in superconductors.''\\
The theory of the strong interaction between quarks (quantum chromodynamics, \emph{QCD})~\cite{huang07}  is approximately invariant under what 
is called charge symmetry. In other words, if we swap an up quark for a down quark, then the strong interaction will 
look almost the same. This symmetry is related to the concept of isospin, and is not the same as charge conjugation 
(in which a particle is replaced by its antiparticle).
Charge symmetry is broken by the competition between two different effects. The first is the small difference in mass 
between up and down quarks, which is about 200 times less than the mass of the proton. The second is their different 
electric charges. The up quark has a charge of $+2/3$ in units of the proton charge, while the down quark has a negative 
charge of $-1/3$.
If the Standard Model of particle 
physics~\cite{huang07,wein04,hoof07} were perfectly symmetric, none of the particles in the model would have any mass. Looked at another way, 
the fact that most fundamental particles have non-zero masses breaks some of the symmetry in the model. Something must 
therefore be generating the masses of the particles and breaking the symmetry of the model. That something - which has 
yet to be detected in an experiment - is called the Higgs field.
The origin of the quark masses is not fully understood. In the Standard Model of particle physics~\cite{huang07,wein04,hoof07}, 
the Higgs mechanism allows the generation of such masses but it cannot predict the actual mass values. 
No fundamental understanding of the mass hierarchy exists. It is clear that the violation of charge symmetry can  be 
used to threat this problem.\\
C. Smeenk~\cite{sme06} called
the Higgs mechanism as an essential but elusive component of the Standard Model of particle physics. 
In his opinion without it Yang--Mills gauge theories would have been little more than a warm-up exercise in the attempt to 
quantize gravity rather than serving as the basis for the Standard 
Model. C. Smeenk focuses on two problems related to the Higgs mechanism, namely:   
i) what is the gauge-invariant content of the Higgs mechanism, 
and ii) what does it mean to break a local gauge symmetry?\\
A more critical view was presented by H. Lyre~\cite{lyre}. He
explored the argument structure of the concept of spontaneous symmetry breaking in the electroweak gauge theory 
of the Standard Model: the so-called Higgs mechanism. As commonly understood, the Higgs argument is designed to 
introduce the masses of the gauge bosons by a spontaneous breaking of the gauge symmetry of an additional field, 
the Higgs field. H. Lyre claimed that the technical derivation of the Higgs mechanism, however, 
consists in a mere re-shuffling of degrees of freedom by transforming the Higgs Lagrangian in a gauge-invariant manner. 
In his opinion, this already raises serious doubts about the adequacy of the entire manoeuvre. He insist  that no 
straightforward ontic interpretation of the Higgs mechanism was tenable since gauge transformations possess no real instantiations. 
In addition, the explanatory value of the Higgs argument was critically examined in that open to question paper.
%
\section{Chiral Symmetry} 
%
Many symmetry principles were known,  a large fraction of them were only approximate. 
The concept of chirality was introduced in the nineteenth century when
 L. Pasteur discovered one of the most interesting and enigmatic asymmetries 
in nature: that the chemistry of life shows a preference for molecules with a particular \emph{handedness}.
 Chirality is a general concept based on the geometric characteristics
of an object. A chiral object is an object which has a mirror-image non superimposable to itself.
Chirality deals with molecules but also with macroscopic objects such as crystals. 
Many chemical and physical systems can occur in two forms distinguished solely by being mirror images of each other. 
This phenomenon, known as chirality, is important in biochemistry~\cite{bon98,meier08}, where reactions involving chiral molecules often require 
the participation of one specific enantiomer (mirror image) of the two possible ones. Chirality 
is an important concept~\cite{petit03} which has
many consequences and applications in many fields of science~\cite{bon98,gold1,gold2,chir09} and especially in 
chemistry~\cite{kond83,kond90,kond06,grid06,kond07}.\\
The problem of homochirality has attracted attention of chemists and physicists since it
was found by Pasteur.  The methods of solid-state physics and statistical thermodynamics were
of use to study this complicated interdisciplinary problem~\cite{kond83,kond90,kond06,kond07}.
A general theory of spontaneous chiral symmetry breaking in chemical systems has been 
formulated by D. Kondepudi~\cite{kond83,kond90,kond06,kond07}. 
The fundamental equations of this theory depend only on the two-fold mirror-image symmetry and not on the details of the 
chemical kinetics. Close to equilibrium, the system will be in a symmetric state in which the amounts of the two 
enantiomers of all chiral molecules are equal. When the system is driven away from equilibrium by a flow of chemicals, 
a point is reached at which the system becomes unstable to small fluctuation in the difference in the amount of the two 
enantiomers. As a consequence, a small random fluctuation in the difference in the amount of the two enantiomers 
spontaneously grows and the system makes a transition to an asymmetric state. The general theory describes this phenomenon 
in the vicinity of the transition point.\\
Amino acids and DNA are the fundamental building blocks of life itself~\cite{bon98,meier08}. They exist in left- and right-handed 
forms that are mirror images of one another. Almost all the naturally occurring amino acids that make up proteins are 
left-handed, while DNA is almost exclusively right-handed~\cite{meier08}.
Biological macromolecules, proteins and nucleic acids are composed exclusively of chirally pure
monomers. The chirality consensus~\cite{salla} appears vital for life and it has even been considered as a prerequisite 
of life.
However the primary cause for the ubiquitous handedness has remained obscure yet. It was conjectured~\cite{salla} that 
the chirality
consensus is a kinetic consequence that follows from the principle of increasing entropy, i.e. the 2nd law of
thermodynamics. Entropy increases when an open system evolves by decreasing gradients in free energy with
more and more efficient mechanisms of energy transduction. The rate of entropy increase can be considered as the 
universal fitness
criterion of natural selection that favors diverse functional molecules and drives the system to the chirality
consensus to attain and maintain high-entropy non-equilibrium states. 
Thus the chiral-pure outcomes have emerged from certain scenarios and understood as consequences of
kinetics~\cite{salla}. It was pointed out
that the principle of increasing entropy, equivalent to
diminishing differences in energy, underlies all kinetic
courses and thus could be a cause of chirality consensus.
Under influx of external energy systems evolve to high entropy
non-equilibrium states using mechanisms of energy
transduction. The rate of entropy increase is the universal
fitness criterion of natural selection among the diverse
mechanisms that favors those that are most effective in
leveling potential energy differences. The ubiquitous
handedness enables rapid synthesis of diverse metastable
mechanisms to access free energy gradients to attain and
maintain high-entropy non-equilibrium states. When the
external energy is cut off, the energy gradient from the
system to its exterior reverses and racemization will
commence toward the equilibrium. Then the mechanisms of
energy transduction have become improbable and will
vanish since there are no gradients to replenish them.
The common consent that a racemic mixture has higher
entropy than a chirally pure solution is certainly true at the
stable equilibrium. Therefore high entropy is often
associated with high disorder. However entropy is not an
obscure logarithmic probability measure but probabilities
describe energy densities and mutual gradients in energy~\cite{salla}.
The local order and structure that associate with the
mechanisms of energy transduction are well warranted when
they allow the open system as a whole to access and level
free energy gradients. Order and standards are needed to
attain and maintain the high-entropy non-equilibrium states.
We expect that the principle of increasing entropy accounts
also for the universal genetic code to allow exchange of
genetic material to thrust evolution toward new more
probable states. The common chirality convention is often
associated with a presumed unique origin of life but it
reflects more the all-encompassing unity of biota on Earth
that emerged from evolution over the eons~\cite{salla}.\\
Many researchers have pointed on the role of the magnetic field for the chiral asymmetry.
 Recently G. Rikken and E. Raupach  have 
demonstrated that a static magnetic field can indeed generate chiral asymmetry~\cite{rikk}. 
Their work reports the first unequivocal use of a static magnetic field to bias a chemical process in favour of one of 
two mirror-image products (left- or right-handed enantiomers).
G. Rikken and E. Raupach used the fact that terrestrial life utilizes 
only the $L$ enantiomers of amino acids, a pattern that is known as the '\emph{homochirality of life}' and which has 
stimulated long-standing efforts to understand its origin. Reactions can proceed enantioselectively if chiral reactants or 
catalysts are involved, or if some external chiral influence is present. But because chiral reactants and catalysts 
themselves require an enantioselective production process, efforts to understand the homochirality of life have focused on 
external chiral influences. One such external influence is circularly polarized light, which can influence the 
chirality of photochemical reaction products. Because natural optical activity, which occurs exclusively in media 
lacking mirror symmetry, and magnetic optical activity, which can occur in all media and is induced by longitudinal 
magnetic fields, both cause polarization rotation of light, the potential for magnetically induced enantioselectivity in 
chemical reactions has been investigated, but no convincing demonstrations of such an effect have been found. The authors 
shown experimentally that magnetochiral anisotropy - an effect linking chirality and magnetism - can give rise to an 
enantiomeric excess in a photochemical reaction driven by unpolarized light in a parallel magnetic field, which suggests 
that this effect may have played a role in the origin of the homochirality of life.
These results 
clearly suggest that there could be a difference between the way the two types of amino acids break down in a strong 
interstellar magnetic field. A small asymmetry produced this way could be amplified through other chemical reactions 
to generate the large asymmetry observed in the chemistry of life on Earth.\\
Studies of chiral crystallization~\cite{kond93} of achiral molecules are of importance for the clarification of the nature of
chiral symmetry breaking. The  study of chiral crystallization of
achiral molecules  focuses on chirality of crystals and more specifically on chiral symmetry
breaking for these crystals. 
Some molecules, although achiral, are able to generate chiral crystals. Chirality is then due to the
crystal structure, having two enantiomorphic forms. Cubic chiral crystals are easily identifiable. Indeed, they 
deviate polarized light. The distribution and the ratio of the two enantiomorphic
crystal forms of an achiral molecule not only in a sample but also in numerous samples
prepared under specific conditions. The
relevance of this type of study is, for instance,
a better comprehension of homochirality. The
experimental conditions act upon the breaking
of chiral symmetry. Enantiomeric excess is not
obviously easy to induce. Nevertheless, a constant
stirring of the solution during the crystallization
will generate a significant rupture of
chiral symmetry in the sample and can offer an
interesting and accessible case study~\cite{kond93}.\\
The discovery of  L. Pasteur
came about 100 years before physicists demonstrated that processes governed by weak-force interactions look different in a 
mirror-image world. The chiral symmetry breaking has been observed in various physical problems, e.g.
chiral symmetry breaking of magnetic vortices, caused by the surface roughness of thin-film magnetic structures~\cite{vans09}. 
Charge-symmetry breaking also manifests itself in the interactions of pions with protons and neutrons in a very interesting 
way that is linked to the neutron-proton (and hence, up and down quark) mass difference. Because the masses of the up 
and down quarks are almost zero, another approximate symmetry of QCD called \emph{chiral}  symmetry 
comes into play~\cite{lsb92,oku85,esk99,cre01,sim07}. 
This symmetry relates to the spin angular momentum of fundamental particles. Quarks can either be \emph{right-handed}  
or \emph{left-handed}, depending on whether their spin is clockwise or anticlockwise with respect to the direction they are 
moving in. Both of these states are treated approximately the same by QCD.\\
Symmetry-breaking terms may appear in the theory because of quantum-mechanical effects. One reason for the presence 
of such terms - known as  \emph{anomalies}  - is that in passing from the classical to the quantum level, because of 
possible operator ordering ambiguities for composite quantities such as Noether charges and currents, it may be that 
the classical symmetry algebra (generated through the Poisson bracket structure) is no longer realized in terms of the 
commutation relations of the Noether charges. Moreover, the use of a regulator  (or  \emph{cut-off} ) required in the 
renormalization procedure to achieve actual calculations may itself be a source of anomalies. It may violate a symmetry 
of the theory, and traces of this symmetry breaking may remain even after the regulator is removed at the end of 
the calculations. Historically, the first example of an anomaly arising from renormalization is the so-called chiral 
anomaly, that is the anomaly violating the chiral symmetry of the strong interaction~\cite{lsb92,esk99,cre01,funa88}.\\
Kondepudi and Durand~\cite{kond01} applied  the ideas of chiral symmetry to astrophysical problem.
They considered the so-called chiral asymmetry in spiral galaxies.
Spiral galaxies are chiral entities when coupled with the direction of their recession velocity. 
As viewed from the Earth, the $S$-shaped and $Z$-shaped spiral galaxies are two chiral forms. 
The authors investigated what is the nature of chiral symmetry in spiral galaxies. 
In the Carnegie Atlas of Galaxies that lists photographs of a total of 1,168 galaxies, there are 540 galaxies, 
classified as normal or barred spirals, that are clearly identifiable as $S$- or $Z$- type. The recession velocities 
for 538 of these galaxies could be obtained from this atlas and other sources. A statistical analysis of this sample 
reveals no overall asymmetry but there is a 
significant asymmetry in certain subclasses: dominance of $S$-type galaxies in the $Sb$ class of normal spiral galaxies 
and a dominance of $Z$-type 
in the $SBb$ class of barred spiral galaxies. Both $S$- and $Z$-type galaxies seem to have similar velocity 
distribution, indicating no spatial segregation of the two chiral forms. Thus the ideas of symmetry and chirality
penetrate deeply into modern science ranging from microphysics to astrophysics.
%
%

%
%
\section{Quantum Protectorate}
%
%
It is well known that there are many branches of physics and
chemistry where phenomena occur which cannot be described in the
framework of   interactions amongst a few particles.
As a rule, these phenomena arise essentially from the cooperative
behavior of a large number of   particles. Such many-body
problems are of   great interest not only because of the nature
of phenomena themselves, but also because of the intrinsic
difficulty in solving problems which involve   interactions of
many particles  in terms of   known Anderson  statement that "more
is different"~\cite{and72}. It is often difficult to formulate a
fully consistent and adequate microscopic theory of complex
cooperative phenomena. R. Laughlin and D. Pines invented
an idea of a quantum protectorate, "a stable state of matter,
whose generic low-energy properties are determined by a
higher-organizing principle and nothing else"~\cite{pnas}. This
idea brings into physics the concept that emphasize   the
crucial role of low-energy and high-energy scales for treating the propertied of the substance. 
It is known that a many-particle  system  (e.g. electron gas) in the low-energy limit can be
characterized by a small set of  \emph{collective}  (or hydrodynamic)
variables and equations of motion corresponding
to these variables. Going beyond the framework
of the low-energy region would require the consideration
of plasmon excitations, effects of electron
shell reconstructing, etc. The existence of two
scales, low-energy and high-energy, in the description
of physical phenomena is used in physics, explicitly or
implicitly.\\
According to R. Laughlin and D. Pines,
''The emergent physical phenomena regulated by higher organizing
principles have a property, namely their insensitivity to
microscopics, that is directly relevant to the broad question of
what is knowable in the deepest sense of the term. The low energy
excitation spectrum of a conventional superconductor,
for example, is completely generic and is characterized by a
handful of parameters that may be determined experimentally
but cannot, in general, be computed from first principles. An
even more trivial example is the low-energy excitation spectrum
of a conventional crystalline insulator, which consists of transverse
and longitudinal sound and nothing else, regardless of
details. It is rather obvious that one does not need to prove the
existence of sound in a solid, for it follows from the existence of
elastic moduli at long length scales, which in turn follows from
the spontaneous breaking of translational and rotational symmetry
characteristic of the crystalline state. Conversely, one
therefore learns little about the atomic structure of a crystalline
solid by measuring its acoustics.
The crystalline state is the simplest known example of a
quantum protectorate, a \emph{stable state of matter whose generic
low-energy properties are determined by a higher organizing
principle and nothing else} \ldots  Other important quantum protectorates
include superfluidity in Bose liquids such as $^{4}He$ and
the newly discovered atomic condensates, superconductivity, band insulation, ferromagnetism, 
antiferromagnetism, and the quantum Hall states. The
low-energy excited quantum states of these systems are particles
in exactly the same sense that the electron in the vacuum of
quantum electrodynamics is a particle  \ldots Yet they are not elementary, and, as in the
case of sound, simply do not exist outside the context of the
stable state of matter in which they live. These quantum protectorates,
with their associated emergent behavior, provide us
with explicit demonstrations that the underlying microscopic
theory can easily have no measurable consequences whatsoever
at low energies. The nature of the underlying theory is unknowable
until one raises the energy scale sufficiently to escape
protection''.
The notion of \emph{quantum protectorate} was
introduced to unify some generic features of complex physical
systems on different energy scales,  and is a complimentary unifying idea
resembling the symmetry breaking concept in a certain sense. \\
The sources of quantum protection in high-$T_c$
superconductivity~\cite{scz99} and low-dimensional systems were discussed in
Refs.~\cite{and2,and3,pin2,amu01,bar06,kop06} According to Anderson~\cite{and2},
"the source of quantum protection is likely to be a collective
state of the quantum field, in which the individual particles are
sufficiently tightly coupled that elementary excitations no
longer involve just a few particles, but are collective
excitations of the whole system. As a result, macroscopic
behavior is mostly determined by overall conservation laws".\\ 
The quasiparticle picture of high-temperature superconductors in the frame of a Fermi liquid with the fermion condensate
was investigated by Amusia and  Shaginyan~\cite{amu01}. In their paper
a model of a Fermi liquid with the fermion condensate  was applied to the consideration of quasiparticle excitations in 
high-temperature superconductors, in their superconducting and normal states. Within that model the appearance of the 
fermion condensate presents a quantum phase transition that separates the regions of normal and strongly correlated 
electron liquids. Beyond the phase transition point the quasiparticle system is divided into two subsystems, one containing 
normal quasiparticles and the other - fermion condensate localized at the Fermi surface and characterized by almost 
dispersionless single-particle excitations. In the superconducting 
state the quasiparticle dispersion in systems with fermion condensate can be presented by two straight lines, characterized 
by  two effective masses  and intersecting near the binding energy, which is of the order of the superconducting gap. 
This same quasiparticle picture persists in the normal state, thus manifesting itself over a wide range of temperatures 
as new energy scales. Arguments were presented that fermion systems with fermion condensate have features 
of a "quantum protectorate".\\ 
Barzykin and Pines~\cite{bar06} formulated a
phenomenological model of protected behavior in the pseudogap state of underdoped cuprate superconductors. 
By extending their previous work on the scaling of low frequency magnetic properties of the $2-1-4$ cuprates 
to the $1-2-3$ materials, 
they arrived at a consistent phenomenological description of protected behavior in the pseudogap state of the magnetically 
underdoped cuprates. Between zero hole doping and a doping level of $ \sim 0.22$, it reflects the presence of a mixture of an insulating spin liquid that produces the 
measured magnetic scaling behavior and a Fermi liquid that becomes superconducting for doping levels $x > 0.06$. 
Their analysis suggests the 
existence of two quantum critical points, at doping levels $x \sim  0.05$ and $x \sim 0.22$, and that $d$-wave superconductivity in 
the pseudogap region arises from quasiparticle-spin liquid interaction, i.e., magnetic interactions between 
quasiparticles in the Fermi liquid induced by their coupling to the spin liquid excitations.\\
Kopec~\cite{kop06} attempted to discover the origin of quantum protection in high-$T_{c}$  cuprates.
The concept of topological excitations and the related ground state degeneracy were employed to establish an effective 
theory of the 
superconducting state evolving from the Mott insulator~\cite{phil09}   for high-$T_{c}$ cuprates. The theory includes the effects of the 
relevant energy scales with the emphasis on the Coulomb interaction $U$ governed by the electromagnetic $U(1)$ compact 
group. The results were obtained for the layered $t-t'-t_{\bot}-U-J$
 system of strongly correlated electrons relevant for cuprates. Casting the Coulomb interaction in terms 
of composite-fermions via the gauge flux attachment facility, it was  shown that instanton events in the Matsubara 
"imaginary time," labeled by topological winding numbers, 
were essential configurations of the phase field dual to the charge. This provides a nonperturbative concept of the 
topological quantization and displays the significance of discrete topological sectors in the theory governed by the 
global characteristics of the phase field. In the paper it was shown 
that for topologically ordered states these quantum numbers play the role of an order parameter in a way similar to 
the phenomenological order parameter for conventionally ordered states. In analogy to the usual phase transition that is characterized 
by a sudden change of the symmetry, the topological phase transitions are governed by a discontinuous change of the 
topological numbers signaled by the divergence of the zero-temperature topological susceptibility. This defines a quantum 
criticality ruled by topologically conserved numbers rather than the 
reduced principle of the symmetry breaking. The author shown that in the limit of strong correlations topological 
charge is linked to the average 
electronic filling number and the topological susceptibility to the electronic compressibility of the system.  
The impact of these nontrivial $U(1)$ instanton phase field configurations    for the cuprate phase diagram was exploited.  
The phase diagram displays the   "hidden" quantum critical point 
covered by the superconducting lobe in addition to a sharp crossover between a compressible normal "strange metal" state 
and a region characterized by a vanishing compressibility, which marks the Mott insulator. It was argued that the 
existence of robust quantum numbers explains the stability against small perturbation of the system and attributes 
to the topological "quantum protectorate" as observed in strongly correlated systems.\\
Some other applications of the idea of the quantum protectorate were 
discussed in Refs.~\cite{kee01,cape02,mar04,bask04,ios05}
%
%
%
\section{Emergent Phenomena}
%
Emergence - macro-level effect from micro-level causes - is an important and profound interdisciplinary notion of modern 
science~\cite{zim03,morri06,pavu07,step08,vol08,aziz09,mina08}.
Emergence is a notorious philosophical term, that was used in the domain of art. A variety of theorists have appropriated 
it for their purposes ever since it was applied to the problems of life 
and mind~\cite{zim03,morri06,pavu07,step08,aziz09,mina08}. It might be roughly 
defined as the shared meaning.  Thus  emergent entities (properties or substances) '\emph{arise}' out of more 
fundamental entities and yet are '\emph{novel}' or '\emph{irreducible}' with respect to them.  
Each of these terms are uncertain in its own right, and their specifications 
yield the varied notions of emergence that have been  discussed 
in literature~\cite{zim03,morri06,pavu07,step08,vol08,aziz09,mina08}. There has been renewed interest in emergence 
within discussions of the behavior of complex systems~\cite{aziz09,mina08} and debates over the 
reconcilability of mental causation, intentionality, 
or consciousness with physicalism. This concept is also at the heart of the numerous discussions on the interrelation of the
reductionism and functionalism~\cite{zim03,morri06,pavu07,step08,mina08}.\\
A vast amount of current researches focuses on the search for the organizing principles responsible for emergent behavior 
in matter~\cite{pnas,pines}, with particular attention to correlated matter, the study of materials in which unexpectedly 
new classes of behavior emerge in response to the strong and competing interactions among their elementary constituents.
As it was   formulated at Ref.~\cite{pines}, ''we call \emph{emergent behavior} \ldots the phenomena that owe their
existence to interactions between many subunits, but whose existence cannot be deduced from a detailed knowledge
of those  subunits alone''.\\ 
Models and simulations of collective behaviors are often based on considering them as interactive particle systems~\cite{mina08}. The 
focus is then on behavioral and interaction rules of particles by using approaches based on artificial agents designed to 
reproduce swarm-like behaviors in a virtual world by using symbolic, sub-symbolic and agent-based models. New approaches 
have been considered in the literature~\cite{mina08} based, for instance, on  topological rather than metric distances and 
on fuzzy systems. Recently a new research approach~\cite{mina08} was  proposed  allowing generalization possibly suitable 
for a general theory of emergence.  The coherence of collective behaviors, i.e., their identity detected by the observer, 
as given by meta-structures, properties of meta-elements, i.e., sets of values adopted by mesoscopic state variables 
describing collective, structural aspects of the collective phenomenon under study and related to a higher level of description 
(meta-description) suitable for dealing with coherence, was considered. Mesoscopic state variables were abductively 
identified by the observer detecting emergent properties, such as sets of suitably clustered distances, speed, directions, 
their ratios and ergodic properties of sets. This research approach is under implementation and validation and may be 
considered to model general processes of collective behavior and establish an possible initial basis for a general theory of 
emergence.\\
Emergence and complexity refer to the appearance of higher-level properties and behaviors of a system that obviously 
comes from the collective dynamics of that system's components~\cite{pnas,pines,wen05,sew02,zim03,aziz09}. These 
properties are not directly deducible from the lower-level motion of that system. Emergent properties are properties 
of the ''whole'' that are not possessed by any of the individual parts making up that whole. Such phenomena 
exist in various domains and can be described, using complexity concepts and thematic knowledges~\cite{zim03,aziz09,mina08}. 
Thus this problematic   is highly pluridisciplinary~\cite{emerws}.  
%
%
\subsection{ Quantum Mechanics And Its Emergent Macrophysics}
%
%
The notion of emergence in quantum physics  was considered by Sewell 
in his book "Quantum Mechanics And Its Emergent Macrophysics"~\cite{sew02}. According to his point of view,
the quantum theory of macroscopic systems is a vast, ever-developing area of science that serves to relate the properties 
of complex physical objects to those of their constituent particles. Its essential challenge is that of finding the 
conceptual structures needed for the description of the various states of organization of many-particle quantum systems. 
In that book,  Sewell proposes a new approach to the subject, based on a "\emph{macrostatistical mechanics}", which 
contrasts sharply with the standard microscopic treatments of many-body problems.\\
According to Sewell,
quantum theory began with Planck's   derivation of the thermodynamics of black body radiation from the hypothesis that 
the action of his oscillator model of matter was quantized in integral multiples of a fundamental constant, $\hbar$. This 
result provided a microscopic theory of a macroscopic phenomenon that was incompatible with the assumption of underlying 
classical laws. In the century following Planck's discovery, it became abundantly clear that quantum theory is essential 
to natural phenomena on both the microscopic and macroscopic scales.\\
As a first step towards contemplating the quantum mechanical basis of macrophysics, Sewell notes the empirical fact that 
macroscopic systems enjoy properties that are radically different from those of their constituent particles. Thus, unlike 
systems of few particles, they exhibit 
irreversible dynamics, phase transitions and various ordered structures, including those characteristic of life. 
These and other macroscopic 
phenomena signify that complex systems, that is, ones consisting of enormous numbers of interacting particles, are 
qualitatively different from the sums of their constituent parts (this point of view was also stressed by 
Anderson~\cite{and72}).\\
Sewell proceeds by presenting the operator algebraic framework for the theory. He then undertakes a macrostatistical treatment 
of both equilibrium and nonequilibrium thermodynamics, which yields a major new characterization of a complete set of 
thermodynamic variables and a nonlinear generalization of the Onsager theory. He focuses especially on 
ordered and chaotic structures that arise in some key areas of condensed matter physics. This includes a general derivation 
of superconductive electrodynamics from the assumptions of off-diagonal long-range order, gauge covariance, and 
thermodynamic stability, which avoids the enormous complications of the microscopic treatments. 
Sewell also re-analyzes a theoretical framework for phase transitions far from thermal equilibrium. 
It gives  a coherent approach to the complicated problem of the emergence of macroscopic phenomena from quantum mechanics 
and clarifies the problem of how macroscopic phenomena can be interpreted from the laws and structures of microphysics.\\ 
Correspondingly, theories of such phenomena must be based not only on the quantum mechanics, but 
also on conceptual structures that serve to represent the characteristic features of 
highly complex systems~\cite{pnas,pines,aziz09,mina08,emerws}. Among the main   concepts 
involved here are ones representing various types of order, or organization, disorder, or chaos, and different levels 
of macroscopicality. 
Moreover, the particular concepts required to describe the ordered structures of superfluids and laser light are 
represented by macroscopic wave functions   that are strictly quantum mechanical, although radically different from 
the Schrodinger wave functions of microphysics.\\
Thus, according to Sewell, to provide a mathematical framework for the conceptual structures required for 
quantum macrophysics, it is clear that one needs to go beyond 
the traditional form of quantum mechanics, since that does not discriminate qualitatively between microscopic and 
macroscopic systems. 
This may be seen from the fact that the traditional theory serves to represent a system of N particles within the standard 
Hilbert space scheme, which takes the same form regardless of whether $N$ is 'small' or 'large'.\\ 
Sewell's approach to the basic problem of how macrophysics emerges from quantum mechanics is centered on macroscopic 
observables. 
The main objective of his approach is to obtain the properties imposed on them by general demands of quantum theory 
and many-particle statistics. 
This approach resembles in a certain sense the Onsager's irreversible thermodynamics, which
bases also  on macroscopic observables and certain general structures of complex systems.\\
The conceptual basis of quantum mechanics which go far beyond its  traditional form was formulated 
by S. L. Adler~\cite{adler04}. According to his view,
quantum mechanics is not a complete theory, but rather is an emergent phenomenon arising from the statistical mechanics of 
matrix models that have a global unitary invariance.
The mathematical presentation  of these ideas is based on dynamical variables that are matrices in complex Hilbert space, 
but many of the ideas carry over to a statistical dynamics of matrix models in real or quaternionic Hilbert space.
Adler starts from a classical dynamics in which the dynamical variables are non-commutative matrices or operators. 
Despite the non-commutativity, a sensible Lagrangian and Hamiltonian dynamics was obtained by forming the Lagrangian and 
Hamiltonian as traces of polynomials in the dynamical variables, and repeatedly using cyclic permutation under the 
trace. It was  assumed that the Lagrangian and Hamiltonian are constructed without use of non-dynamical matrix 
coefficients, so that there is an invariance under simultaneous, identical unitary transformations of all the dynamical 
variables, that is, there is a global unitary invariance. The author supposed that the complicated dynamical equations 
resulting from this system rapidly reach statistical equilibrium, and then shown that with suitable approximations, 
the statistical thermodynamics of the canonical ensemble for this system takes the form of quantum field theory.  
The requirements for the underlying trace dynamics to yield quantum theory at the level of thermodynamics are stringent, 
and include both the generation of a mass hierarchy and the existence of boson-fermion balance.   From the equilibrium 
statistical mechanics of trace dynamics, the rules of quantum mechanics \emph{emerge} as an approximate thermodynamic 
description of the behavior of low energy phenomena. "Low energy" here means small relative to the natural energy scale 
implicit in the canonical ensemble for trace dynamics, which author identify with the Planck scale, and by "equilibrium" 
he means local equilibrium, permitting spatial variations associated with dynamics on the low energy scale. Brownian 
motion corrections to the thermodynamics of trace dynamics then lead to fluctuation corrections to quantum mechanics 
which take the form of stochastic modifications of the Schrodinger equation, that can account in a mathematically precise 
way for state vector reduction with Born rule probabilities~\cite{adler04}.\\
Adler emphasizes~\cite{adler04} that he have not identified a candidate for the specific matrix model that realizes his 
assumptions; there may be only one, which could then provide the underlying unified theory of physical phenomena that 
is the goal of current researches in high-energy physics and cosmology.\\  
He admits  the possibility also that the underlying dynamics may be discrete, and this could naturally be implemented 
within his framework of basing an underlying dynamics on trace class matrices. 
The ideas of the Adler's book suggest, that one should seek a common origin for both gravitation and quantum field theory 
at the deeper level of physical phenomena from which quantum field theory 
emerges~\cite{adler04} (see also Ref.~\cite{adler06}). \\ Recently, in Ref.~\cite{grin09}, the causality as an emergent
macroscopic phenomenon was analyzed within the Lee-Wick $O(N)$ model. In quantum mechanics the deterministic property of classical
physics is an emergent phenomenon appropriate only on macroscopic scales. Lee and Wick introduced Lorenz invariant
quantum theories where causality is an emergent phenomenon appropriate for macroscopic time scales. In Ref.~\cite{grin09},
authors analyzed a Lee-Wick version of the $O(N)$ model. It was argued that in the large-$N$ limit this theory
has a unitary and Lorenz invariant $S$ matrix and is therefore free of paradoxes of scattering experiments.
%
%
%
%
\subsection{ Emergent Phenomena in Quantum Condensed Matter Physics }
%
%
Statistical physics and condensed matter physics supply us with many examples of emergent phenomena.
For example, taking a macroscopic approach to the problem, 
and identifying the right degrees of freedom of a many-particle system, the equations of motion of interacting particles 
forming a fluid can be described by the Navier-Stokes equations 
for fluid dynamics from which complex new behaviors arise such as turbulence. This is  the clear example of 
an emergent phenomenon in classical physics.\\
Including quantum mechanics into the consideration leads to even more complicated situation. 
In 1972 P. W. Anderson published his essay "More is Different"
which describes how new concepts, not applicable in ordinary classical or
quantum mechanics, can arise from the consideration of aggregates of large numbers of
particles~\cite{and72} (see also Ref.~\cite{zur08}). 
Quantum mechanics is a basis of macrophysics. However 
macroscopic systems have the properties that are radically different from those of their constituent particles. Thus, 
unlike systems of few particles, they exhibit irreversible dynamics, phase transitions and various ordered structures, 
including those characteristic of life. These and other macroscopic phenomena signify that complex systems, that is, 
ones consisting of huge numbers of interacting particles, are qualitatively different from the sums of their 
constituent parts~\cite{and72}.\\
Many-particle systems where the interaction 
is strong have often complicated behavior, and require nonperturbative approaches to treat their properties.  
Such  situations  are often arise in condensed matter systems.
Electrical, magnetic and mechanical properties of materials are \emph{emergent collective behaviors}  of the underlying quantum mechanics of 
their electrons and constituent atoms. A principal aim of solid state physics and materials science is to elucidate this emergence. 
A full achievement of this goal would imply the ability to engineer a material that is optimum for any particular 
application. The current understanding of electrons in solids uses simplified but workable picture
 known as the Fermi liquid theory. This theory explains why 
electrons in solids can often be described in a simplified manner which appears to ignore the large repulsive forces 
that electrons are known 
to exert on one another. There is a growing appreciation that this theory probably fails for entire classes of possibly 
useful materials and 
there is the suspicion that the failure has to do with unresolved competition between different possible emergent 
behaviors.\\ 
Strongly correlated electron materials manifest emergent phenomena by the remarkable range of quantum ground states that 
they display, e.g., 
insulating, metallic, magnetic, superconducting, with apparently trivial, or modest changes in chemical composition, 
temperature or pressure. 
Of great recent interest are the behaviors of a system poised between two stable zero temperature ground states, i.e. at 
a quantum critical point. These behaviors intrinsically support non-Fermi liquid (NFL) phenomena, including the electron 
fractionalization that is characteristic of thwarted ordering in a one-dimensional interacting electron gas.\\ 
In spite of the difficulties, a substantial progress has been made in understanding strongly interacting 
quantum systems~\cite{pwa84,and72,kuznc02,wen04}, and this is the main scope of the quantum condensed matter physics. 
It was speculated that a strongly 
interacting system can be roughly understood in terms of weakly interacting   quasiparticle excitations. 
In some of the  cases, the quasiparticles bear almost no resemblance to the underlying degrees of freedom of the 
system - they have \emph{emerged} as a complex collective effect.
In the last three decades there has been the emergence of the new profound concepts associated with fractionalization, topological 
order, emergent gauge bosons and fermions, and string condensation~\cite{wen04}. These new physical concepts are so 
fundamental that they may even influence 
our understanding of the origin of light and electrons in the universe~\cite{wen05}.  Other systems of interest are 
dissipative quantum systems, 
Bose-Einstein condensation, symmetry breaking and gapless excitations, phase transitions, Fermi liquids, spin density 
wave states, Fermi and 
fractional statistics, quantum Hall effects, topological/quantum order, spin liquid and string condensation~\cite{wen04}. 
The typical example of emergent phenomena is in fractional quantum Hall systems~\cite{bala08} - two dimensional systems of electrons at 
low temperature and in high magnetic fields. In this case, the underlying degrees of freedom are the electron, but the emergent quasiparticles 
have charge which is only a fraction of that of the electron. The fractionalization of the elementary electron is one of
the  remarkable discoveries of quantum physics, and is purely a collective emergent effect. It is quite interesting   
that the quantum properties of these fractionalized quasiparticles are unlike any ever found elsewhere 
in nature~\cite{wen04}. 
In non-Abelian topological phases of matter, the existence of a degenerate ground state subspace suggests the possibility of using this 
space for storing and processing quantum information~\cite{simo07}. 
In topological quantum computation~\cite{simo07} quantum information is stored in exotic states of matter which are intrinsically 
protected from decoherence, and quantum operations are carried out by dragging particle-like excitations (quasiparticles) around one 
another in two space dimensions. The resulting quasiparticle trajectories define world-lines in three dimensional space-time, and the 
corresponding quantum operations depend only on the topology of the braids formed by these world-lines. Authors~\cite{simo07} described recent work showing 
how to find braids which can be used to perform arbitrary quantum computations using a specific kind of quasiparticle (those described by 
the so-called Fibonacci anyon model) which are thought to exist in the experimentally 
observed $\nu = 12/5$ fractional quantum Hall state.\\
In Ref.~\cite{wen05}  Levine and Wen proposed to consider photons and electrons as emergent phenomena.
Their arguments are based on recent advances in condensed-matter theory~\cite{wen04} which have revealed that new 
and exotic phases of matter can exist in spin models (or more precisely, local bosonic models) via a simple physical 
mechanism, known as ''\emph{string-net condensation}''.  These new phases of matter have the unusual property 
that their collective excitations are gauge bosons and fermions. In some cases, the collective excitations can behave just like the photons, 
electrons, gluons, and quarks in the relevant vacuum. This suggests that photons, electrons, and other elementary particles may have a 
unified origin-string-net condensation in that  vacuum. In addition, the string-net picture indicates how to make artificial photons,
artificial electrons, and artificial quarks and gluons in condensed-matter systems.\\
In paper~\cite{xgwen05}, Hastings and Wen analyzed the quasiadiabatic continuation of quantum states. They considered  the 
stability of topological ground-state degeneracy and emergent gauge invariance for quantum many-body systems.
The continuation is valid when the Hamiltonian 
has a gap, or else has a sufficiently small low-energy density of states, and thus is away from a quantum phase transition. 
This continuation takes local operators into local operators, while approximately preserving the ground-state 
expectation values. They applied 
this continuation to the problem of gauge theories coupled to matter, and propose the distinction of perimeter law versus "zero law" to 
identify confinement. The authors also applied the continuation to local bosonic models with emergent gauge theories. It was 
shown that local gauge invariance is topological and cannot be broken by any local perturbations in the bosonic models 
in either continuous or discrete gauge groups. Additionally they shown 
that the ground-state degeneracy in emergent discrete gauge theories is a robust property of the bosonic model, 
and the arguments were given that the robustness of local gauge invariance in the continuous case protects the 
gapless gauge boson.\\
Pines and co-workers~\cite{hfpin04} carried out a theory of scaling in the emergent behavior of heavy-electron materials.
It was shown that the NMR Knight shift anomaly exhibited by a large number of heavy electron materials can be 
understood in terms of the different hyperfine couplings of probe nuclei to localized spins and to conduction electrons. 
The onset of the anomaly is at a temperature $T^{*}$, below which an itinerant component of the magnetic susceptibility 
develops. This second component characterizes the polarization of the conduction electrons by the local moments and is a 
signature of the emerging heavy electron state. The heavy electron component grows as log  T below T* , and scales universally for all measured 
Ce , Yb and U based materials. Their results suggest that $T^{*}$ is not related to the single ion Kondo 
temperature, $T_{K}$ (see Ref.~\cite{mack08}), but rather 
represents a correlated Kondo temperature that provides a measure of the strength of the intersite coupling between the 
local moments.\\ 
The complementary questions concerning the emergent symmetry and dimensional reduction at a quantum critical point were
investigated at Refs.~\cite{bati05,bati08}
Interesting discussion of the emergent physics  which was only partially reviewed here
may be found in  the paper of Volovik~\cite{vol08}.
%
%
%
%
\section{ Magnetic Degrees of Freedom and Models of Magnetism}
%
%
The development of the quantum theory of magnetism was concentrated on the right
definition of the fundamental "magnetic" degrees of freedom and
their correct model description for complex magnetic systems~\cite{dmat06,tyab,kuz09}. We
shall first describe   the phenomenology of the magnetic
materials to look at the physics involved. The problem of
identification of the fundamental "magnetic" degrees of freedom
in complex materials is rather nontrivial. Let us discuss
briefly, to give a flavor only, the very intriguing problem of
the electron dual behavior. The existence and properties of
localized and itinerant magnetism in insulators, metals, oxides
and alloys and their interplay in complex materials is an
interesting and not yet fully understood problem of quantum
theory of magnetism~\cite{tyab,kuznc02,kuz02,kuz09}. The central problem of
recent efforts is to investigate the interplay and competition of
the insulating, metallic, superconducting, and heavy fermion
behavior versus the magnetic behavior, especially in the
vicinity of a transition to a magnetically ordered state. The
behavior and the true nature of the electronic and spin states
and their quasiparticle dynamics are of central importance  to
the understanding of the physics of strongly correlated systems
such as magnetism and metal-insulator transition  in metals and
oxides, heavy fermion states , superconductivity and their
competition with magnetism. The strongly correlated
electron systems are systems in which electron correlations
dominate. An important problem in understanding the physical
behavior of these systems was the connection between relevant
underlying chemical, crystal and electronic structure, and the
magnetic and transport properties which continue to be the
subject of intensive debates. Strongly correlated
$d$ and $f$ electron systems are   of special interest~\cite{kuznc02,kuz02,kuz09}. 
In these materials electron  correlation effects are essential
and, moreover, their spectra are complex, {\it i.e.}, have many
branches.   Importance of the studies on strongly correlated
electron systems are concerned with a fundamental problem of
electronic solid state theory, namely, with a tendency of $3(4)
d$ electrons in transition metals and compounds and  $4(5) f$
electrons in rare-earth metals and compounds and alloys to
exhibit both localized and delocalized behavior. Many electronic and magnetic features of
these substances relate intimately to this dual behavior of the
relevant electronic states. For example, there are some alloy
systems in which radical changes in physical properties occur
with relatively modest changes in chemical composition or
structural perfection of the crystal lattice.  Due to
competing interactions of comparable strength, more complex
ground states than usually supposed may be realized. The strong
correlation effects among electrons, which lead to the formation
of the heavy fermion state take part to some extent in formation
of a magnetically ordered phase, and thus imply that the very
delicate competition and interplay of interactions exist in these
substances. For most of the heavy fermion
superconductors, cooperative magnetism, usually some kind of
antiferromagnetic ordering was observed in the "vicinity" of
superconductivity. In the case of $U$-based compounds, the two
phenomena, antiferromagnetism and superconductivity coexist on a
microscopic scale, while they seem to compete with each other in
the Ce-based systems. For a Kondo lattice system~\cite{kuzsf99,kuze04,kuzsf05}, the
formation of a Neel state via the RKKY intersite interaction
compete with the formation of a local Kondo singlet~\cite{mack08}. Recent data
for many heavy fermion $Ce$- or $U$-based compounds and alloys
display a pronounced non-Fermi-liquid behavior. A number of
theoretical scenarios have been proposed and they can be broadly
classified into  two categories which deal  with the localized
and extended states of $f$-electrons. Of special interest is the
unsolved controversial problem of the reduced magnetic moment in
Ce- and U-based alloys and the description of the heavy fermion
state in the presence of the coexisting magnetic state. In other
words, the main interest is in  the understanding of the
competition of intra-site (Kondo screening) and inter-site (RKKY
exchange) interactions. Depending on the relative magnitudes of
the Kondo and  RKKY scales, materials with different
characteristics are found which are classified as non-magnetic
and magnetic concentrated Kondo systems.  These features reflect the
very delicate interplay and competition of interactions  and
changes in a chemical composition. As a rule, very little
intuitive insight could be gained from this very complicated
behavior.\\ 
Magnetism in materials such as iron
and nickel results from the cooperative alignment of the
microscopic magnetic moments of electrons in the material. The
interactions between the microscopic magnets are described
mathematically by the form of the Hamiltonian of the system. The
Hamiltonian depends on some parameters, or coupling constants,
which measure the strength of different kinds of interactions. The
magnetization, which is measured experimentally, is related to
the average or mean alignment of the microscopic magnets. It is
clear that some of the parameters describing the transition to
the magnetically ordered state do depend on the detailed nature
of the forces between the microscopic magnetic moments. The
strength of the interaction will be reflected in the critical
temperature which is high if the aligning forces are strong and
low if   they are weak. In quantum theory of magnetism, the
method of model Hamiltonians has proved to be very effective~\cite{dmat06,tyab,kuz09,dmat03,rscom08,lee07}.
Without exaggeration, one can say that the great advances in the
physics of magnetic phenomena are to a considerable extent due to
the use of   very simplified and schematic model representations
for the theoretical interpretation~\cite{dmat06,tyab,kuz09,dmat03,rscom08,lee07}.
%
%
%
\subsection{ Ising Model}
%
 One can regards
the Ising model~\cite{dmat06,dmat03} as the first model of the quantum
theory of magnetism. In this model, formulated by
W. Lenz   in 1920 and studied by E. Ising, it was assumed that the spins are
arranged at the sites of a regular one-dimensional lattice.
Each spin can obtain the values $\pm \hbar/2$:
\begin{equation}
\label{i1} H = -  \sum_{ij}  J(i-j)  S_{i}  S_{j}
-g\mu_{B}H\sum_{i}S_{i}. 
\end{equation}
This Hamiltonian was one of the first attempts to describe the
magnetism as a cooperative effect. It is interesting that
the one-dimensional Ising model 
\begin{equation}
\label{i2} H = -  J \sum_{i=1}^{N}     S_{i}  S_{i+1}
\end{equation}
was solved by Ising in 1925, while the exact solution of the Ising model on a
two-dimensional square lattice was obtained by
L. Onsager  only in 1944.\\
Ising model with no external magnetic field have a global discrete symmetry, namely the symmetry
under reversal of spins $S_{i} \rightarrow - S_{i}$. We recall that the symmetry is spontaneously broken if there is
a quantity (the order parameter) that is not invariant under the symmetry operation and has a nonzero expectation value.
For Ising model the order parameter is equal to $ M  = \sum_{i=1} S_{i}. $ It is not invariant under the symmetry operation.
In principle, there schould not be any spontaneous symmetry breaking as it is clear from the consideration of the  
thermodynamic average
$m = \langle M \rangle = \textrm{Tr} \left ( M \rho (H) \right ) =0. $
We have
\begin{equation}
\label{i3} m = \langle N^{-1} \sum_{i=1}      S_{i}  \rangle =   
\frac{1}{N \cdot Z_{N}}\sum_{i} \sum_{S_{i} = \pm 1}  S_{i} \exp \Bigl ( - \beta H(S_{i}) \Bigr ) = 0
\end{equation}
Thus to get the spontaneous symmetry breaking one should take the thermodynamic limit $(N \rightarrow \infty).$
But this is not enough. In addition, one needs the symmetry breaking field $h$ which lead to extra term in the
Hamiltonian $\mathcal{H} = H - h \cdot M.$ It is important to note that
\begin{equation}
\label{i4} 
\lim_{h \rightarrow 0} \lim_{N \rightarrow \infty} =  \langle M \rangle_{h, N} = m \neq 0.
\end{equation}
In this equation limits cannot be interchanged.\\
Let us remark that for Ising model energy cost to rotate one spin is equal to $E_{g} \propto J.$ 
Thus every excitation costs finite
energy.   As a consequence, long-wavelength spin-waves cannot happen with discrete broken symmetry.\\
In  one-dimensional case $(D = 1)$ the average value $\langle M \rangle = 0,$ i.e. there is no 
spontaneously symmetry breaking for all $T > 0.$ In  two-dimensional case $(D = 2)$ the average value 
$\langle M \rangle \neq 0,$ i.e. there is spontaneously symmetry breaking and the phase transition. In other words,
for  two-dimensional case for $T$ small enough, the system will prefer the ordered phase, whereas 
for  one-dimensional case no matter how small $T$,  the system will prefer the disordered phase
(for the number of flipping  neighboring spins large enough).
%
%
%
\subsection{ Heisenberg Model}
%
 The Heisenberg model~\cite{dmat06,dmat03} is based on
the assumption that the wave functions of magnetically active
electrons in crystals differ little from the atomic orbitals. The
physical picture can be represented by a model in which the
localized magnetic moments originating from ions with incomplete
shells interact through a short-range interaction. Individual
spin moments form a regular lattice. The  model of a system of
spins on  a lattice is termed the Heisenberg
ferromagnet~\cite{dmat06} and establishes the origin of the coupling
constant as the exchange energy. The Heisenberg ferromagnet in  a
magnetic field $H$ is described by the Hamiltonian
\begin{equation}
\label{he1} H = -  \sum_{ij}  J(i-j) \vec S_{i} \vec S_{j}
-g\mu_{B}H\sum_{i}S_{i}^{z}
\end{equation}
The coupling coefficient $J(i-j)$ is the measure of the  exchange
interaction between  spins at the lattice sites $i$ and $j$ and
is defined usually to have the property $J(i - j = 0) = 0.$ This
constraint means that only the inter-exchange interactions are
taken into account. The coupling, in principle, can be of a more
general type (non-Heisenberg terms). For crystal lattices in
which every ion is at the centre of symmetry, the exchange
parameter has the property  $ J(i-j) = J(j-i).$\\  
We can rewrite
then the Hamiltonian (\ref{he1}) as
\begin{equation}
\label{he2} H = -  \sum_{ij} J(i-j) ( S^z_{i}S^z_{j} +
 S^+_{i}S^-_{j})
\end{equation}
Here $S^{\pm} = S^x \pm iS^y$ are the raising and lowering spin
angular momentum operators. The complete set of spin commutation
relations is
\begin{eqnarray} \nonumber
[S^{+}_{i},S^{-}_{j}]_{-} = 2S^{z}_{i} \delta_{ij}; \quad
[S^{+}_{i},S^{-}_{i}]_{+} = 2S(S + 1) - 2(S^{z}_{i})^{2};  \\
\nonumber [S^{\mp}_{i},S^{z}_{j}]_{-} = \pm
S^{\mp}_{i}\delta_{ij}; \quad S^{z}_{i} = S(S + 1) -
(S^{z}_{i})^{2} - S^{-}_{i}S^{+}_{i}; \\ \nonumber
(S^{+}_{i})^{2S+1} = 0, \quad (S^{-}_{i})^{2S+1} = 0 \nonumber
\end{eqnarray}
We  omit the term of interaction of the spin with an external
magnetic field for the brevity of notation. The statistical
mechanical problem involving this Hamiltonian was not  exactly
solved, but many approximate solutions were obtained~\cite{tyab}.\\
To proceed further, it is important to note that for the isotropic
Heisenberg model, the total $z$-component of spin  $S^z_{tot} =
\sum_{i}S^z_{i}$ is a constant of motion, i.e. 
\begin{equation}
\label{he3} [H,S^z_{tot}] = 0
\end{equation}
  There are cases when the total spin is not a constant of motion,
as, for instance, for the Heisenberg model with the dipole terms
added.\\Let us define the eigenstate $|\psi_{0}>$  so that
$S^+_{i}|\psi_{0}> = 0$ for all lattice sites $R_{i}$. It is
clear that $|\psi_{0}>$ is a state in which all the spins are
fully aligned and for which $S^z_{i}|\psi_{0}> = S|\psi_{0}>$. We
also have $$ J_{\vec k} = \sum_{i}e^{(i\vec k \vec R_i)} J(i) =
J_{-\vec k},$$  where the reciprocal vectors $\vec k$ are defined
by cyclic boundary conditions. Then we obtain $$ H |\psi_{0}> = -
\sum_{ij} J(i-j)S^2 = - NS^2 J_{0}$$  Here $N$ is the total number
of ions in the crystal. So, for the isotropic Heisenberg
ferromagnet, the ground state $|\psi_{0}>$ has an energy $-NS^2
J_{0}$.\\ The state $|\psi_{0}>$ corresponds to a total spin
$NS$. \\ Let us consider now the first excited state. This state
can be constructed by creating one unit of spin deviation in the
system. As a result, the total spin is $NS - 1$. The state
$$|\psi_{k}> = \frac {1}{\sqrt {(2SN)}}\sum_{j}e^{(i\vec k \vec R_j)}S^-_{j}|\psi_{0}>$$ 
is
an eigenstate of $H$ which corresponds to a single magnon of the
energy
\begin{equation}
\label{he4} E  (q) = 2S (J_{0} - J_{q}).
\end{equation}
Note that the role of translational symmetry, i.e. the regular
lattice of spins, is essential, since the state $|\psi_{k}>$ is
constructed from the fully aligned state by decreasing the spin
at each site and summing over all spins with the phase factor
$e^{i\vec k \vec R_j}$ (we consider the 3-dimensional case only).
It is easy to verify that
$$<\psi_{k}|S^z_{tot} |\psi_{k}> = NS - 1.$$   
Thus the Heisenberg model possesses the continuous symmetry
under rotation of spins $ \vec S_{i} \rightarrow \mathcal{R} \vec S_{i}.$
Order parameter $M^{z} \sim  \sum_{i}  S^{z}_{i}$ is not invariant under
this transformation. Spontaneously symmetry breaking of continuous symmetry
is manifested by new excitations --
Goldstone modes which cost little energy. 
Let us rewrite the Heisenberg Hamiltonian in the following form $(|S_{i} | = 1):$
\begin{equation}
\label{he5} H = -  J \sum_{\langle ij \rangle}  \vec S_{i} \vec S_{j} = -  J \sum_{ij} \cos (\theta_{ij})
\end{equation}
In the ground state all spins are aligned in one direction (ferromagnetic state).
The energy cost to rotate one spin is equal $E_{g} \propto J (1 - \cos \theta)$,
where $\theta $ is infinitesimal small angle.
Thus the energy cost to rotate all spins is very small
due to continuous  symmetry  of the Hamiltonian.
As a result the long-wavelength spin-waves exist in the Heisenberg model.\\
The above
consideration was possible because we knew the exact ground state
of the Hamiltonian. There are many models where this is not the
case. For example, we do not know the exact ground state of a
Heisenberg ferromagnet with dipolar forces and the ground state
of the Heisenberg antiferromagnet.\\
The isotropic Heisenberg
ferromagnet (\ref{he1}) is often used as an example of a system
with spontaneously broken symmetry. This
means that the Hamiltonian symmetry, the invariance
with respect to rotations, is no longer the symmetry of
the equilibrium state. Indeed the ferromagnetic states
of the model are characterized by an axis of the preferred
spin alignment, and, hence, they have a lower
symmetry than the Hamiltonian itself. 
The essential role of the physics of magnetism
 in the development of symmetry ideas
was noted in the paper   by  Y. Nambu~\cite{namb07}, devoted to the development of the
elementary particle physics and the origin of the concept
of spontaneous symmetry breakdown. Nambu
points out that back at the end of the 19th century
P. Curie used symmetry principles in the
physics of condensed matter. 
Nambu also notes:
"More relevant examples for us, however, came
after Curie. The ferromagnetism is the prototype of
today's spontaneous symmetry breaking, as was
explained by the works of Weiss, Heisenberg,
and others. Ferromagnetism has since served us as a
standard mathematical model of spontaneous symmetry
breaking".\\
This statement by Nambu should be understood in
light of the clarification made by Anderson~\cite{ander90} (see also Ref.~\cite{zur08}).
He claimed that there is
"the false analogy between broken symmetry and ferromagnetism".
According to Anderson~\cite{ander90}, "in ferromagnetism,
specifically, the ground state is an eigenstate of the relevant
continuous symmetry (that of spin rotation), and and as a result
the symmetry is unbroken and the low-energy excitations have no new
properties. Broken symmetry proper occurs when the ground state is not an
eigenstate of the original group, as in antiferromagnetism or superconductivity; only then does one have
the concepts of quasidegeneracy and of Goldstone bosons and the 'Higgs' phenomenon". 
%
%
%
\subsection{  Itinerant Electron Model} 
%
E. Stoner~\cite{ston51} has
proposed an alternative, phenomenological band model of magnetism
of the transition metals in which the bands for electrons of
different spins are shifted in energy in a way that is favourable
to ferromagnetism~\cite{dmat06,hunt}. 
E. P. Wohlfarth ~\cite{wohl53} developed further the Stoner ideas
by considering in greater detail the quantum-mechanical and statistical-mechanical foundations of the 
collective electron theory and   by analyzing   a wider range of relevant experimental results. 
Wohlfarth considered the difficulties of a rigorous quantum mechanical 
derivation of the internal energy of a ferromagnetic metal at absolute zero. In order to determine  the form 
of the expressions, he carried out a calculation based on the tight binding approximation for a crystal containing $N$ 
singly charged ions, which are fairly widely separated, 
and $N$ electrons. The forms of the Coulomb and exchange contributions to the energy were discussed in the two 
instances of maximum and minimum multiplicity. The need for   correlation corrections were stressed, and the effects 
of these corrections were discussed with special reference to the state of affairs at infinite ionic separation. 
The fundamental difficulties involved in calculating the energy as 
function of magnetization were considered as well; it was shown that they are probably less serious for tightly bound 
than for free electrons, so that the approximation of neglecting them in the first instance is not too unreasonable. 
The dependence of the exchange energy on the relative magnetization $m$ was corrected.\\
The Stoner model promoted the subsequent development of the itinerant model of magnetism.
It was established that the band shift effect is a consequence of
strong intra-atomic correlations. The itinerant-electron picture
is the alternative conceptual picture for
magnetism~\cite{dmat06}. It must be noted that the problem
of band antiferromagnetism is a much more complicated
subject~\cite{kuz99}. The antiferromagnetic state is characterized
by a spatially changing component of magnetization which varies
in such a way that the net magnetization of the system is zero.
The concept of antiferromagnetism of localized spins,  which is
based on the Heisenberg model and two-sublattice Neel ground
state, is relatively well founded contrary to the
antiferromagnetism of delocalized or itinerant electrons .  In
relation to the duality of localized and itinerant electronic
states, G.Wannier~\cite{wan37} showed the importance of the description of the
electronic states which reconcile the band and local (cell)
concept as a matter of principle. 
Wannier functions $\phi(\vec{r} - \vec{R}_{n} )$ form a complete set of mutually orthogonal functions localized around each lattice site $\vec{R}_{n}$
within any band or group of bands. They permit one to formulate  an effective Hamiltonian for electrons
in periodic potentials and span the space of a singly energy band. However, the real computation of Wannier functions in
terms of sums over Bloch states is a difficult task. 
A method for determining the optimally localized set of generalized Wannier functions associated with a set of Bloch
bands in a crystalline solid was discussed in Ref.~\cite{vand} Thus, in the condensed matter theory, the Wannier functions play an important role in the
theoretical description of transition metals, their compounds and disordered alloys, impurities and imperfections,
surfaces, etc.
P.W. Anderson~\cite{andimp61} proposed a model
of transition metal impurity in the band of a host metal. All these
and many others works have led to formulation of the narrow-band model
of magnetism.
%
\subsection{  Hubbard Model } 
%
There are big difficulties in the
description of the complicated problem of magnetism in a metal
with the $d$ band electrons which  are really neither "local" nor
"itinerant" in a full sense. 
The Wannier functions basis set is the background of the widely used Hubbard model.
The Hubbard model~\cite{hub63} is in a
certain sense an intermediate model (the narrow-band model) and
takes into account the specific features of transition metals and
their compounds by assuming that the $d$ electrons form a band,
but are subject to a strong Coulomb repulsion  at one lattice
site.  The Hubbard   Hamiltonian is of the
form
\begin{equation}
\label{hu1} H =
\sum_{ij\sigma}t_{ij}a^{\dagger}_{i\sigma}a_{j\sigma} +
U/2\sum_{i\sigma}n_{i\sigma}n_{i-\sigma}.
\end{equation}
It includes the intra-atomic Coulomb repulsion $U$ and the
one-electron hopping energy $t_{ij}$. The electron correlation
forces electrons to localize in the atomic orbitals which are
modelled here by a complete and orthogonal set of the Wannier wave
functions $[\phi({\vec r} -{\vec R_{j}})]$. On the other hand,
the kinetic energy is reduced when electrons are delocalized. The
band energy of Bloch electrons $\epsilon_{\vec k}$ is defined as
follows: 
\begin{equation} \label{hu2}
  t_{ij} = N^{-1}\sum_{\vec k}\epsilon_{
k} \exp[i{\vec k}({\vec R_{i}} -{\vec R_{j}}], 
\end{equation}
where $N$ is the number of lattice sites. This conceptually simple
model is mathematically very
complicated~\cite{kuznc02,kuz09}.  The Pauli exclusion
principle~\cite{pauli05} which does not allow two electrons of common spin to be
at the same site, plays a crucial role. It can be shown, that
under transformation $\mathcal{R}  H \mathcal{R}^{\dag}$, where $\mathcal{R}$ is the spin rotation
operator 
\begin{equation} \label{hu3} \mathcal{R} = \bigotimes_{j}\exp ({1
\over 2} i \phi \vec \sigma_{j} \vec n),
\end{equation}
the Hubbard Hamiltonian is invariant under spin rotation, {\em
i.e.,} $\mathcal{R}  H \mathcal{R}^{\dag} = H$. Here $\phi$ is the angle of rotation
around the unitary axis $\vec n$ and $\vec \sigma$ is the Pauli
spin vector; symbol $\bigotimes_{j}$ indicates a tensor product
over all site subspaces. The summation over $j$ extends to all
sites.\\ 
The equivalent expression for the Hubbard model that
manifests the property of   rotational invariance explicitly can
be obtained with the aid of the transformation
\begin{equation}
\label{hu4} \vec S_{i} = {1 \over 2} \sum_{\sigma \sigma'}
a^{\dagger}_{i\sigma} \vec \sigma_{\sigma \sigma'} a_{j\sigma'}.
\end{equation}
Then the second term in (\ref{hu1}) takes the following form
$$n_{i \uparrow}n_{i \downarrow} =
\frac{n_{i}}{2} - \frac{2}{3} \vec S_{i}^{2}.$$
As a result we get
\begin{equation}
\label{hu5} H =
\sum_{ij\sigma}t_{ij}a^{\dagger}_{i\sigma}a_{j\sigma} + U \sum_{i
} (\frac{n_{i}^{2}}{4} - \frac{1}{3} \vec S_{i}^{2}).
\end{equation}
The total $z$-component $S^z_{tot}$ commutes with Hubbard
Hamiltonian and the relation $[H,S^z_{tot}] = 0$ is valid.
%
%
%
\subsection{Multi-Band Models. Model with $s-d$ Hybridization}
%
The Hubbard model is the single-band model. It is necessary, in
principle, to take into account the multi-band structure, orbital
degeneracy, interatomic effects and electron-phonon interaction.
The band structure calculations and the experimental studies
showed that for noble, transition and rare-earth metals the
multi-band effects are essential. An important generalization of
the single-band Hubbard model is the so-called model with $s-d$
hybridization~\cite{smit,kiso}. For transition $d$ metals,
investigation of the energy band structure reveals that $s-d$
hybridization processes play an important part. Thus, among the
other generalizations of the Hubbard model that correspond more
closely to the real situation in transition metals, the model
with $s-d$ hybridization serves as an important tool for
analyzing of the multi-band effects. The system is described by a
narrow $d$-like band, a broad $s$-like band and a $s-d$ mixing
term coupling the two former terms. The model Hamiltonian reads
\begin{equation}
\label{mb1}
   H = H_{d} + H_{s} + H_{s-d}.
\end{equation}
The Hamiltonian $H_{d}$ of tight-binding electrons is the Hubbard
model (\ref{hu1}).
\begin{equation}
\label{mb2} H_{s} =
\sum_{k\sigma}\epsilon^{s}_{k}c^{\dagger}_{k\sigma}c_{k\sigma}
\end{equation}
is the Hamiltonian of a broad $s$-like band of electrons.
\begin{equation}
\label{mb3} H_{s-d} = \sum_{k\sigma} V_{k} (
c^{\dagger}_{k\sigma}a_{k\sigma} +
 a^{\dagger}_{k\sigma}c_{k\sigma})
\end{equation}
is the interaction term which represents a mixture of the
$d$-band and $s$-band electrons. The model Hamiltonian
(\ref{mb1}) can be interpreted also in terms of a series of
Anderson impurities~\cite{andimp61} placed regularly in each site (the so-called
periodic Anderson model ). The model (\ref{mb1}) is rotationally
invariant also.
%
\subsection{ Spin-Fermion Model } 
%
Many magnetic and electronic
properties of rare-earth metals and compounds ({\it e.g.,}
magnetic semiconductors) can be interpreted in terms of a combined
spin-fermion model~\cite{kuzsf99,kuze04,kuzsf05}  that includes the
interacting localized spin and itinerant charge subsystems. The
concept of the $s(d)-f$ model plays an important role in the
quantum theory of magnetism, especially the generalized $d-f$
model, which describes the localized $4f(5f)$-spins interacting
with $d$-like tight-binding itinerant electrons and takes into
consideration the electron-electron interaction.  The total
Hamiltonian of the model is given by
\begin{equation}
\label{sf1}
   H = H_{d} + H_{d-f}.
\end{equation} 
The Hamiltonian $H_{d}$ of tight-binding electrons is the
Hubbard model (\ref{hu1}).
  The term $H_{d-f}$ describes the
interaction of the total $4f(5f)$-spins with the spin density of
the itinerant electrons
\begin{equation} \label{sf2}
 H_{d-f} = \sum_{i}J{\vec \sigma_{i}}{\vec S_{i}}
 = - J
N^{-1/2}\sum_{kq}\sum_{\sigma}[S^{-\sigma}_{-q}a^{\dagger
}_{k\sigma} a_{k+q-\sigma} + z_{\sigma}S^{z}_{-q}a^{\dagger
}_{k\sigma}a_{k+q\sigma}],
\end{equation}
where sign factor $z_{\sigma}$ is given by
\begin{equation} 
z_{\sigma} = (+ ,-); - \sigma =  (\uparrow  ,  \downarrow); \quad
S^{-\sigma}_{-q} =  \begin{cases}
S^{-}_{-q},  & - \sigma = +, \\
S^{+}_{-q} & - \sigma = -.
\end{cases}
\end{equation}  
In general the indirect
exchange integral $J$ strongly depends on the wave vectors
$J(\vec k; \vec k+ \vec q)$ having its maximum value at $k=q=0$.
We omit this dependence for the sake of brevity  of notation. To
describe the magnetic semiconductors the Heisenberg interaction
term  (\ref{he1}) should be added~\cite{kuzsf99,kuze04,kuzsf05} ( the
resulting model is called the modified Zener model ).\\
These   model Hamiltonians (\ref{he1}), (\ref{hu1}), (\ref{mb1}), (\ref{sf1})  (and their simple modifications and
combinations) are the most commonly used models in quantum theory
of magnetism. In our previous paper~\cite{kuz2}, where the
detailed analysis of the neutron scattering experiments on
magnetic transition metals and their alloys and compounds was
made, it was concluded that at the level of low-energy
hydrodynamic excitations one cannot distinguish between the
models. The reason for that is the spin-rotation symmetry. In
terms of Ref.\cite{pnas}, the spin waves (collective waves of the order parameter) are in a quantum
protectorate~\cite{kuz02} precisely in this sense. 
%
\subsection{ Symmetry and Physics of Magnetism }
%
In many-body interacting systems, the symmetry is important in
classifying different phases and   understanding the phase
transitions between them~\cite{rose,symm79,bir64,jos91,chat08,pei91,pei92,symcm08,stern64,bar78,andre80,march87}. To penetrate
at the nature of the magnetic properties of the materials it is necessary to establish the symmetry properties
and corresponding conservation laws of the microscopic models of
magnetism.  
For ferromagnetic materials, the laws describing it are invariant under spatial rotations. Here, the order 
parameter is the magnetization, which measures the magnetic dipole density. Above the Curie temperature, the order 
parameter is zero, which is spatially invariant and there is no symmetry breaking. Below the Curie temperature, however, 
the magnetization acquires a constant (in the idealized situation where we have full equilibrium; otherwise, translational 
symmetry gets broken as well) nonzero value which points in a certain direction. The residual rotational symmetries 
which leaves the orientation of this vector invariant remain unbroken but the other rotations get spontaneously broken.\\
In the context of the condensed matter physics
the qualitative explanations for the Goldstone theorem~\cite{lang65,lange,wagn66} is that for
a Hamiltonian with a continuous symmetry many different degenerate
ordered states can be realized (e.g. a Heisenberg ferromagnet in which
all directions of the magnetization are possible). The collective mode
with $k \rightarrow 0$ describes a very slow transition of one such state to another
(e.g. an extremely slow rotation of the total magnetization of the sample
as a whole). Such a very slow variation of the magnetization should cost
no energy and hence the dispersion curve $E(k)$ starts from $E = 0$ when
$k \rightarrow 0$, i,e, there exists a gapless excitation.
An important point is that for the Goldstone modes to appear the
interactions need to be short ranged. In the so called   Lieb-Mattis modes~\cite{dmat06} the
interactions between spins are effectively infinitely long ranged, as in
the model a spin on a certain sublattice interacts with all spins on the
other sublattice, independent of the "distance" between the spins. Thus
there are no Goldstone modes in the Lieb-Mattis model~\cite{dmat06} and the spin excitations
are gapped. A physically more relevant example is the plasmon: the
electromagnetic interactions are very long ranged, which leads to a gap
in the excitation spectrum of bulk plasmons. It is possible to show that in
a $2D$ sheet of electrons the dispersion $E(k) \sim \sqrt{k}$, thus in this case the
Coulomb interaction is  not long ranged enough   to induce a gap in the
excitation spectrum. Also in the case of the breaking of gauge invariance
there is an important distinction between charge neutral systems,
e.g. a Bose condensate of $He^{4}$, where there is a Goldstone mode called
the Bogoliubov sound excitations~\cite{bb}, whereas in a charge condensate, e.g.
a superconductor, the elementary excitations  
are gapped due to the long range character of Coulomb interactions.
These considerations on the elementary excitations in symmetry broken
systems are important in order to establish whether or not long
range order is possible at all.\\
The Goldstone theorem~\cite{lang65,lange,wagn66} states that, in a system with
broken continuous symmetry ( {\it i.e.,} a system such that the
ground state is not invariant under the operations of a
continuous unitary group whose generators commute with the
Hamiltonian ), there exists a collective mode with frequency
vanishing as the momentum goes to zero. For many-particle systems
on a lattice, this statement needs a proper adaptation. In the
above form, the Goldstone theorem is true only if the condensed
and  normal phases have the same translational properties. When
translational symmetry is also broken, the Goldstone mode appears
at   zero frequency but at nonzero momentum, {\it e.g.}, a
crystal and a helical spin-density-wave (SDW) ordering.\\ 
All the four models considered above, the
Heisenberg model, the Hubbard model,  the  Anderson and spin-fermion models, are spin
rotationally invariant, $ \mathcal{R} H \mathcal{R^{\dag}} = H$. The spontaneous
magnetization of the spin or fermion system on a lattice that
possesses  the spin rotational invariance, indicate on a broken
symmetry  effect, {\it i.e.,} that the physical ground state is
not an eigenstate of the time-independent generators of symmetry
transformations on the original Hamiltonian of the system. As a
consequence, there must exist an excitation mode, that is an
analog of the Goldstone mode for the continuous case (referred to
as "massless" particles). It was shown that both the models, the
Heisenberg model and the band or itinerant electron model of a
solid, are capable of describing the theory of spin waves for
ferromagnetic insulators and metals~\cite{kuz2}. In their
paper~\cite{kit51}, Herring and Kittel showed that in simple
approximations the spin waves can be described equally well in
the framework of the model of localized spins or the model of
itinerant electrons. Therefore the study of, for example, the
temperature dependence of the average moment in magnetic
transition metals in the framework of low-temperature spin-wave
theory does not, as a rule, give any indications in favor of a
particular model.  Moreover, the itinerant electron model (as
well as the localized spin model) is capable of accounting for the
exchange stiffness determining the properties of the transition
region, known as the Bloch wall, which separates adjacent
ferromagnetic domains with different directions of magnetization.
The spin-wave stiffness constant $D_{sw}$ is defined so~\cite{kit51,bar78} that the
energy of a spin wave with a small wave vector $\vec q$ is $E
\sim D_{sw}q^2$. To characterize the dynamic behavior of the magnetic
systems in terms of the quantum many-body theory, the generalized
spin susceptibility (GSS) is a very useful tool~\cite{low92}. The
GSS is defined by
\begin{equation}
\label{sm1} \chi (\vec q, \omega) = \int \langle\langle S^{-}_{q}(t),S^{+}_{-q} \rangle\rangle \exp{(-i\omega t)}  dt
\end{equation}
Here
$ \langle\langle A(t),B \rangle\rangle$  is the retarded two-time temperature
Green function~\cite{bb,tyab} defined by
\begin{eqnarray}
\label{sm2} G^{r} = \langle\langle A(t), B(t') \rangle\rangle^{r} = -i\theta(t -
t')\langle[A(t)B(t')]_{\eta}\rangle, \eta = \pm 1.
\end{eqnarray}
where $\langle \ldots \rangle$ is the average over a grand canonical ensemble,
$\theta(t)$ is a step function, and square brackets represent the
commutator or anticommutator
\begin{equation} \label{sm3}
[A,B]_{\eta} = AB - \eta BA
\end{equation}
The Heisenberg representation is given by:
\begin{equation}
\label{sm4} A(t) =\exp (i H t) A(0) \exp (- i H t).
\end{equation}
For the Hubbard model $ S^{-}_{i} =
a^{\dag}_{i\downarrow}a_{i\uparrow}$. This GSS satisfies the
important sum rule
\begin{equation}
\label{sm5} \int \textrm{Im} \chi (\vec q, \omega)d\omega =  \pi (
n_{\downarrow} - n_{\uparrow}) = - 2\pi \langle S^{z} \rangle
\end{equation}
It is possible to check that~\cite{kuz2}
\begin{equation}
\label{sm6} \chi (\vec q, \omega) = -\frac{2\langle S^{z} \rangle}{\omega} +
\frac{q^2}{\omega^{2}} \{ \Psi(\vec q, \omega) - \frac{1}{q}
\langle [ Q^{-}_{q}, S^{+}_{-q}] \rangle \}.
\end{equation}
Here the following notation was used for $q Q^{-}_{q} = [
S^{-}_{q}, H ] $ and $\Psi(\vec q, \omega) = \langle\langle Q^{-}_{q} |
Q^{+}_{-q} \rangle\rangle_{\omega}.$ 
It is clear from (\ref{sm5}) that  for
$q = 0$ the GSS (\ref{sm6}) contains only the first term
corresponding to the spin-wave pole for $q = 0$ which exhausts
the sum rule (\ref{sm5}). For small $q$, due to the continuation
principle, the GSS $\chi (\vec q, \omega)$ must be dominated by
the spin wave pole with the energy
\begin{equation}
\label{sm7} \omega = Dq^2 = \frac{1}{2\langle S^{z} \rangle}\{ q  \langle[
Q^{-}_{q}, S^{+}_{-q}]\rangle - q^{2} \lim_{\omega \rightarrow 0}
\lim_{q \rightarrow 0}
\Psi(\vec q, \omega) \}
\end{equation}
This result is the direct consequence of the spin rotational
invariance and is valid for all the four models considered
above.
%
%
%
\subsection{Quantum Protectorate and Microscopic Models of Magnetism}
%
The main problem of  the quantum theory of magnetism lies in a
choosing of the most adequate microscopic model of magnetism of
materials. The essence of this problem is related with the duality of localized
and itinerant behavior of electrons. In describing of that duality the microscopic theory
meets the most serious difficulties~\cite{kuz09}. This is the central issue of  the quantum theory of magnetism.\\ 
The idea of quantum protectorate~\cite{pnas}  reveals the essential
difference in the behavior of the complex many-body systems at the low-energy
and high-energy scales. There are many examples of the quantum protectorates~\cite{pnas}.
According this point of view  the nature of the underlying theory is unknowable
until one raises the energy scale sufficiently to escape
protection. The existence of two scales, the low-energy and
high-energy scales, relevant to the description of magnetic
phenomena was stressed by the author of this
report in the papers~\cite{kuz02,kuz09} devoted to comparative
analysis  of localized and band models of
quantum theory of magnetism.
It was suggested by us~\cite{kuz02} that the
difficulties in the formulation of   quantum theory of magnetism
at the  microscopic level,  that are related to the choice of
relevant models,  can be understood better in the light of the quantum protectorate
concept~\cite{kuz02}. We argued that the difficulties in the formulation of
adequate microscopic models of electron and magnetic properties of
materials are intimately related to dual, \textbf{itinerant} and \textbf{localized}
behavior of electrons. We formulated a criterion of what basic
picture describes best  this dual behavior. The main suggestion
is that quasi-particle excitation spectra might provide
distinctive signatures and good criteria for the appropriate
choice of the relevant model. It was shown there~\cite{kuz02}, that
the low-energy spectrum of magnetic excitations in the
magnetically-ordered solid bodies corresponds to a
hydrodynamic pole ($\vec{k}, \omega \rightarrow 0 $) in the generalized
spin susceptibility  $\chi$, which is present in the Heisenberg,
Hubbard, and the combined $s-d$ model. In
the Stoner band model the hydrodynamic pole is
absent, there are no spin waves there. At the same time,
the Stoner single-particle's excitations are absent in the
Heisenberg model's spectrum. The Hubbard model  
with narrow energy bands contains both types
of excitations: the collective spin waves (the low-energy
spectrum) and Stoner single-particle's excitations
(the high-energy spectrum). This is a big advantage
and flexibility of the Hubbard model in comparison
to the Heisenberg model. The latter, nevertheless, is
a very good approximation to the realistic behavior in
the limit $\vec{k}, \omega \rightarrow 0,$
the domain where the hydrodynamic description is
applicable, that is, for long wavelengths and low energies.
The quantum protectorate concept was applied to
the quantum theory of magnetism by the author of this
report in the paper~\cite{kuz02}, where a criterion of applicability of models
of the quantum theory of magnetism  to
description of concrete substances was formulated. The
criterion is based on the analysis of the model's low-energy
and high-energy spectra. 
%
%
%
%
\section{Bogoliubov's Quasiaverages in Statistical Mechanics}
%
%
Essential progress in the understanding of the spontaneously
broken symmetry concept is connected with
Bogoliubov's fundamental   ideas of quasiaverages~\cite{nnb61,bb,bs75,petr95,nnb60}. 
In the work of N. N. Bogoliubov
"Quasiaverages in Problems of Statistical Mechanics" the innovative notion of \emph{quasiaverege}~\cite{nnb61}
was introduced and applied to various problem of statistical physics. In particular,
quasiaverages of Green's functions constructed from ordinary averages, degeneration 
of statistical equilibrium states, principle of weakened correlations, and particle pair states were considered. 
In this framework the $1/q^{2}$-type properties in the theory of the superfluidity of Bose and Fermi systems, the properties of basic 
Green functions for a Bose system in the presence of condensate, and a model with separated condensate 
were analyzed.\\
The method of quasiaverages is a constructive workable scheme for studying systems with spontaneous symmetry breakdown. 
A quasiaverage is a thermodynamic (in statistical mechanics) or vacuum (in quantum field theory) average of dynamical 
quantities in a specially modified averaging procedure, enabling one to take into account the effects of the influence of 
state degeneracy of the system.
The method gives the so-called macro-objectivation of the degeneracy in the domain of quantum statistical mechanics and in quantum physics.
In statistical mechanics, under spontaneous symmetry breakdown one can, by using the method of quasiaverages, describe 
macroscopic observable within the framework of the microscopic approach.\\
In considering problems of findings the eigenfunctions in quantum mechanics it is well known that the theory of 
perturbations should be modified substantially for the degenerate systems. In the problems of statistical mechanics
we have always the degenerate case due to existence of the additive conservation laws.
The traditional approach to quantum statistical mechanics~\cite{petr95,uhle66} is based on the unique canonical
quantization of classical Hamiltonians for systems with finitely many degrees of freedom together with the
ensemble averaging in terms of traces involving a statistical operator $\rho$.
For an operator $\hat{A}$ corresponding to some physical quantity $A$ the average value of $A$ will be given as
\begin{equation}\label{q1}
\langle A \rangle_{H}  =   \textrm{Tr} \rho A ; \quad  \rho  = \exp ^{- \beta H} / \textrm{Tr} \exp ^{- \beta H}.         
\end{equation}
where $H$ is the Hamiltonian of the system, $\beta = 1/kT$ is the reciprocal of the temperature.\\
In general, the statistical operator~\cite{zub71} or density matrix $\rho$ is defined by its matrix elements in 
the $\varphi_{m}$-representation:
\begin{equation}\label{q2}
 \rho_{nm}  = \frac{1}{N} \sum_{i=1}^{N}c_{n}^{i}(c_{m}^{i})^{*}.         
\end{equation}
In this notation the average value of $A$ will be given as
\begin{equation}\label{q3}
 \langle A \rangle = \frac{1}{N} \sum_{i=1}^{N}   \int  \Psi_{i}^{*} A \Psi_{i}d\tau.         
\end{equation}
The averaging in Eq.(\ref{q3}) is both over the state of the $i$th system and over all the systems in the ensemble.
The Eq.(\ref{q3}) becomes
\begin{equation}\label{q4}
 \langle A \rangle = \textrm{Tr} \rho A; \quad     \textrm{Tr} \rho = 1.     
\end{equation}
Thus an ensemble of quantum mechanical systems is described by a density matrix~\cite{zub71}. In a suitable representation, a density matrix $\rho$ takes the form
$$ \rho = \sum_k p_k |\psi_k \rangle \langle \psi_k|$$
where $p_k$ is the probability of a system chosen at random from the ensemble will be in the microstate $  |\psi_k \rangle.$ 
So the trace of $\rho$, denoted by $ \textrm{Tr}(\rho),$ is $1.$ This is the quantum mechanical analogue of the fact that the accessible region of the classical phase 
space has total probability $1.$
It is also assumed that the ensemble in question is stationary, i.e. it does not change in time. Therefore, by Liouville  
theorem, $[\rho, H] = 0,$ i.e. 
$\rho H = H \rho $ where $H$ is the Hamiltonian of the system. Thus the density matrix describing $\rho$ is diagonal 
in the energy representation.\\
Suppose that
$$ H = \sum_i E_i |\psi_i \rangle \langle \psi_i|$$
where $ E_i$ is the energy of the $i$-th energy eigenstate. If a system $i$-th energy eigenstate has $n_i$ number of 
particles, the corresponding 
observable, the number operator, is given by
$$ N = \sum_i n_i |\psi_i \rangle \langle \psi_i|.$$
It is known~\cite{zub71},  that the state
$   |\psi_i \rangle $ 
has (unnormalized) probability
$$  p_i = e^{-\beta (E_i - \mu n_i)}.$$
Thus the grand canonical ensemble is the mixed state
\begin{eqnarray}\label{q5}
 \rho = \sum_i p_i |\psi_i \rangle \langle \psi_i| = \\ \nonumber
 \sum_i e^{-\beta (E_i - \mu n_i)} |\psi_i \rangle \langle \psi_i| = e^{- \beta (H - \mu N)}. 
\end{eqnarray}
The grand partition, the normalizing constant for $\textrm{Tr}(\rho)$ to be $1,$ is
$$  {\mathcal Z} =\mathbf{Tr} [ e^{- \beta (H - \mu N)} ]. $$
Thus we obtain~\cite{zub71}
\begin{equation}\label{q6}
 \langle A \rangle = \textrm{Tr} \rho A  =  \textrm{Tr} e^{ \beta (\Omega  - H + \mu N)} A  .     
\end{equation}
Here $\beta = 1/k_{B}T$ is the reciprocal temperature and $\Omega $ is the normalization factor.\\
It is known~\cite{zub71} that the averages $\langle A \rangle$ are unaffected by a change of representation.  The most important is the
representation in which $\rho$ is diagonal $\rho_{mn} = \rho_{m} \delta_{mn}$. We then have
$\langle \rho \rangle = Tr \rho^{2} = 1.$ It is clear then that  $Tr \rho^{2} \leq 1$ in any representation.
The core of the problem lies in establishing the existence of a thermodynamic limit (such as $N/V = $ const, 
$V \rightarrow \infty$, $N$ = number of degrees of freedom, $V$ = volume) and its evaluation for the quantities of interest.\\
The evolution equation for the density matrix is a quantum analog of the Liouville equation in classical mechanics. A related equation describes the time evolution of 
the expectation values of observables, it is given by the Ehrenfest theorem.
Canonical quantization yields a quantum-mechanical version of this theorem. This procedure, often used to devise quantum analogues of 
classical systems, involves describing a classical system using Hamiltonian mechanics. Classical variables are then re-interpreted as quantum 
operators, while Poisson brackets are replaced by commutators. In this case, the resulting equation is
\begin{equation}\label{q7}
  \frac{\partial}{\partial t}\rho = -\frac{i}{\hbar}[H,\rho]
\end{equation}
where $\rho$ is the density matrix. 
When applied to the expectation value of an observable, the corresponding equation is given by Ehrenfest  theorem, and takes the form
\begin{equation}\label{q8}
  \frac{d}{dt}\langle A\rangle = \frac{i}{\hbar}\langle [H,A] \rangle
\end{equation}
where $A$ is an observable.
Thus in the statistical mechanics the average $\langle A \rangle$ of any dynamical quantity $A$
is defined in a single-valued way.\\
In the situations with degeneracy  the specific problems appear. In quantum mechanics, if two linearly independent
state vectors (wavefunctions in the Schroedinger picture) have the same energy, there is a degeneracy~\cite{hill40}.
In this case more than one independent state of the system corresponds to a single energy level.  
If the statistical equilibrium state of the system 
possesses lower symmetry than the Hamiltonian of the system (i.e. the situation with the spontaneous symmetry breakdown), 
then it is necessary to supplement the averaging procedure  (\ref{q6}) by a rule forbidding  irrelevant   averaging over the values of macroscopic quantities considered 
for which a change is not accompanied by a change in energy.\\
This is achieved by introducing quasiaverages, that is, averages over the Hamiltonian $H_{\nu \vec{e}}$ 
supplemented by infinitesimally-small 
terms that violate the additive conservations laws 
$H_{\nu \vec{e}} = H + \nu (\vec{e}\cdot \vec{M})$, ($\nu \rightarrow 0$). Thermodynamic 
averaging may turn out to be unstable with respect to such a change of the original 
Hamiltonian, which is another indication of degeneracy of the equilibrium state.\\
According to Bogoliubov~\cite{nnb61}, the quasiaverage of a dynamical quantity  $A$ for the system with the
Hamiltonian $H_{\nu \vec{e}}$  
is defined as the limit
\begin{equation}
\label{q9}
\curlyeqprec A \curlyeqsucc = \lim_{\nu \rightarrow 0} \langle A \rangle_{\nu \vec{e}},
\end{equation}
where $\langle A \rangle_{\nu \vec{e}}$  denotes the ordinary average taken over the Hamiltonian $H_{\nu \vec{e}}$, containing the small symmetry-breaking terms introduced by the inclusion 
parameter $\nu$, which vanish as $\nu \rightarrow 0$ after passage to the thermodynamic limit $V \rightarrow \infty$. 
Thus the existence of degeneracy is reflected directly in the quasiaverages by their dependence upon the arbitrary
unit vector $\vec{e}$.
It is also clear that
\begin{equation}
\label{q10}
\langle A \rangle = \int \curlyeqprec A \curlyeqsucc d \vec{e}.
\end{equation}
According to definition (\ref{q10}), the ordinary thermodynamic average is obtained by extra averaging of the 
quasiaverage over the symmetry-breaking group. Thus to describe the case of a degenerate state of statistical
equilibrium quasiaverages are more convenient, more physical, than ordinary averages~\cite{petr95,uhle66}. The latter are the same 
quasiaverages only averaged over all the directions $\vec{e}$.\\
It is necessary to stress,
that the starting point for Bogoliubov's work~\cite{nnb61} was
an investigation of additive conservation laws and
selection rules, continuing and developing the  approach by P. Curie for derivation of
selection rules for physical effects. Bogoliubov demonstrated
that in the cases when the state of statistical
equilibrium is degenerate, as in the case of the Heisenberg ferromagnet (\ref{he1}),
one can remove the degeneracy of equilibrium
states with respect to the group of spin rotations by
including in the Hamiltonian $H$ an additional noninvariant
term $\nu M_{z} V$ with an infinitely small $\nu$. Thus the quasiaverages do not follow the same selection rules as
those which govern the ordinary averages. For the Heisenberg ferromagnet the ordinary averages must be invariant
with regard to the spin rotation group. The corresponding quasiaverages possess only the property of covariance.
 It is clear that  the unit vector $\vec{e}$, i.e., the
direction of the magnetization $\vec{M}$ vector, characterizes the degeneracy of the considered state of statistical
equilibrium. In order to remove the degeneracy one should fix the direction  of the unit vector $\vec{e}$. It can be
chosen to be along the $z$ direction. Then all the quasiaverages will be the definite numbers. This is the kind 
that one usually deals with in the theory of ferromagnetism.\\
The value of the quasi-average (\ref{q9}) may depend on the concrete structure of the additional 
term  $\Delta H = H_{\nu} - H$, if the dynamical 
quantity to be averaged is not invariant with respect to the symmetry group of the original Hamiltonian $H$. For a degenerate 
state the limit of ordinary averages (\ref{q10}) as the inclusion parameters $\nu$  of the sources tend to zero in an arbitrary 
fashion, may not exist. For a complete definition of quasiaverages it is necessary to indicate the manner in which these 
parameters tend to zero in order to ensure convergence~\cite{nnbj}. On the other hand, in order to remove degeneracy it 
suffices, in the construction of $H$, to violate only those additive 
conservation laws whose  switching  lead  to instability of the ordinary average. 
Thus  in terms of quasiaverages the selection rules for 
the correlation functions~\cite{petr95,nnb61w} that are not relevant are those that are restricted by these conservation 
laws.\\
By using $H_{\nu}$, we define the state $\omega(A) = \langle A \rangle_{\nu}$ and then let $\nu$ tend to 
zero (after passing to the thermodynamic limit). If all averages $\omega(A)$ get infinitely small increments under 
infinitely small perturbations $\nu$, this means that
the state of statistical equilibrium under consideration is nondegenerate~\cite{petr95,nnb61w}. However, if some states 
have finite increments
as $\nu \rightarrow 0$, then the state is degenerate. In this case, instead of ordinary averages $\langle A \rangle_{H}$, one should introduce 
the quasiaverages (\ref{q9}), for which the usual selection rules do not hold.\\
The method of quasiaverages is directly related to the principle  weakening of the  correlation~\cite{petr95,nnb61w} in 
many-particle systems. According to this principle, the notion of the weakening of the  correlation, known in
statistical mechanics~\cite{bb,petr95}, in the case of state degeneracy must be interpreted in the sense 
of the quasiaverages~\cite{nnb61w}.\\
The quasiaverages may be obtained from the ordinary averages by using the cluster property which was formulated
by Bogoliubov~\cite{nnb61w}. This was first done when deriving the Boltzmann equations from the chain of equations
for distribution functions, and in the investigation of the model Hamiltonian in the theory 
of superconductivity~\cite{nnb61,bb,bs75,petr95,nnb60}.  To demonstrate this
let us   consider averages (quasiaverages) of the form
\begin{equation}
\label{q11}
F(t_{1},x_{1}, \ldots t_{n},x_{n}) = \langle \ldots \Psi^{\dag}(t_{1},x_{1}) \ldots \Psi(t_{j},x_{j}) \ldots \rangle,
\end{equation}
where the number of creation operators $\Psi^{\dag}$ may be not equal to the number of annihilation operators $\Psi$. We 
fix times and split the arguments $(t_{1},x_{1}, \ldots t_{n},x_{n})$ into several 
clusters $( \ldots , t_{\alpha},x_{\alpha}, \ldots ), \ldots ,$
$( \ldots , t_{\beta},x_{\beta}, \ldots ).$  Then it is reasonably to assume that the
distances between all clusters $|x_{\alpha} - x_{\beta}|$ tend to infinity. Then, according to the cluster property, the
average value (\ref{q11}) tends to the product of averages of collections of operators with the arguments
$( \ldots , t_{\alpha},x_{\alpha}, \ldots ), \ldots ,$  $( \ldots , t_{\beta},x_{\beta}, \ldots )$ 
\begin{equation}
\label{q12}
\lim_{|x_{\alpha} - x_{\beta}| \rightarrow \infty}  F(t_{1},x_{1}, \ldots t_{n},x_{n}) = 
F( \ldots , t_{\alpha},x_{\alpha}, \ldots )  \ldots   F( \ldots , t_{\beta},x_{\beta}, \ldots ).
\end{equation}
For equilibrium states with small densities and short-range potential, the validity of this property can be proved~\cite{petr95}.
 For the general case, the validity of the cluster property has not yet been proved. Bogoliubov
formulated it not only for ordinary averages but also for quasiaverages, i.e., for anomalous averages, too. It works
for many important models, including the models of superfluidity and superconductivity.\\
To illustrate  this statement consider Bogoliubov's theory of a Bose-system with separated condensate, which is given by the 
Hamiltonian~\cite{bb,petr95}
\begin{eqnarray}\label{q13}
H_{\Lambda}  = \int_{\Lambda}\Psi^{\dag}(x)(- \frac{\Delta}{2m})\Psi(x)dx - \mu \int_{\Lambda}\Psi^{\dag}(x)\Psi(x)dx \\ \nonumber
+ \frac{1}{2} \int_{\Lambda^{2}}\Psi^{\dag}(x_{1}) \Psi^{\dag}(x_{2}) \Phi (x_{1} - x_{2})   \Psi(x_{2})\Psi(x_{1})dx_{1} dx_{2}.
\end{eqnarray} 
This Hamiltonian can be written also in the following form
\begin{eqnarray}\label{q14}
H_{\Lambda}  =  H_{0} + H_{1} = \int_{\Lambda}\Psi^{\dag}(q)(- \frac{\Delta}{2m})\Psi(q)d q \\ \nonumber
+ \frac{1}{2} \int_{\Lambda^{2}}\Psi^{\dag}(q) \Psi^{\dag}(q') \Phi (q - q')   \Psi(q')\Psi(q)d q d q'.
\end{eqnarray} 
Here, $\Psi(q)$, and $\Psi^{\dag}(q)$  are the operators of annihilation and creation of bosons. They satisfy the 
canonical commutation relations
\begin{equation}
\label{q15}
 [\Psi(q),\Psi^{\dag}(q')]  = \delta (q - q'); \quad [\Psi(q),\Psi(q')]  = [\Psi^{\dag}(q),\Psi^{\dag}(q')]  = 0.
\end{equation}
The system of bosons is contained in the cube $A$ with the edge $L$ and volume $V$. It was assumed that it satisfies
periodic boundary conditions and the potential $\Phi(q)$ is spherically symmetric and proportional 
to the small parameter.
It was also assumed that, at temperature zero, a certain macroscopic number of particles having a nonzero
density is situated in the state with momentum zero.\\
The operators $\Psi(q)$, and $\Psi^{\dag}(q)$  are represented in the form
\begin{equation}
\label{q16}
  \Psi(q)  = a_{0}/\sqrt{V};  \quad \Psi^{\dag}(q) = a^{\dag}_{0}/\sqrt{V},
\end{equation}
where $a_{0}$ and $a^{\dag}_{0}$ are the operators of annihilation and creation of particles with momentum zero.\\
To explain the phenomenon of
superfluidity, one should calculate the spectrum of the Hamiltonian, which is quite a difficult problem. Bogoliubov
suggested the idea of approximate calculation of the spectrum of the ground state and its elementary excitations
based on the physical nature of superfluidity. His idea consists of a few assumptions. The main assumption
 is   that at temperature zero the macroscopic number of particles (with nonzero density) has the
momentum zero. Therefore, in the thermodynamic limit, the operators $a_{0}/\sqrt{V}$ and $a^{\dag}_{0}/\sqrt{V}$ commute
\begin{equation}
\label{q17}
\lim_{V \rightarrow \infty} \left [ a_{0}/\sqrt{V}, a^{\dag}_{0}/\sqrt{V} \right ] = \frac{1}{V} \rightarrow 0
\end{equation}
and are $c$-numbers. Hence, the operator of the number of particles $N_{0} =a^{\dag}_{0}a_{0}$  is a $c$-number, too.
It is worth noting that the Hamiltonian (\ref{q14}) is invariant under the gauge 
transformation $\tilde{a}_{k} = \exp (i \varphi) a_{k}$, $\tilde{a}^{\dag}_{k} = \exp ( - i \varphi) a^{\dag}_{k}$, where $\varphi$ is an
arbitrary real number. Therefore, the averages $\langle a_{0}/\sqrt{V} \rangle$ and  
$\langle a^{\dag}_{0}/\sqrt{V} \rangle$ must vanish. But this contradicts to the
assumption  that $ a_{0}/\sqrt{V}$ and $a^{\dag}_{0}/\sqrt{V}$ must become $c$-numbers in the thermodynamic limit. 
In addition it must be taken into account that $a^{\dag}_{0}a_{0}/V = N_{0}/V \neq 0$
which implies that $a_{0}/\sqrt{V} = N_{0} \exp (i \alpha)/\sqrt{V} \neq 0$ and 
$a^{\dag}_{0}/\sqrt{V} = N_{0}\exp (- i \alpha)/\sqrt{V} \neq 0$, where $\alpha$ is an arbitrary
real number. This contradiction may be overcome if we assume that the eigenstates of the Hamiltonian are degenerate
and not invariant under gauge transformations, i.e., that a spontaneous breaking of symmetry takes place.\\
Thus the averages $\langle a_{0}/\sqrt{V} \rangle$ and ($\langle a^{\dag}_{0}/\sqrt{V} \rangle$, which are nonzero 
under spontaneously broken gauge invariance,
are called anomalous averages or \emph{quasiaverages}. This innovative idea of Bogoliubov penetrate deeply into the 
modern quantum physics. The systems with spontaneously
broken symmetry are studied by use of the transformation of the operators of the form 
\begin{equation}
\label{q18}
 \Psi(q)  = a_{0}/\sqrt{V} + \theta(q);  \quad \Psi^{\dag}(q) = a^{\dag}_{0}/\sqrt{V} + \theta^{*}(q),
\end{equation}
where $a_{0}/\sqrt{V}$ and $a^{\dag}_{0}/\sqrt{V}$ are the numbers first introduced by Bogoliubov in 1947 in his investigation of the
phenomenon of superfluidity~\cite{bb,nnb61,petr95,bogo47}. The main conclusion was made that for the systems with spontaneously broken symmetry, the 
quasiaverages should be studied instead of the ordinary averages. It turns out that the long-range order appears not only in the system of
Bose-particles but also in all systems with spontaneously broken symmetry. Bogoliubov's papers outlined above
anticipated the methods of investigation of systems with spontaneously broken symmetry for many years.\\ 
As mentioned above, in order to explain the phenomenon of superfluidity, Bogoliubov assumed that the operators
$a_{0}/\sqrt{V}$ and $a^{\dag}_{0}/\sqrt{V}$ become $c$-numbers in the thermodynamic limit. This statement was rigorously 
proved in the papers by Bogoliubov and some other authors.
Bogoliubov's proof was based on the study of the equations for two-time Green's functions and on the assumption
that the cluster property holds. It was proved that the solutions of equations for Green's functions for the system
with Hamiltonian (\ref{q14})  coincide with the solutions of the equations for the system with the same Hamiltonian in
which the operators $a_{0}/\sqrt{V}$ and $a^{\dag}_{0}/\sqrt{V}$ are replaced by numbers. These numbers should be determined from
the condition of minimum for free energy. Since all the averages in both systems coincide, their free energies coincide, too.\\
It is worth noting that the validity of the replacement of the operators 
$a_{0}$ and $a^{\dag}_{0}$ by $c$-numbers in the thermodynamic limit was confirmed  in the
numerous subsequent publications of various authors~\cite{gin68,lieb04,chin06}. Lieb, Seiringer and Yngvason~\cite{lieb04} analyzed justification of $c$-number
substitutions in bosonic Hamiltonians. The validity of substituting a $c$-number $z$ for the $k = 0$  mode 
operator $a_{0}$ was established rigorously in
full generality, thereby verifying that aspect of Bogoliubov's 1947 theory. The authors shown that this substitution not only yields
the correct value of thermodynamic quantities such as the pressure or ground state energy, but also the
value of $|z|^{2}$ that maximizes the partition function equals the true amount of condensation in the presence
of a gauge-symmetry-breaking term. This point had previously been elusive.
Thus Bogoliubov's 1947 analysis of the many-body Hamiltonian by means of a
$c$-number substitution for the most relevant operators in
the problem, the zero-momentum mode operators, was justified rigorously.
Since the Bogoliubov's 1947 analysis is one of the key developments in the theory of the Bose
gas, especially the theory of the low density gases currently
at the forefront of experiment, this result is of importance for the legitimation of that theory. Additional arguments
were given in Ref.~\cite{chin06}, where the 
Bose-Einstein condensation and spontaneous $U(1)$ symmetry breaking were investigated.
Based on Bogoliubov's truncated Hamiltonian $H_{B}$ for a weakly interacting Bose system, and adding a $U(1)$
symmetry breaking term $ \sqrt{V} ( \lambda a_{0} +  \lambda^{*} a^{\dag}_{0})$
 to $H_{B}$, authors shown by using the coherent state theory and the mean-field
approximation rather than the $c$-number approximations, that the Bose-Einstein condensation   occurs if and only
if the $U(1)$ symmetry of the system is spontaneously broken. The real ground state energy and the justification of the
Bogoliubov $c$-number substitution were given by solving the Schroedinger eigenvalue equation and using the self-consistent
condition. Thus the Bogoliubov
$c$-number substitutions were fully correct and   the symmetry breaking causes the displacement of the condensate
state.\\
The concept of quasiaverages was introduced by Bogoliubov on the basis of an analysis of many-particle systems with
a degenerate statistical equilibrium state. Such states are inherent to various physical many-particle systems~\cite{bb,petr95}. Those are
liquid helium in the superfluid phase, metals in the superconducting state, magnets in the ferromagnetically ordered state,
liquid crystal states, the states of superfluid nuclear matter, etc. (for a review, see Refs.~\cite{kuz09,kov06}).
In case of superconductivity, the 
source 
$$\nu \sum_{k}v(k)(a^{\dag}_{k\uparrow}a^{\dag}_{-k\downarrow} + a_{-k\downarrow}a_{k\uparrow})$$ 
was inserted 
in the BCS-Bogoliubov Hamiltonian, and the quasiaverages were defined by use of the Hamiltonian $H_{\nu}$. 
In the general case, the sources are introduced to remove degeneracy. If infinitesimal sources give infinitely small contributions
to the averages, then this means that the corresponding degeneracy is absent, and there is no reason to insert
sources in the Hamiltonian. Otherwise, the degeneracy takes place, and it is removed by the sources. The ordinary
averages can be obtained from quasiaverages by averaging with respect to the parameters that characterize the
degeneracy.\\
N. N. Bogoliubov, Jr.~\cite{nnbj} considered some features of quasiaverages for model systems with four-fermion interaction.
He discussed the treatment of certain three-dimensional model systems which can be solved exactly. For this aim
a new effective way of defining quasiaverages for the systems under consideration was proposed.\\
Peletminskii and Sokolovskii~\cite{peso74} have found general expressions for the operators of the flux densities of physical
variables in terms of the density operators of these variables. The method of quasiaverages and the expressions 
found for the flux operators were used to obtain the averages of these operators in terms of the thermodynamic
potential in a state of statistical equilibrium of a superfluid liquid. \\ Vozyakov~\cite{voz79} reformulated the theory
of quantum crystals in terms of quasiaverages. He analyzed a Bose system with periodic distribution of particles
which simulates an ensemble in which the particles cannot be regarded as vibrating independently about a position of 
equilibrium lattice sites. With allowance for macroscopic filling of the states corresponding to the
distinguished symmetry, a calculation was made of an excitation spectrum in which there exists a collective
branch of gapless type.\\
Peregoudov~\cite{pere97} discussed
the effective potential method, used in quantum field theory to study spontaneous symmetry breakdown,  
 from the point of view of Bogoliubov's quasiaveraging procedure. It was shown that the effective
potential method is a disguised type of this procedure. The catastrophe theory approach to the study of
phase transitions was discussed and the existence of the potentials used in that approach was proved from the
statistical point of view. It was shown that in the ease of broken symmetry, the nonconvex effective potential
is not a Legendre transform of the generating functional for connected Green's functions. Instead, it is a
part of the potential used in catastrophe theory. The relationship between the effective potential and the
Legendre transform of the generating functional for connected Green's functions is given by  Maxwell's rule.
A rigorous rule for evaluating quasiaveraged quantities within the framework of the effective  potential
method was established.\\
N. N. Bogoliubov, Jr. with M. Yu. Kovalevsky and co-authors~\cite{kov09} developed a statistical approach for solving 
the problem of classification of equilibrium states in condensed media with spontaneously broken
symmetry based on the quasiaverage concept. 
Classification of equilibrium states of condensed media with spontaneously broken symmetry was
carried out. The generators of residual and spatial symmetries
were introduced and equations of classification for the order
parameter has been found.
Conditions of residual symmetry and spatial symmetry were formulated. The connection between
these symmetry conditions and equilibrium states of various media with tensor order parameter was found out.
An analytical solution of the problem of classification of
equilibrium states for superfluid media, liquid crystals and
magnets with tensor order parameters was obtained.
Superfluid $^{3}He$, liquid crystals, quadrupolar magnetics were considered in detail. Possible homogeneous and
heterogeneous states were found out. Discrete and continuous thermodynamic parameters, which define an
equilibrium state, allowable form of order parameter, residual symmetry, and spatial symmetry generators
were established.
This approach, which is alternative to the well-known Ginzburg-Landau method, 
does not contain any model assumptions concerning the form of the free energy as functional of
the order parameter and does not employ the requirement of
temperature closeness to the point of phase transition. For all
investigated cases they found the structure of the order parameters and the explicit forms of generators of residual and
spatial symmetries.
Under the certain restrictions they established the form of
the order parameters in case of spins $0$, $1/2$, $1$ and proposed
the physical interpretation of the studied degenerate states of
condensed media.\\
To summarize, the Bogoliubov's
quasiaverages concept plays an important
role in equilibrium statistical mechanics of many-particle systems. According to
that concept, infinitely small perturbations can trigger
macroscopic responses in the system if they break some
symmetry and remove the related degeneracy (or quasidegeneracy)
of the equilibrium state. As a result, they
can produce macroscopic effects even when the perturbation
magnitude is tend to zero, provided that happens
after passing to the thermodynamic limit. 
%
%
%
\subsection{Bogoliubov Theorem on the  Singularity of $1/q^{2}$} 
%
Spontaneous symmetry breaking in 
a nonrelativistic theory is manifested in a nonvanishing value of a certain macroscopic parameter 
(spontaneous polarization, density of a superfluid component, etc.). In this sense it is intimately related to 
the problem of phase transitions. These problems
were discussed intensively from different points of view in literature~\cite{bb,nnb61,petr95,sado73}. In particular, there
has been an extensive discussion of the conjecture that the spontaneous symmetry breaking corresponds under
a certain restriction on the nature of the interaction to a branch of collective excitations of zero-gap type 
$(\lim_{q \rightarrow 0} E(q) = 0)$. It was shown in the previous sections that some ideas have here been borrowed from the theory of elementary particles, in
which the ground state (vacuum) is noninvariant under a group of continuous transformations that leave
the field equations invariant, and the transition from one vacuum to the other can be described in terms of
the excitation of an infinite number of zero-mass particles (Goldstone bosons).\\
It should be stressed here that the main questions of this kind have already been resolved by Bogoliubov in his
paper~\cite{nnb61}  on models of Bose and Fermi systems of many particles with a gauge-invariant interaction.
Ref.~\cite{sado73} reproduces the line of arguments of the corresponding section in Ref.~\cite{nnb61}, in which 
the inequality for the mass operator  $M$  of a boson system, which is expressed in terms of "normal" and "anomalous" 
Green's functions, made it possible, under the assumption of its regularity for $E = 0$ and $q =0$, to obtain "acoustic"
nature of the energy of the low-lying excitation $(E = \sqrt{s q})$. It is also noted in Ref.~\cite{nnb61} that a "gap" 
in the spectrum of elementary excitations may be due either to a discrepancy in the approximations that are used (for the
mass operator and the free energy) or to a certain choice of the interaction potential, (i. e., essentially to
an incorrect use of quasiaverages).This Bogoliubov's   remark is still important, especially in connection
with the application of different model Hamiltonians to concrete systems.\\
It was demonstrated above that Bogoliubov's fundamental concept  of quasiaverages  is an effective method of
investigating problems related to degeneracy of a state of statistical equilibrium due to the presence of additive
conservation laws or alternatively invariance of the Hamiltonian of the system under a certain group of
transformations. 
The mathematical apparatus of the method of quasi-averages includes the Bogoliubov theorem~\cite{bb,nnb61,petr95} on 
singularities of type $1/q^{2}$  and the Bogoliubov 
inequality for Green and correlation functions as a direct consequence of the method. It includes algorithms 
for establishing non-trivial estimates for equilibrium quasiaverages, enabling one to study the problem of ordering in statistical systems 
and to elucidate the structure of the energy spectrum of the underlying excited states. 
In that sense the mathematical scheme proposed by Bogoliubov~\cite{bb,nnb61,petr95} is a workable tool for 
establishing nontrivial inequalities for equilibrium mean values (quasiaverages)  
for the commutator Green's functions and also the inequalities that majorize it. Those inequalities
enable  one to investigate questions relating
to the specific ordering in models of statistical mechanics and to consider the structure of the energy
spectrum of low-lying excited states in the limit $(q \rightarrow 0)$. \\
Let us consider the the proof of Bogoliubov's theorem  on singularities of $1/q^{2}$-type. For this aim consider the
retarded, advanced, and causal Green's functions of the following form~\cite{bb,nnb61,petr95,kuz02,kuz09}
\begin{eqnarray}
\label{bt1} G^{r}(A,B; t-t') = \langle \langle A(t), B(t')\rangle \rangle^{r} = -i\theta(t -
t')\langle [A(t),B(t')]_{\eta}\rangle, \eta = \pm,
 \\ \label{bt2}
G^{a}(A,B; t-t') = \langle \langle A(t), B(t')\rangle \rangle^{a} =
i\theta(t' - t)\langle [A(t),B(t')]_{\eta}\rangle, \eta = \pm,  \\
G^{c}(A,B; t-t') = \langle \langle A(t), B(t')\rangle \rangle^{c} = i T \langle A(t)B(t')\rangle =   \\ \nonumber
i\theta(t - t')\langle A(t)B(t')\rangle + \eta i\theta (t'- t) \langle B(t')A(t)\rangle ,
\eta = \pm. \label{bt3}
\end{eqnarray}
It is well known~\cite{bb,nnb61,petr95,kuz02,kuz09} that the 
Fourier transforms of the retarded and advanced Green's functions are different limiting values (on the real axis) of
the same function that is holomorphic on the complex ${\mathcal E}$-plane with cuts along the real axis
\begin{eqnarray}
\label{bt4}
\langle \langle A|B \rangle \rangle_{{\mathcal E}} = 
\int^{ + \infty}_{ - \infty} d \omega \frac{J (B, A; \omega)(\exp( \beta \omega) - \eta)}{{\mathcal E} - \omega}
\end{eqnarray}
Here, the function $J (B, A; \omega)$ possesses the properties
\begin{equation}
\label{bt5}
J (A^{\dag}, A; \omega) \geq 0; \quad J^{*} (B, A; \omega) = J (A^{\dag}, B^{\dag}; \omega).
\end{equation}
Moreover, it is a bilinear form of the operators $A = A (0)$ and $B = B (0)$. This implies that the bilinear form
\begin{equation}
\label{bt6}
- \langle \langle A|B \rangle \rangle_{{\mathcal E}} = Z(A,B) =
\int^{ + \infty}_{ - \infty} d \omega \frac{J (B, A; \omega)(\exp( \beta \omega) - \eta)}{\omega}
\end{equation}
possesses the similar properties
\begin{equation}
\label{bt7}
 Z( A, A^{\dag}) \geq 0;\quad  Z^{*}( A, B ) = Z( B^{\dag}, A^{\dag} ).
\end{equation}
Therefore, the bilinear form $Z(A, B)$ possesses all properties of the scalar product~\cite{petr95} in linear space whose elements
are operators $A, B \ldots $ that act in the Fock space of states. This scalar product can be introduced as follows:
\begin{equation}
\label{bt8}
  \left( A,B \right) = Z \left( A^{\dag},B \right ).
\end{equation}
Just as this is proved for the scalar product in a Hilbert space~\cite{petr95}, we can establish the inequality
\begin{equation}
\label{bt9}
 |  \left( A,B \right )|^{2} \leq  (A, A^{\dag}) (B^{\dag},B).
\end{equation}
This implies that $(A, B ) = 0$ if $(A, A ) = 0$ or $(B, B ) = 0.$ If we introduce a factor-space with respect to the collection
of the operators for which $(A, A ) = 0,$ then we obtain an ordinary Hilbert space whose elements are linear
operators, and the scalar product is given by (\ref{bt8}).\\
To illustrate  this line of reasoning consider Bogoliubov's theory of a Bose-system with separated condensate~\cite{bb,petr95}, which is given by the 
Hamiltonian (\ref{q14})
In the system with separated condensate, the anomalous 
averages $\langle a_{0}/\sqrt{V} \rangle$  and  $\langle a^{\dag}_{0}/\sqrt{V} \rangle$ are nonzero.
This indicates
that the states of the Hamiltonian are degenerate with respect to the number of particles. In order to remove
this degeneracy, Bogoliubov inserted infinitesimal terms of the form $\nu \sqrt{V} (a_{0} + a^{\dag}_{0})$ in the Hamiltonian. 
As a result, we obtain the Hamiltonian
\begin{equation}
\label{bt10}
 H_{\nu}  = H - \nu \sqrt{V} \left (  a_{0} + a^{\dag}_{0}   \right )
\end{equation}
For this Hamiltonian, the fundamental theorem "\emph{on singularities of $1/q^{2}$-type}" was proved for Green's 
functions~\cite{bb,nnb61,petr95,sado73}. 
In its simplest version  this theorem   consists in the fact that
the Fourier components of the Green's  functions corresponding to energy $E = 0$ satisfy the inequality
\begin{equation}
\label{bt11}
|\langle \langle a_{q}, a_{q}^{\dag} \rangle \rangle_{E = 0}| \geq \frac{\textrm{const}}{q^{2}} \quad \textrm{as} \quad q^{2} \rightarrow 0.
\end{equation}
Here $\langle \langle a_{q}, a_{q}^{\dag} \rangle \rangle_{E = 0}$
is the two-time temperature commutator Green function in the energy representation, and $a_{q}^{\dag}$, $a_{q}$ are the
creation and annihilation operators of a particle with momentum $\vec{q}$. A more detailed consideration gives the
following result~\cite{bb,nnb61,petr95,sado73}
\begin{eqnarray}\label{bt12}
\langle \langle a_{q}, a_{q}^{\dag} \rangle \rangle_{E = 0} \geq 
\frac{N_{0}}{4 \pi \left ( N \frac{q^{2}}{2m} + \nu N_{0} V^{1/2} \right )} = \\ \nonumber
\frac{N_{0} 2m}{4 \pi \left ( N  q^{2}  + \nu 2m N_{0} V^{1/2} \right )} =
\frac{\rho_{0}  m}{4 \pi \left ( \rho  q^{2}  + \nu 2m \sqrt{\rho_{0}}   \right )},\\ \nonumber
\frac{N}{V} = \rho, \, \frac{N_{0}}{V} = \rho_{0}.
\end{eqnarray} 
Finally, by passing here to the limit as $\nu \rightarrow 0,$ we obtain the required inequality
\begin{equation}
\label{bt13}
\langle \langle a_{q}, a_{q}^{\dag} \rangle \rangle_{E = 0}  \geq \frac{\rho_{0} m}{4 \pi \rho}\frac{1}{q^{2}}.     
\end{equation}
The concept of quasiaverages is indirectly related to the theory of phase transition. The instability of thermodynamic 
averages with respect to perturbations of the Hamiltonian by a breaking of the invariance with respect to a certain 
group of transformations means that in the system transition to an extremal state occurs.
In quantum field theory, for a number of model systems it has been proved that there is a phase transition, and the 
validity of the Bogoliubov theorem on singularities of type $1/q^{2}$  has been established~\cite{bb,nnb61,petr95}. In addition, the possibility has been 
investigated of local instability of the vacuum and the appearance of a changed structure in it.\\
In summary, the main achievement of the method of quasiaverages is the fundamental Bogoliubov theorem~\cite{bb,nnb61,petr95,sado73} on the
singularity of $1/q^{2}$ for Bose and Fermi systems with gauge-invariant interaction between particles.
The singularities in the Green functions specified in Bogoliubov's theorem which appear when correspond to 
elementary excitations in the physical system under study. Bogoliubov's theorem also predicts the asymptotic behavior 
for small momenta of macroscopic properties of the system which are connected with Green functions by familiar theorems.
The theorem establishes   the asymptotic behavior of Green functions  in the limit of small momenta $(q \rightarrow 0)$
for systems of  interacting particles in the case of a degenerate statistical equilibrium state.\\  
The appearance of singularities in the Green functions as $(q \rightarrow 0)$  is connected with the presence of a branch of 
collective excitations in the energy spectrum of the system that corresponds with spontaneous symmetry breaking 
under certain restrictions on the interaction potential.\\
The nature of the energy spectrum of elementary excitations may be studied with the aid of the 
mass (or self-energy) operator $M$
inequality constructed for Green functions of type (\ref{bt11}). In the case of Bose systems, for a finite temperature , this 
inequality has the form:
\begin{equation}
\label{bt14}
|M_{11} (0,q) - M_{12} (0,q)| \leq \frac{\textrm{const}}{q^{2}}.
\end{equation}
For $(q = 0)$, formula (\ref{bt13}) yields a generalization of the so-called Hugenholtz-Pines 
formula~\cite{hpin59} to finite temperatures. If one 
assumes that the mass operator is regular in a neighbourhood of the point $(E = 0, \, q = 0)$, then one can use (\ref{bt11}) 
to prove the absence of a gap in the (phonon-type) excitation energy spectrum.\\
In the case of zero temperatures the inequality (\ref{bt13}) establishes a connection between the density of the continuous 
distribution of the particle momenta and the minimum energy of an excited state. 
Relations of type (\ref{bt13}) should also be valid in quantum field theory, in which a spontaneous symmetry breaking 
(at a transition between two ground states) results in an infinite number of particles of zero mass (Goldstone's theorem), 
which are interpreted as singularities for small momenta in the quantum field Green functions. Bogoliubov's theorem has 
been applied to a numerous statistical and quantum-field-theoretical  models with a spontaneous symmetry breaking.
In particular, S. Takada~\cite{staka} investigated the relation between the long-range order in the ground state and the 
collective mode, namely, 
the Goldstone particle, on the basis of Bogoliubov's $1/q^{2}$ theorem. It was pointed out that Bogoliubov's inequality 
rules out the long-range orders in the ground states
of the isotropic Heisenberg model, the half-filled Hubbard model and the interacting Bose system for one dimension while 
it admits the long-range orders for two dimensions. The Takada's proof was based on the fact that the lowest-excited state 
that can be regarded as the 
Goldstone particle has the energy $E(q) \propto |q|$ for small $q$. This energy spectrum was exactly given in the 
one-dimensional models and was shown to 
be proven in the ordered state on a reasonable assumption except for the ferromagnetic case.
Baryakhtar and Yablonsky~\cite{bar76} applied Bogoliubov's theorem on $1/q^{2}$ law to quantum theory of magnetism 
and studied the asymptotic behavior of the correlation functions of magnets in the long-wavelength limit.\\
These papers and also
some others   demonstrated the strength of the $1/q^{2}$ theorem for obtaining rigorous proofs of the absence
of specific ordering in one- and two-dimensional systems, in which spontaneous
symmetry is broken in completely different ways: ferro- ferri-, and antiferromagnets, systems that
exhibit superfluidity and superconductivity, etc.   All that  indicates that $1/q^{2}$ theorem  provides the workable
and very useful tool for rigorous
investigation of the problem of specific ordering in various concrete systems of  interacting particles.
%
%
%
\subsection{Bogoliubov's Inequality and  the Mermin-Wagner Theorem} 
%
One of the most interesting features of an interacting system is the
existence of a macroscopic order which breaks the underlying symmetry of
the Hamiltonian. It was shown above, that the continuous rotational symmetry (in
three-dimensional spin space) of the isotropic Heisenberg ferromagnet is
broken by the spontaneous magnetization that exists in the limit of vanishing
magnetic field for a three-dimensional lattice. For systems of restricted
dimensionality  it has  been argued  long ago that there is no macroscopic
order, on the basis of heuristic arguments.  For instance, because the
excitation spectrum for systems with continuous symmetry has no gap,   the
integral of the occupation number over momentum will diverge in one and
two dimensions for any nonzero temperature. 
The heuristic arguments have been supported by a rigorous ones by using of
an operator inequality due to Bogoliubov~\cite{nnb61,sado73}.\\
The Bogoliubov inequality can be introduced by the following arguments.
Let us consider a scalar product $\left ( A,B \right )$ of two operators  $A$ and $B$ defined in the previous section
\begin{equation}
\label{bi1}
 \left ( A,B \right ) =  \frac{1}{Z}  \sum_{n \neq m}  \langle n|A^{\dag}|m \rangle  \langle m|B|n \rangle      
\Bigl ( \frac{\exp - (E_{m}/k_{B}T) -   \exp - (E_{n}/k_{B}T)}{E_{n} - E_{m}} \Bigr ).
\end{equation} 
We have obvious inequality
\begin{equation}
\label{bi2}
 \left ( A,B \right ) \leq
 \frac{1}{2 k_{B}T}  \langle A A^{\dag} + A^{\dag} A  \rangle.
\end{equation} 
Then we make use the Cauchy-Schwartz inequality (\ref{bt9})  which has the form
\begin{equation}
\label{bi3}
| \left ( A,B \right )|^{2} \leq  \left ( A,A \right )  \left ( B,B \right ).
\end{equation} 
If we take $B = [C^{\dag},H]_{-}$ we arrive at the Bogoliubov inequality
\begin{equation}
\label{bi4}
|\langle [C^{\dag},A^{\dag}]_{-} \rangle |^{2} \leq  
\frac{1}{2 k_{B}T}  \langle  A^{\dag} A + A A^{\dag} \rangle \langle [C^{\dag}, [H,C]_{-}]_{-}\rangle. 
\end{equation} 
In a more formal language we can formulate this as follows.
Let us suppose that $H$ is a symmetrical operator in the
Hilbert space $L$. For an operator $X$  in $L$  let us define
\begin{equation}
\label{bi5}
\langle X  \rangle  = \frac{1}{Z} \textrm{Tr} X \exp ( - H/k_{B}T); \quad
 Z = \textrm{Tr}   \exp ( - H/k_{B}T).
\end{equation}
The Bogoliubov inequality for operators $A$ and $C$  in $L$  has the form
\begin{equation}
\label{bi6}
\frac{1}{2 k_{B}T}  \langle A A^{\dag} + A^{\dag} A  \rangle \langle [[C,H]_{-}, C^{\dag} ]_{-}\rangle 
\geq |\langle [C,A]_{-}  \rangle |^{2}.
\end{equation}
The Bogoliubov inequality can be rewritten in a  slightly different form
\begin{equation}
\label{bi7}
k_{B}T |\langle [C,A]_{-}  \rangle |^{2}/ \langle [[C,H]_{-}, C^{\dag} ]_{-}\rangle 
\leq \frac{\langle [A,A^{\dag}]_{+} \rangle}{2}.
\end{equation}
It is valid for arbitrary operators $A$ and $C$,  provided the Hamiltonian is
Hermitian and the appropriate thermal averages exist.
The operators $C$ and $A$ are chosen in
such a way that the numerator on the left-hand side reduces to the order
parameter and the denominator approaches zero in the limit of a vanishing
ordering field. Thus the upper limit placed on the order parameter vanishes
when the symmetry-breaking field is reduced to zero.\\
The very elegant piece of work by Bogoliubov~\cite{nnb61} stimulated  numerous investigations on
the upper and lower bounds for thermodynamic 
averages~\cite{wagn66,sado73,ahar67,wong72,plec76,roep77,wres81,pit91,mw66,hoh67,weg67,mer69,mer68,frol81,frol86,ches69}.
A. B. Harris~\cite{ahar67} analyzed the upper and lower bounds for thermodynamic averages
of the form $\langle [A,A^{\dag}]_{+} \rangle.$ From the lower bound he derived a special
case of the Bogoliubov inequality of the form
\begin{equation}
\label{bi8}
\langle A^{\dag}A \rangle \geq \langle [A,A^{\dag}]_{-} \rangle/ \left ( \exp (\beta \langle \omega \rangle)  - 1\right )
\end{equation}
and a few additional weaker inequalities.\\     
The rigorous consideration of the Bogoliubov inequality was carried out by 
Garrison and Wong~\cite{wong72}. They pointed rightly that
in the conventional Green's function approach to statistical mechanics
all relations are first derived for strictly finite systems; the thermodynamic
limit is taken at the end of the calculation. Since the original derivation  
of the Bogoliubov inequalities was carried out within this framework, the
subsequent applications had to follow the same prescription. It was applied by a number of authors  to
show the impossibility of various kinds of long-range order in one- and
two-dimensional systems. In the latter class of problems, a special
difficulty arises from the fact that finite systems do not exhibit the broken
symmetries usually associated with long-range order. This has led to the
use of Bogoliubov's quasiaveraging method in which the finite-system
Hamiltonian was modified by the addition of a symmetry breaking term,
which was set equal to zero only after the passage to the thermodynamic
limit. Authors emphasized that this approach has never been shown to be equivalent to the more
rigorous treatment of broken symmetries provided by the theory of
integral decompositions of states on $C^{*}$-algebras; furthermore,
for some problems (e.g. Bose condensation and antiferromagnetism) the
symmetry breaking term has no clear physical interpretation. Garrison and Wong~\cite{wong72}
 shown how these difficulties can be
avoided by establishing the Bogoliubov inequalities directly in the
thermodynamic limit. 
In their work the Bogoliubov inequalities were derived for the infinite volume states
describing the thermodynamic limits of physical systems. The only property of the states
required is that they satisfy the Kubo-Martin-Schwinger boundary condition. 
Roepstorff~\cite{roep77} investigated a stronger version of Bogoliubov's inequality
and the Heisenberg model. He
derived a rigorous upper bound for the magnetization in the ferromagnetic
quantum Heisenberg model with arbitrary spin and dimension $D \geq 3$ 
on the basis of general inequalities in quantum statistical mechanics.\\
Further generalization was carried out by  L. Pitaevskii and S. Stringari~\cite{pit91} who
carefully reconsidered the interrelation of the  uncertainty principle, quantum fluctuations, and
broken symmetries for many-particle interacting systems. 
At zero temperature the Bogoliubov inequality provides significant
information on the static polarizability, but not directly on the fluctuations
occurring in the system. Pitaevskii and Stringari~\cite{pit91} presented a different
inequality yielding, at low temperature, relevant information on the fluctuations
of physical quantities  
\begin{equation}
\label{bi9}
\int d \omega J(A^{\dag},A;\omega) \coth \frac{\beta \omega}{2}
\int d \omega J(B^{\dag},B;\omega) \tanh \frac{\beta \omega}{2} \geq \,
      \vline \int d \omega J(A^{\dag},B;\omega)\vline \, ^{2}.
\end{equation}
They shown also that the following inequality holds
\begin{equation}
\label{bi10}
\langle [A^{\dag},A]_{+} \rangle \langle [B^{\dag},B]_{+} \rangle \geq \,
 \vline \langle [A^{\dag},B]_{-} \rangle\vline \, ^{2}.
\end{equation} 
The inequality  (\ref{bi9}) can be applied to both hermitian
and non-hermitian operators and can be consequently regarded as a natural
generalization of the Heisenberg uncertainty principle.  Its determination
is based on the use of the Schwartz inequality for auxiliary operators related
to the physical operators through a linear transformation.
 The inequality  (\ref{bi9}) was employed to derive useful
constraints on the behavior of quantum fluctuations in problems with continuous
group symmetries. Applications to Bose superfluids, antiferromagnets
and crystals at zero temperature were discussed as well. In particular, a
simple and direct proof of the absence of long range order at zero temperature
in the $1D$ case was formulated. 
Note that inequality  (\ref{bi9}) does not coincide, except at $T = 0$, with inequality (\ref{bi10})
because of the occurrence of the $\tanh$ factor instead of the $\coth$ one in the
integrand of the left-hand side containing $J(B^{\dag},B;\omega)$. However the inequality  (\ref{bi10})
follows immediately  from inequality  (\ref{bi9}) using the inequality~\cite{pit91}
\begin{equation}
\label{bi11}
J(B^{\dag},B;\omega)\coth \frac{\beta \omega}{2} \geq \,  J(B^{\dag},B;\omega)\tanh \frac{\beta \omega}{2}.
\end{equation}
The Bogoliubov inequality
\begin{equation}
\label{bi12}
\langle [A^{\dag},A]_{+} \rangle \langle [B^{\dag},[H,B]_{-}]_{-} \rangle \geq \,
\frac{2}{\beta} \vline \langle [A^{\dag},B]_{-} \rangle \vline \, ^{2} 
\end{equation} 
can be  obtained  from (\ref{bi9}) using  the inequality (\ref{bi11}).
Pitaevskii and Stringari~\cite{pit91} noted, however,  that in general their inequality (\ref{bi10}) for the
fluctuations of the operator $A$ differs from the Bogoliubov inequality (\ref{bi12})
in an important way. In fact result (\ref{bi12}) provides particularly strong conditions
when $k_{B}T$ is larger than the typical excitation energies induced by the
operator $A$ and explains in a simple way the divergent $k_{B}T/q^{2}$ behavior
exhibited by the momentum distribution of Bose superfluids as well as from
the transverse structure factor in antiferromagnets. Vice-versa, inequality
(\ref{bi10}) is useful when $k_{B}T$ is smaller than the typical excitation energies and
consequently emphasizes the role of the zero point motion of the elementary
excitations which is at the origin of the $1/q^{2}$ behavior. The general inequality
(\ref{bi9}) provides in their opinion the proper interpolation between the two different regimes.\\
Thus Pitaevskii and Stringari  proposed a zero-temperature analogue of the
Bogoliubov inequality, using the uncertainty relation of quantum mechanics.
They presented a method for showing the absence of breakdown of continuous
symmetry in the ground state. T. Momoi ~\cite{momo96} developed their ideas further.
He discussed conditions for the absence of spontaneous breakdown of continuous
symmetries in quantum lattice systems at $T = 0$. His analysis was based on
Pitaevskii and Stringari's idea that the uncertainty relation can be employed to
show quantum fluctuations. For one-dimensional systems, it was shown that the
ground state is invariant under a continuous transformation if a certain uniform
susceptibility is finite. For the two- and three-dimensional systems, it was shown
that truncated correlation functions cannot decay any more rapidly than $|r|^{- d + 1}$
 whenever the continuous symmetry is spontaneously broken. Both of
these phenomena occur owing to quantum fluctuations. The Momois's results cover a
wide class of quantum lattice systems having not-too-long-range interactions.\\
An important aspect of the later use of Bogoliubov's results was their application to obtain rigorous
proofs of the absence of specific ordering in one- and two-dimensional systems of many particles interacting
through binary potentials with a definite restriction on the interaction~\cite{bb,nnb61,petr95,sado73}. The problem of the
presence or absence of phase transitions in systems with short-range interaction has been discussed for
quite a long time. The physical reasons why specific ordering cannot occur in one- and two-dimensional
systems is known. The creation of a macroscopic region of disorder with characteristic scale $\sim L$
requires negligible energy ($\sim L^{d-2} $ if the interaction has a finite range). However, a unified approach to
this problem was lacking and few rigorous results were obtained~\cite{sado73}.\\ 
Originally the Bogoliubov inequality was
applied to exclude ordering in isotropic Heisenberg ferromagnets
and antiferromagnets by Mermin and Wagner~\cite{mw66} and
in one or two dimensions in superconducting
and superfluid systems by Hohenberg~\cite{hoh67} (see also Refs.~\cite{weg67,mer69,mer68,frol81,frol86,ches69}).
 The physics behind the Mermin-Wagner theorem  is based
on the conjecture that the excitation of spin waves can destroy the magnetic order 
since the density of states of the excitations depends on the dimensionality of the system.
In  $D = 2 $ dimensions  thermal  excitations  of
spin waves  destroy  long-range order.
The number of thermal spin excitations is
\begin{equation}
\label{bi22a}
\mathcal{N} = \sum_{k} \langle N_{k} \rangle = \sum_{k} \frac{1}{\exp (\beta E_{k}) - 1} \sim
\int \frac{k^{D - 1}  d k}{\exp (\beta D_{\textrm{sw}} k^{2}) - 1} \sim \int \frac{k^{D}  dk}{k^{3}}
\end{equation}
This expression diverges  for $D = 2 $.  Thus the ground
 state is unstable to thermal excitation
The reason for the absence of magnetic order under the above assumptions is   that
at finite temperatures spin waves are  easily   excitable, what destroys magnetic
order.\\
In their paper, exploiting a thermodynamic
inequality due to Bogoliubov~\cite{nnb61}, Mermin and
Wagner~\cite{mw66} formulated the  statement that
for one- or two-dimensional Heisenberg
systems with isotropic interactions of the form
\begin{equation}
\label{bi13} H = \frac{1}{2}  \sum_{i,j}  J_{ij} \vec S_{i}\cdot \vec S_{j} -h S^{z}_{\vec{q}}
\end{equation}
and
such that the interactions are short-ranged,
namely which satisfy the condition  
\begin{equation}
\label{bi14} \mathcal{J} = \frac{1}{2 N}  \sum_{i,j} | J_{ij}|| \vec r_{i} - \vec r_{j}|^{2} < \infty,
\end{equation}
cannot be ferro- or antiferromagnetic. 
Here $S^{z}_{\vec{q}}$ is the Fourier component of $S^{z}_{i},$ $N$ is the number of spins.
Consider the inequality (\ref{bi6}) and take $C = S^{z}_{\vec{k}}$ and $A = S^{y}_{- \vec{k} - \vec{q}}.$
It   follows from (\ref{bi6}) that
\begin{equation}
\label{bi15}
\frac{\langle S^{z}_{\vec{q}} \rangle}{N} \leq \frac{1}{\hbar^2 k_{B}T} S_{yy}(\vec{k} + \vec{q})\frac{1}{N}
\langle [S^{x}_{-\vec{k}},[H,S^{x}_{\vec{k}}]_{-}]_{-} \rangle. 
\end{equation}
Here $S_{yy}(\vec{q}) = \langle S^{y}_{\vec{q}} S^{y}_{-\vec{q}} \rangle / N.$
The direct calculation leads to the equality
\begin{eqnarray}
\label{bi16}
\Lambda (k) = \frac{1}{N} \langle [S^{x}_{-\vec{k}},[H,S^{x}_{\vec{k}}]_{-}]_{-} \rangle = \\ \nonumber
\hbar^2 \Bigl ( h \frac{\langle S^{z}_{\vec{q}} \rangle}{N} + \frac{1}{N} \sum_{j,j'} | J_{jj'} 
\left ( \cos \vec{k} (\vec r_{j'} - \vec r_{j}) - 1 \right ) \langle S^{y}_{j} S^{y}_{j'}  + S^{z}_{j} S^{z}_{j'}\rangle \Bigr ).
\end{eqnarray}
Thus we have
\begin{equation}
\label{bi17} 
\Lambda (k) \leq \hbar^2 \left ( h \frac{\langle S^{z}_{\vec{q}} \rangle}{N}  + S (S + 1) \mathcal{J} k^{2} \right ).
\end{equation}
It follows from   the Eqs. (\ref{bi15}) and (\ref{bi17}) that
\begin{equation}
\label{bi18} 
S_{yy}(\vec{k} +  \vec{q}) \geq \frac{k_{B}T h 
 \frac{\langle S^{z}_{\vec{q}} \rangle^{2}}{N^{2}}}{h \frac{\langle S^{z}_{\vec{q}} \rangle}{N} + S (S + 1) \mathcal{J} k^{2}}.
\end{equation}
To proceed, it is necessary to sum up ($1/N \sum_{k}$) on the both sides of the inequality (\ref{bi18}). After doing that
we obtain
\begin{equation}
\label{bi19}
\frac{k_{B}T  \frac{\langle S^{z}_{\vec{q}} \rangle^{2}}{N^{2}}}{(2 \pi)^{D}} \int^{\tilde{K}}_{0}
\frac{F_{D} k^{D - 1} d k}{h \frac{\langle S^{z}_{\vec{q}} \rangle}{N} + S (S + 1) \mathcal{J} k^{2} } \leq S (S + 1).
\end{equation}
The following notation were introduced
\begin{equation}
\label{bi20}
F_{D} = \frac{2 \pi^{D/2}}{\Gamma (D/2)}.  
\end{equation}
Here $\Gamma (D/2)$  is the gamma function. Considering the two-dimensional case
we find that
\begin{equation}
\label{bi21}
h \frac{\langle S^{z}_{\vec{q}} \rangle}{N} \leq  \textrm{const}  \frac{S (S + 1) \sqrt{\mathcal{J}} }{\sqrt{T}}
\frac{1}{\sqrt{\ln |h|}}.   
\end{equation}
Thus, at any non-zero temperature, a one- or two-dimensional isotropic spin-$S$
Heisenberg model with finite-range exchange interaction cannot be neither ferromagnetic
nor antiferromagnetic. \\
In other words, according to the Mermin-Wagner theorem
there can be no long range order at any non-zero temperature
in one- or two-dimensional systems whenever this ordering would correspond
to the breaking of a continuous symmetry and the interactions fall off
sufficiently rapidly with inter-particle distance~\cite{swiec}.  The Mermin-
Wagner theorem follows from the fact that in one and two-dimensions a
diverging number of infinitesimally low energy excitations is created at
any finite temperature and therefore in these cases the assumption of
there being a non-vanishing order parameter is not self consistent.
The proof does not apply to $T = 0$, thus the ground state itself may
be ordered. Two dimensional  ferromagnetism is possible strictly
at $T = 0$. In this case quantum fluctuations oppose, but do not prevent
a finite order parameter  to appear in  a ferromagnet. In contrast, for one dimensional systems quantum
fluctuations tend to become so strong that they prevent ordering,
even in the ground state~\cite{momo96}.\\ 
Note that the basic assumptions of the  Mermin-Wagner theorem (isotropic and short-range~\cite{swiec,wres76} interaction) are usually not strictly 
fulfilled  in   real systems. Thorpe~\cite{mft71} applied the method of Mermin and Wagner to show that 
one- and two-dimensional  spin systems interacting with a general isotropic interaction
\begin{equation}
\label{bi22} H = \frac{1}{2}  \sum_{i j n}  J^{(n)}_{ij} \left ( \vec S_{i}\cdot \vec S_{j}  \right )^{n},
\end{equation}
where the exchange interactions $J^{(n)}_{ij}$ are of finite range, cannot order in the sense that $\langle O_{i} \rangle = 0$
for all traceless operators $O_{i}$ defined at a single site $i$. Mermin and Wagner have proved the above for the case
$n = 1$ with $  O_{i}   = \vec S_{i},$ i.e. for the Heisenberg Hamiltonian (\ref{bi13}). The Thorpe's results shown
that a small isotropic biquadratic exchange $\left ( \vec S_{i}\cdot \vec S_{j}  \right )^{2} $ cannot induce
ferromagnetism or antiferromagnetism in a two-dimensional Heisenberg system. The proof utilizes the Bogoliubov
inequality (\ref{bi12}). Further discussion of the results of Mermin and Wagner and Thorpe was carried out in 
Ref.~\cite{fan78}   The Hubbard identity was used to show the absence of magnetic phase transitions in Heisenberg spin systems in one and two dimensions, 
generalizing Mermin and Wagner's next term result in an alternative way as Thorpe has done.\\
The results of Mermin and Wagner and Thorpe shown that the isotropy of the Hamiltonian plays the essential role. However it is clear
that although one- and two-dimensional systems exist in nature that may be very nearly isotropic, they all have a small
amount of anisotropy. Experiments suggested that a small amount of anisotropy can induce a spontaneous magnetization in 
two dimension.
Froehlich and Lieb~\cite{frli77} proved  the existence of phase transitions for anisotropic Heisenberg models.
They shown rigorously  that
the two-dimensional anisotropic, nearest-neighbor Heisenberg model on a square lattice, both quantum and classical, 
 have a phase transition in the sense that the spontaneous magnetization is positive at low temperatures. This is so 
for all anisotropies. An analogous result (staggered polarization) holds for the antiferromagnet in the classical case; 
in the quantum case it holds if the anisotropy is large enough (depending on the single-site spin).\\
Since then, this method has been applied to show the absence of crystalline order in classical 
systems~\cite{mer68,frol81,frol86,ches69},
the absence of an excitonic insulating state~\cite{walk68},
to rule out long-range spin density waves in an electron gas~\cite{ham67} and  magnetic ordering in metals~\cite{ghosh,kis72}.
The systems considered include not only one- and two-dimensional lattices, but also three-dimensional
systems of finite cross section or thickness~\cite{ches69}.\\ 
In this way
the inequalities have been applied by Josephson~\cite{jose67} to derive rigorous
inequalities for the specific heat  in either one- or two-dimensional systems.
A rigorous inequality was derived relating the specific heat of a system, the
temperature derivative of the expectation value of an arbitrary operator and the
mean-square fluctuation of the operator in an equilibrium ensemble. The class of
constraints for which the theorem was shown to hold includes most of those of practical
interest, in particular the constancy of the volume, the pressure, and (where applicable)
the magnetization and the applied magnetic field.\\
Ritchie and  Mavroyannis ~\cite{mavro} investigated the
ordering in systems with quadrupolar interactions and proved the
absence of ordering in quadrupolar systems of restricted dimensionality.
The Bogoliubov inequality was applied
to the isotropic model to show that there is no ordering in one- or two-dimensional
systems. Some properties of the anisotropic model were presented.
Thus in this paper it was shown that an isotropic quadrupolar model
does not have macroscopic order in one or two dimensions.\\
The statements above on the impossibility of magnetic order   or other long-range order in one
and two spatial dimensions can be generalized to other symmetry broken
states and to other geometries, such as  fractal systems~\cite{cas92,dcas92,buri99},  Heisenberg~\cite{garri72} 
 thin films, etc. In Ref.~\cite{garri72} 
thin films were described as idealized systems having finite extent in one direction but infinite extent in the other two. 
For systems of particles interacting through smooth potentials (e.g., no hard cores), it was shown~\cite{garri72} that an equilibrium 
state for a homogeneous thin film is necessarily invariant under any continuous internal symmetry group generated by a 
conserved density. For short-range interactions it was also shown that equilibrium states are necessarily translation 
invariant. The absence of long-range order follows from its relation to 
broken symmetry. The only properties of the state required for the proof are local normality, spatial translation 
invariance, and the Kubo-Martin-Schwinger boundary condition. The argument employs the Bogoliubov inequality and the 
techniques of the algebraic approach to statistical mechanics. For inhomogeneous systems, the same argument shows that 
all order parameters defined by anomalous averages must vanish. Identical results can be obtained for systems with 
infinite extent in one direction only.\\ 
In the  case of thin films the Mermin-Wagner theorem provides an important leading idea and gives a qualitative
explanation~\cite{bres90} why the ordering temperature $T_{c}$ is usually reduced for thinner films.
Two models of magnetic bilayers were considered in Ref.~\cite{nsad06},  both based on the Heisenberg model. In the first 
case of ferromagnetically ordered ferromagnetically coupled planes of $S = 1$ the anisotropy is of easy plane/axis type, 
while in the study of antiferromagnetically 
ordered antiferromagnetically coupled planes of $S = 1/2$, the anisotropy is of $XXZ$ type. Both systems were treated 
by Green's function method, 
which consistently applied within random phase approximation. The calculations lead to excitation energies and the 
system of equations for order parameters 
which can be solved numerically and which satisfies both Mermin-Wagner   and Goldstone   theorem   in the corresponding 
limit and also agrees with the mean field results. The basic result was that the transition temperature for magnetic 
dipole order parameter is unique for both planes. 
Nonexistence of magnetic order in the Hubbard model of thin films was shown in Ref.~\cite{suk72} 
Introduction of the Stoner molecular field approximation is responsible for the appearance of magnetic order
in the Hubbard model of thin films.\\
The Mermin-Wagner theorem was strengthened by Bruno~\cite{brun01} so as to rule out magnetic long-range order at $T > 0$
in one- or two-dimensional Heisenberg and $ XY$  systems with long-range interactions decreasing as $R^{- \alpha}$
with a sufficiently large exponent $\alpha$. For oscillatory interactions, ferromagnetic long-range order at
$T > 0$ is ruled out if $\alpha \geq 1$ ($D = 1$) or $\alpha > 5/2$ ($D = 2$). For systems with monotonically decreasing
interactions, ferro- or antiferromagnetic long-range order at $T > 0$ is ruled out if $\alpha \geq 2D$.
In view of the fact that most magnetic ultrathin films investigated experimentally
consist of metals and alloys these results are of great importance.\\
The Mermin-Wagner theorem   states that at non-zero temperatures the two
dimensional Heisenberg model has no spontaneous magnetization. 
A global rotation of spins in a plane means that we can not have a long-range magnetic 
ordering at non-zero temperature. Consequently
the spin-spin correlation function decays to zero at large distances, although the
Mermin-Wagner theorem gives no indication of the rate of decay.
Martin~\cite{mar82} shown that
the Goldstone theorem in any dimension and the absence of symmetry breaking in two dimensions result from a simple use of 
the Bogoliubov inequality. 
Goldstone theorem is the statement that an equilibrium phase which breaks
spontaneously a continuous symmetry must have a slow (non-exponential)
clustering. The classical arguments about the absence of symmetry breakdown
in two dimensions were formulated in a few earlier studies, where
it was proved that in any dimension a phase of a lattice system which
breaks a continuous internal symmetry cannot have an integrable clustering.
Classical continuous systems were also studied in all dimensions   with
the result that the occurrence of crystalline   or orientational   order implies
a slow clustering. The same property holds for Coulomb systems.
In particular, the rate of clustering of particle correlation functions in a 3-dimensional classical crystal 
is necessarily than slower or equal to $|x|^{-1}$ (see also Refs.~\cite{kle81,bona82,koma07}).\\
Landau, Peres and Wreszinski~\cite{wres81}
 proved a Goldstone-type theorem for a wide class of lattice and continuum
quantum systems, both for the ground state and at nonzero temperature. For the
ground state (T = 0) spontaneous breakdown of a continuous symmetry implies
no energy gap. For nonzero temperature, spontaneous symmetry breakdown
implies slow clustering (no $L^{1} $ clustering). The methods applied also to nonzero-temperature
classical systems. They showed that for
a physical system with short-range forces and a continuous symmetry,
if the ground state is not invariant under the symmetry the Goldstone
theorem states that the system possesses excitations of arbitrarily low
energy.  In the case of the ground state (vacuum) of local quantum field
theory, the existence of an energy gap is equivalent to exponential clustering.
For general ground states of non-relativistic systems, the two properties
(energy gap and clustering) are, however, independent and, in particular,
the assumption that the ground state is the unique vector invariant under
time translations does not necessarily follow from the assumption of
spacelike clustering.  Another related aspect, of
greater relevance to their discussion, is the fact that the rate of clustering is
not expected to be related to symmetry breakdown and absence of an
energy gap, since for example the ground state of the Heisenberg
ferromagnet   has a broken symmetry and no energy gap, but is exponentially
clustering (for the ground state is a product state of spins pointing in
a fixed direction). On the other hand, for $T > 0 $ no energy gap is expected
to occur, at least under general timelike clustering assumptions.
These assumptions may be verified for the free Bose gas.  
At nonzero temperature it is the cluster properties that are important
in connection with symmetry breakdown. At nonzero temperature the authors
  formulated the Goldstone theorem as follows. Given a system with
short-range forces and a continuous symmetry, if the equilibrium state is
not invariant under the symmetry, then the system does not possess
exponential clustering.\\
Landau, Peres and Wreszinski~\cite{wres81}    explored the validity of the Goldstone theorem for a
wide class of spin systems and many-body systems, both for the ground
state and at nonzero temperature. The main tool that was used at nonzero
temperature was the Bogoliubov inequality, which is valid for both classical
and quantum systems. 
Their results apply to states which are invariant with respect to spatial
translations by some discrete set which is sufficiently dense. (For lattice
systems this could be a sublattice, and for continuum systems, a lattice
imbedded in the continuum.) 
They proved that for interactions which are not too long range,  
 for the ground state $(T = 0)$  spontaneous
breakdown of a continuous symmetry implies no energy gap. For nonzero
temperature $(T > 0)$ spontaneous symmetry breakdown implies no exponential
clustering (in fact no $L^{1} $  clustering).\\
Rastelli and  Tassi~\cite{rast89} pointed that
the theorem of Mermin and Wagner excludes long-range order in one- and two-dimensional Heisenberg models at any finite 
temperature if the exchange interaction is short ranged. In their opinion strong but nonrigorous indications exist about 
the absence of long-range order even in 
three-dimensional Heisenberg models when suitable competing exchange interactions are present. They argued, as a rigorous 
consequence of the Bogoliubov inequality, that this expectation may be true. It was found that for models where the 
exchange competition concerns at least two over three 
dimensions, a surface of the parameter space exists where long-range order is absent. This surface meets at vanishing 
temperature the continuous 
phase-transition line which is the border line between the ferromagnetic and helical configuration.
They investigated also the spherical model~\cite{rast90} and concluded that
the spherical model is a unique model for which an exact solution at finite temperature exists in three dimensions. 
In that paper 
they proved that this model may show an absence of long-range order   in three dimensions if a suitable competition between 
exchange couplings was assumed. 
In particular they found an absence of long-range order in wedge-shaped regions around the ferromagnet- 
or antiferromagnet-helix transition line or in the 
vicinity of a degeneration line, where infinite nonequivalent isoenergetic helix configurations are possible. They evaluated 
explicitly the phase 
diagram of a tetragonal antiferromagnet with exchange couplings up to third neighbors but their conclusions apply as well 
to any Bravais lattice.\\ 
The problem of generalization of the Mermin-Wagner theorem for the Heisenberg spin-glass order was discussed 
in Refs.~\cite{ozek88,comm90,ferna77} Using the Bogoliubov inequality, Fernandez~\cite{ferna77} considered   the
isotropic Heisenberg Hamiltonian
\begin{equation}
\label{bi23} H = -   \sum_{\langle ij \rangle} J_{ij} \vec S_{i} \vec S_{j},  
\end{equation}
which was used to model spin-glass behavior. The purpose of the model being to produce 
$ |\langle S_{i}^{z} \rangle | \neq 0$  below a certain temperature without the presence of long-range spatial order.
Fernandez showed that there can be no spin-glass in one or two dimensions for isotropic Heisenberg Hamiltonians for 
$T \neq 0$ if
\begin{equation}
\label{bi24} \lim_{N \rightarrow \infty} N^{-1} \sum_{\langle ij \rangle} |J_{ij} |(\vec r_{i} - \vec r_{j})^{2} < \infty.
\end{equation}
In summary, the Mermin-Wagner theorem, which excludes the breaking
of a continuous symmetry in two dimensions at finite
temperatures, was  established in 1966. Since then various more precise and more general versions have
been considered (see Refs.~\cite{koma07,romer75,krze77,ss1,ss2,ss3,verb09}). 
These considerations  of symmetry broken
systems are important in order to establish whether or not long-range order is possible in various concrete situation.\\
The fact that the zero magnetism which is
enforced by the Mermin-Wagner theorem is compatible
with various types of phase transitions was noted by many authors.
For example, 
 Dyson,  Lieb and Simon~\cite{dyson76,dyson78} proved the existence of a phase transition at non-zero
temperature for the Heisenberg model with nearest neighbor coupling. The proof was
essentially relied on some new inequalities involving two-point functions. Some of
these inequalities are quite general and, therefore, apply to any quantum system in
thermal equilibrium. Others rest on the specific structure of the model (spin system,
simple cubic lattice, nearest neighbor coupling, etc.) and have limited applicability.\\
The low dimensional systems show large fluctuations for continuous symmetry~\cite{koma07}.
The Hohenberg-Mermin-Wagner theorem    states that the corresponding
spontaneous magnetizations are vanishing at finite temperatures in one and two dimensions.
Since their articles appeared, their method has been applied to various systems, including
classical and quantum magnets, interacting electrons in a metal and Bose gas. The theorem was extended to
the models on a class of generic lattices with the fractal dimension by Cassi~\cite{cas92,dcas92,buri99}. In a stronger
sense, it was also proved for a class of low-dimensional systems that the equilibrium states are
invariant under the action of the continuous symmetry group.  Even at zero temperature, the
same is true  if the corresponding one- or two-dimensional system satisfies conditions  such as
boundedness of susceptibilities.  Since a single spin shows the spontaneous magnetization at zero
temperature, the absence of the spontaneous symmetry breaking implies that the strong fluctuatiuons
due to the interaction destroy the ordering and lead to the finite susceptibilities. In other
words, one cannot expect the absence of spontaneous symmetry breaking at zero temperature in a
generic situation~\cite{koma07}.\\
Some other applications of the Bogoliubov inequality to various problems of statistical physics were 
discussed in Refs.~\cite{su97,uhrig,nsad05,gelf00,gelf01,sigr07}
%
%
%
%
\section{Broken  Symmetries  and Condensed Matter Physics} 
%
Studies of symmetries and the consequences of breaking them have led to deeper understanding in many areas
of science.
Condensed matter physics is the field of physics that deals with the macroscopic physical properties of matter. 
In particular, it is concerned with the  \emph{condensed}  phases that appear whenever the number of constituents in a 
system is extremely large and the interactions between the constituents are strong. The most familiar examples of 
condensed phases are solids and liquids, which arise from the electric force between atoms. More exotic condensed 
phases include the superfluid and the Bose-Einstein condensate found in certain atomic systems~\cite{pit03,peth02,grif09,leg91}.
Symmetry has always played an important role in condensed matter physics~\cite{leg91}, from fundamental formulations of basic 
principles to concrete applications~\cite{symcm08,andre80,march87,ston67,sim78,lov80,slov80,tur80,cap89,new91,bunk,har07}. 
In condensed matter physics, the symmetry is important in
classifying different phases and   understanding the phase transitions between them.
The phase transition is
a physical phenomenon that occurs in macroscopic systems and consists in the following. In certain equilibrium 
states of the system an arbitrary small influence leads to a sudden change of its properties: the system passes from 
one homogeneous phase to another. Mathematically, a phase transition is treated as a sudden change of the structure and 
properties of the  Gibbs distributions describing the equilibrium states of the system, for arbitrary 
small changes of the parameters determining the equilibrium~\cite{cmp07}. The crucial concept here is the order parameter.\\
In statistical physics the question of interest is to understand how the order of phase
transition in a system of many identical interacting subsystems depends on the degeneracies of
the states of each subsystem and on the interaction between subsystems.
In particular, it is important to investigate  the role of the symmetry and uniformity of the
degeneracy and the symmetry of the interaction. Statistical mechanical theories of the
system composed of many interacting  identical  subsystems have been developed frequently for the case
of ferro- or antiferromagnetic spin system, in which the phase transition is usually found to be
one of  second order unless it is accompanied with such an additional effect as spin-phonon interaction.
The phase transition of first order is also occurs in a variety of systems, such as ferroelectric transition,
orientational transition and so on. Second order phase transitions are frequently, if not always,
associated with  spontaneous breakdown of a global symmetry. It is then possible to find a corresponding
order parameter which vanishes in the disordered phase and is nonzero in the ordered phase. Qualitatively the
transition is understood as condensation of the broken symmetry \emph{charge} carriers. The critical region
is effectively described by a local Lagrangian involving the order parameter field~\cite{itz}.\\
Combining many elementary particles into a single interacting
system may result in collective behavior that qualitatively
differs from the properties allowed by the physical
theory governing the individual building blocks as was stressed by Anderson~\cite{and72}.
It is known that the description of spontaneous symmetry breaking that underlies the connection between classically ordered
objects in the thermodynamic limit and their individual quantum-mechanical building blocks is one of the
cornerstones of modern condensed-matter theory and has found applications in many different areas of physics.
The theory of spontaneous symmetry breaking, however, is inherently an equilibrium theory, which does not
address the dynamics of quantum systems in the thermodynamic limit. 
J. van Wezel~\cite{vwez08} investigated the quantum dynamics  of many particle system in the thermodynamic limit.
Author used the example of a
particular antiferromagnetic model system to show that the presence of a so-called thin spectrum of collective
excitations with vanishing energy - one of the well-known characteristic properties shared by all symmetry breaking
objects - can allow these objects to also spontaneously break time-translation symmetry in the thermodynamic
limit. As a result, that limit is found to be able, not only to reduce quantum-mechanical equilibrium
averages to their classical counterparts, but also to turn individual-state quantum dynamics into classical
physics. In the process, van Wezel found that the dynamical description of spontaneous symmetry breaking can also be
used to shed some light on the possible origins of Born's rule. The work was concluded by describing an experiment on a
condensate of exciton polaritons which could potentially be used to experimentally test the proposed
mechanism.\\
There is an important distinction between the case where the broken
symmetry is continuous (e.g. translation, rotation, gauge invariance)
or discrete (e.g. inversion, time reversal symmetry). The Goldstone theorem
states that when a continuous symmetry is spontaneously broken
and the interactions are short ranged a collective mode (excitation) exists
with a gapless energy spectrum (i.e. the energy dispersion curve
starts at zero energy and is continuous). Acoustical phonons in a crystal
are prime examples of such so-called gapless Goldstone modes. Other
examples are the Bogoliubov sound modes in (charge neutral) Bose condensates~\cite{pit03,grif09}
and spin waves (magnons) in ferro- and antiferromagnets. 
On the same ground one can consider the existence of magnons
in spin systems at low temperatures~\cite{umez84}, acoustic
and optical vibration modes in regular lattices or in
multi-sublattice magnets, as well as the vibration spectra
of interacting electron and nuclear spins in magnetically-ordered crystals~\cite{tur80}.\\  
It was claimed by some authors that there exists a certain class of systems with broken symmetry,
whose condensed state and ensuring macroscopic theory are quite analogous to those of superfluid helium.
These systems are Heisenberg magnetic lattices, both ferro- and antiferromagnetic, for which the
macroscopic modes associated with the \emph{quasi-conservation law} are long-wavelength spin waves.
In contrast to liquid helium, those systems are amenable to a fully microscopic analysis, at least in
the low-temperature limit. However there are also differences in the nature of antiferromagnetism and
superconductivity for many-particle systems on a lattice.
It is therefore of interest to look carefully to the specific features of
the magnetic, superconducting and Bose systems in some detail, both for its own sake, and to gain insight
into the general principles of their behavior.
%
%
%
\subsection{Superconductivity} 
%
%
The BCS-Bogoliubov model of superconductivity is one of few examples in the many particle
system that can be solved (asymptotically) exactly~\cite{bb,nnb58,bts58,nnb71,lieb61,haag62,chen65,wax94,bcs08}.
In the limit of infinite volume, the BCS-Bogoliubov theory of superconductivity provides
an exactly soluble model\cite{bb,nnb58,bts58,nnb71} wherein the phenomenon of spontaneous symmetry breakdown 
occurs explicitly. The symmetry that gets broken being the gauge invariance.\\
It was shown in previous sections that the concept of spontaneously
broken symmetry  is one of the most important
notions in statistical physics, in the quantum field theory and elementary
particle physics. This is especially so as far as creating
a unified field theory, uniting all the different forces of
nature~\cite{nnb85,huang07,hoof07,fwilc08}, is concerned. One should stress that the
notion of spontaneously broken symmetry came to the
quantum field theory from solid-state physics~\cite{pwa84}. It was
originated in quantum theory of  magnetism~\cite{namb07,namb09}, and later was
substantially developed and found wide applications in
the gauge theory of elementary particle physics~\cite{huang07,fstr05,grib}. 
It was in the quantum field theory where the ideas
related to that concept were quite substantially developed
and generalized. The analogy between the Higgs
mechanism giving mass to elementary particles and the
Meissner effect in the Ginzburg-Landau superconductivity
theory is well known~\cite{pwa84,ander58,ander75,ander63,weinb86,weinb08}.\\  
The Ginzburg-Landau model is a special
form of the mean-field theory~\cite{andre80,march87}. 
The superconducting state has lower entropy than the normal
state and is therefore the more ordered state. A general theory
based on just a few reasonable assumptions about the order
parameter is remarkably powerful~\cite{andre80,march87}. It describes not just BCS-Bogoliubov
superconductors but also the high-$T_{c}$ superconductors,
superfluids, and Bose-Einstein condensates.  
The Ginzburg-Landau model operates with
a pseudo-wave function $\Psi(\vec{r})$, which plays the role of a
parameter of complex order, while the square of this
function modulus $|\Psi(\vec{r})|^{2}$ should describe the local density
of superconducting electrons. 
It was conjectured that $\Psi(\vec{r})$
behaves in many respects like a
macroscopic wavefunction but without certain properties
associated with linearity: superposition and normalization.
It is well known, that
the Ginzburg-Landau theory is applicable if the temperature
of the system is sufficiently close to its critical
value $T_{c}$, and if the spatial variations of the functions $\Psi$
and of the vector potential $\vec{A}$ are not too large. The
main assumption of the Ginzburg-Landau approach is
the possibility to expand the free-energy density $f$ in a
series under the condition, that the values of $\Psi$ are
small, and its spatial variations are sufficiently slow.
The Ginzburg-Landau equations follow from an applications
of the variational method to the proposed
expansion of the free energy density in powers of $|\Psi|^{2}$ and $|\nabla  \Psi|^{2}$,
which leads to a pair of coupled differential
equations for $\Psi(\vec{r})$ and the vector potential $\vec{A}$.
The Lawrence-Doniach model was formulated in
the paper~\cite{ldm71} for analysis of the role played by layered
structures in superconducting materials~\cite{kkuz03,kkuz02,hu07}.  
The model considers a stack of parallel two dimensional
superconducting layers separated by an
isolated material (or vacuum), with a nonlinear interaction
between the layers. It was also assumed that an
external magnetic field is applied to the system. In
some sense, the Lawrence-Doniach model can be
considered as an anisotropic version of the Ginzburg-
Landau model. More specifically,
an anisotropic Ginzburg-Landau model can be considered
as a continuous limit approximation to the
Lawrence-Doniach model. However, when the
coherence length in the direction perpendicular to
the layers is less than the distance between the layers,
these models are difficult to compare.\\ 
Both effects, Meissner effect and Higgs effect are consequences of spontaneously
broken symmetry in a system containing two interacting
subsystems. 
According to F. Wilczek~\cite{wil05}, ''the most fundamental phenomenon of superconductivity is the
Meissner effect, according to which magnetic fields are expelled from
the bulk of a superconductor. The Meissner effect implies the possibility
of persistent currents. Indeed, if a superconducting sample is
subjected to an external magnetic field, currents of this sort must
arise near the surface of a sample to generate a cancelling field.
An unusual but valid way of speaking about the phenomenon of
superconductivity is to say that within a superconductor the photon
acquires a mass. The Meissner effect follows from this.''
This is a mechanism by which gauge fields acquire mass:
''the gauge particle 'eats' a Goldstone boson and thereby
becomes massive''.
This general idea has been applied to the more complex problem
of the weak interaction which is mediated by the $W$-bosons.
Essentially, the initially massless $W$-gauge particles become
massive below a symmetry breaking phase transition through a
generalized form of the Anderson-Higgs mechanism. This
symmetry breaking transition is analogous in some sense to
superconductivity with a high transition temperature.\\
A similar situation is encountered in
the quantum solid-state theory~\cite{pwa84}. Analogies
between the elementary particle and the solid-state theories
have both cognitive and practical importance for
their development~\cite{pwa84,ander75,heis69}. We have already discussed the
analogies with the Higgs effect playing an important
role in these theories. However, we have every
reason to also consider analogies with the Meissner
effect in the Ginzburg-Landau superconductivity
model~\cite{kovn92,du05,glm98,rosen}, because the Higgs model is, in fact, only a relativistic
analogue of that model.\\
Gauge symmetry breaking in superconductivity was investigated by
W. Kolley~\cite{kol99,kol01}.
The breakdown of the $U (1)$ gauge invariance in conventional superconductivity was thoroughly reexamined
by drawing parallels between the BCS-Bogoliubov   and Abelian Higgs   models. The global and local
$U (1)$ symmetries were broken spontaneously and explicitly in view of the Goldstone and Elitzur
 theorems, respectively. The approximations at which spontaneity comes into the symmetry-
breaking condensation, that are differently interpreted in the literature, were clarified.
A relativistic version of the Lawrence-Doniach model~\cite{ldm71,kkuz03,kuz09}  was formulated to break the local
$U (1)$ gauge symmetry in analogy to the Anderson-Higgs mechanism. Thereby the global
$U (1)$ invariance is spontaneously broken via the superconducting condensate. The resulting
differential-difference equations for the order parameter, the in-plane and inter-plane components
of the vector potential are of the Klein-Gordon, Proca and sine-Gordon type, respectively. A
comparison with the standard sine-Gordon equation for the superconducting phase difference was
given in the London limit. The presented dynamical scheme is applicable to high-$T_{c}$ cuprates  
with one layer per unit cell and weak interlayer Josephson tunnelling. The role of the layered structure
for the superconducting and normal properties of the correlated metallic systems is the subject of intense
discussions~\cite{kkuz03,kkuz02} and studies.
N.N. Bogoliubov and then Y. Nambu in in their works  shown that the general features of superconductivity 
are in fact model independent consequences of the spontaneous breakdown of electromagnetic gauge invariance. 
S. Weinberg wrote an interesting essay~\cite{weinb86} on superconductivity, whose inspiration 
comes from experience with broken symmetries in particle theory, which was formulated by Nambu. He emphasized that the 
high-precision predictions about superconductors actually follow not only from the microscopic models themselves, but
more generally from the fact that these models exhibit a spontaneous breakdown electromagnetic gauge invariance in a 
superconductor. The importance of broken symmetry in superconductivity has been especially emphasized by Anderson~\cite{ander58,pwa58}.
One needs detailed models like that BCS-Bogoliubov to explain the mechanism for the spontaneous symmetry
breakdown, and as a basis for approximate quantitative calculations, but not to derive the most important exact 
consequences of this breakdown. 
To demonstrate
this, let us assume that, for whatever reason, electromagnetic gauge invariance in a superconductor was broken. The specific
mechanism by which the symmetry breakdown occurs will not be specified for the moment. For this case the electromagnetic
gauge group is $U(1)$, the group of multiplication of fields $\psi(x)$ of charge $q$ with the phases
\begin{equation} \label{e136a}
\psi(x) \rightarrow  e^{i q \Lambda/\hbar}    \psi(x).
\end{equation}
It is possible to assume that all charges $q$ are integer multiples of the electron charge $- e$, so
phases $\Lambda$ and $\Lambda + 2 \pi \hbar/e$ are to be regarded as identical~\cite{weinb86}. This $U(1)$
is spontaneously broken to $Z_{2}$, the subgroup consisting of $U(1)$ transformations with 
$\Lambda = 0$ and $\Lambda =  \pi \hbar/e$. The assumption that $Z_{2}$ is unbroken arises from the physical 
picture that, while pairs of electron operators can have non-vanishing expectation value, individual electron
operators do not. \\ In terms of the  BCS-Bogoliubov theory of superconductivity~\cite{bb} this means that
the averages $\langle a_{k\sigma} a_{k-\sigma} \rangle $ and $\langle a^{\dagger}_{k-\sigma} a^{\dagger}_{k\sigma} \rangle $ will be
of non-zero value. 
It is important to emphasize that the BCS-Bogoliubov theory of
superconductivity~\cite{bb,nnb71}  was formulated on the
basis of a trial Hamiltonian  which  consists of a quadratic form
of creation and annihilation operators, including "anomalous" (off-diagonal) averages~\cite{bb}.
The  strong-coupling BCS-Bogoliubov theory of superconductivity was  formulated for the Hubbard model
in the localized Wannier representation in Refs.~\cite{kuz09,kuznc02,khp83}
Therefore, instead of the algebra of the normal state's
operator $a_{i\sigma}, a^{\dagger}_{i\sigma}$ and $n_{i\sigma}$, for description of superconducting
states, one has to use a more general algebra,
which includes the operators $a_{i\sigma}, a^{\dagger}_{i\sigma}, n_{i\sigma}$
and $a_{i\sigma}a_{i-\sigma}$, $a^{\dagger}_{i\sigma}a^{\dagger}_{i-\sigma}$.
The relevant generalized one-electron Green function will have  
the following form~\cite{kuz09,kuznc02,khp83}:
\begin{eqnarray}
\label{e136}
 G_{ij} (\omega) =
\begin{pmatrix} 
G_{11}&G_{12}\cr 
G_{21}&G_{22}\cr 
\end{pmatrix}    =  \begin{pmatrix}
\langle \langle a_{i\sigma}\vert a^{\dagger}_{j\sigma} \rangle \rangle & \langle \langle a_{i\sigma}\vert
a_{j-\sigma} \rangle \rangle \cr \langle \langle a^{\dagger}_{i-\sigma}\vert
a^{\dagger}_{j\sigma} \rangle \rangle & \langle \langle a^{\dagger}_{i-\sigma}\vert
a_{j-\sigma} \rangle \rangle \cr    
\end{pmatrix}.
\end{eqnarray}
As it was  discussed in Ref.~\cite{kuz09,kuznc02}, the off-diagonal (anomalous)
entries of the above matrix select the vacuum state of
the system in the BCS-Bogoliubov form, and they are
responsible for the presence of anomalous averages.
For treating the problem we follow  the general scheme of the irreducible Green functions 
method~\cite{kuz09,kuznc02}.
In this approach we start from the equation of motion for the Green function $G_{ij} (\omega)$ (normal and anomalous components)
\begin{eqnarray}
\label{e137} \sum_{j}(\omega \delta_{ij} - t_{ij})\langle \langle a_{j\sigma}
\vert a^{\dagger}_{i'\sigma} \rangle \rangle = \delta_{ii'} + \\  \nonumber
U \langle \langle a_{i\sigma} n_{i-\sigma} \vert a^{\dagger}_{i'\sigma} \rangle \rangle +
\sum_{nj} V_{ijn} \langle \langle a_{j\sigma} u_{n} \vert a^{\dagger}_{i'\sigma} \rangle \rangle, \\
\label{e138} \sum_{j}(\omega \delta_{ij} +
t_{ij}) \langle \langle a^{\dagger}_{j-\sigma} \vert a^{\dagger}_{i'\sigma} \rangle \rangle =
\\  \nonumber
-U \langle \langle a^{\dagger}_{i-\sigma} n_{i\sigma} \vert
a^{\dagger}_{i'\sigma} \rangle \rangle + \sum_{nj} V_{jin}
\langle \langle a^{\dagger}_{j-\sigma} u_{n} \vert a^{\dagger}_{i'\sigma} \rangle \rangle.
\end{eqnarray}
The irreducible Green functions are introduced by definition
\begin{eqnarray}
\label{e139}
(^{(ir)}\langle \langle a_{i\sigma}a^{\dagger}_{i-\sigma}a_{i-\sigma} \vert
a^{\dagger}_{i'\sigma}\rangle \rangle_ {\omega} )   =
\langle \langle a_{i\sigma}a^{\dagger}_{i-\sigma}a_{i-\sigma}\vert
a^{\dagger}_{i'\sigma} \rangle \rangle_{\omega} - \\ \nonumber
- \langle n_{i-\sigma}\rangle G_{11} + \langle a_{i\sigma}a_{i-\sigma}\rangle
\langle \langle a^{\dagger}_{i-\sigma} \vert a^{\dagger}_{i'\sigma} \rangle \rangle_{\omega},\\
\nonumber
(^{(ir)}\langle \langle a^{\dagger}_{i\sigma}a_{i\sigma}a^{\dagger}_{i-\sigma}
\vert a^{\dagger}_{i'\sigma} \rangle \rangle_ {\omega} )  =
\langle \langle a^{\dagger}_{i\sigma}a_{i\sigma}a^{\dagger}_{i-\sigma}\vert
a^{\dagger}_{i'\sigma} \rangle \rangle_{\omega} - \\ \nonumber
- \langle n_{i\sigma}\rangle G_{21} +
\langle a^{\dagger}_{i\sigma}a^{\dagger}_{i-\sigma}\rangle \langle \langle a_{i\sigma} \vert
a^{\dagger}_{i'\sigma} \rangle \rangle_{\omega}. \nonumber
\end{eqnarray}
The self-consistent system of
superconductivity equations follows from the Dyson
equation~\cite{kuz09,kuznc02}
\begin{equation}
\label{e140} \hat G_{ii'}(\omega) = \hat G^{0}_{ii'}(\omega) +
\sum_{jj'} \hat G^{0}_{ij} (\omega) \hat M_{jj'}( \omega) \hat
G_{j'i'} (\omega).
\end{equation}
The mass operator $M_{jj'}( \omega)$ describes the processes of
inelastic electron scattering on lattice vibrations.
The elastic processes are described by the quantity
\begin{equation} \label{eq.64}
\Sigma^{c}_{\sigma} =  U 
\begin{pmatrix}  
\langle a^{\dagger}_{i-\sigma}a_{i-\sigma} \rangle & -\langle a_{i\sigma} a_{i-\sigma} \rangle \cr
- \langle a^{\dagger}_{i-\sigma} a^{\dagger}_{i\sigma} \rangle &
- \langle a^{\dagger}_{i\sigma} a_{i\sigma} \rangle \cr 
\end{pmatrix}.
\end{equation}
Thus the "anomalous" off-diagonal terms fix the relevant
BCS-Bogoliubov vacuum and select the appropriate set of
solutions. The functional of the generalized mean
field   for the superconducting single-band Hubbard model  is of
the  form $\Sigma^{c}_{\sigma}$. 
Bogoliubov and Moskalenko~\cite{nnbmo}
have developed an alternative approach to treat the problem of
superconductivity for one-band Hubbard model within a diagrammatic technique.\\
A remark about the BCS-Bogoliubov mean-field
approach is instructive.  Speaking in physical terms, this theory
involves a condensation correctly, in spite that such a
condensation cannot be obtained by an expansion in the effective
interaction between electrons. Other mean field theories, {\it
e.g.} the Weiss molecular field theory~\cite{cmp07} and the van der Waals
theory of the liquid-gas transition are  much less reliable. The
reason why a mean-field theory of the superconductivity in the
BCS-Bogoliubov form is successful would appear to be that the
main correlations in metal are governed by the extreme degeneracy
of the electron gas. The correlations due to the pair
condensation, although they have dramatic effects, are weak (at
least in the ordinary superconductors) in comparison with the
typical electron energies, and may be treated in an average way
with a reasonable accuracy. All above remarks have relevance to
ordinary low-temperature superconductors. In high-$T_c$
superconductors, the corresponding degeneracy temperature is much
lower, and the transition temperatures are much higher. In
addition, the relevant interaction  responsible for the pairing
and its strength are unknown yet. From this point of view, the
high-$T_c$ systems are more complicated~\cite{mom03,jar05,scal09}. It should be clarified
what governs the scale of temperatures, {\it i.e.} critical
temperature, degeneracy temperature, interaction strength or
their complex combination, {\it etc.} In this way a useful insight
into this extremely complicated problem would be gained. 
It should be emphasized that
the high-temperature superconductors, discovered two decades ago, motivated an intensification  of
research in superconductivity, not only because applications are promising, but because they also represent a new state 
of matter that breaks certain fundamental symmetries~\cite{sig98,kirt00,herb02,quin09,bar09}. 
These are the broken symmetries of gauge (superconductivity), reflection ($d$-wave superconducting order parameter),
and time-reversal (ferromagnetism). Note that general discussion of decay of superconducting and magnetic correlations in one-
and two-dimensional Hubbard model was carried out in Ref.~\cite{kotas}  \\
Studies of the high-temperature superconductors confirmed and clarified many important fundamental aspects of
superconductivity theory.
Kadowaki,  Kakeya  and   Kindo~\cite{kado98} reported about
observation of the Nambu-Goldstone mode in the layered
high-temperature superconductor $Bi_{2}Sr_{2}CaCu_{2}O_{8+\delta}.$
The Josephson plasma resonance  (for rewiev see Ref.~\cite{rpp10}) has been observed in a microwave frequency at
35 $GHz$ in magnetic fields up to 6 $T$.
Making use of the different dispersion relations between two Josephson plasma modes predicted
by the recent theories, the longitudinal mode, which is the Nambu-Goldstone mode in a superconductor,
was separated out from the transverse one experimentally. This experimental result
directly proves the existence of the Nambu-Goldstone mode in a superconductor with a finite
energy gap $\hbar \omega_{p} = \hbar c/\lambda_{c} \sqrt{\varepsilon}.$
Such a finite energy gap implies the mass of the Nambu-Goldstone
bosons in a superconductor, supporting the mass formation mechanism proposed by Anderson~\cite{ander58,pwa58,ander63}.\\
Matsui and co-authors~\cite{mats03}
performed high-resolution angle-resolved photoemission spectroscopy on triple-layered high-$T_{c}$ 
cuprate  $Bi_{2}Sr_{2}Ca_{2}Cu_{3}O_{10+\delta}.$   
They have observed the full energy dispersion (electron and hole branches) of Bogoliubov quasiparticles and determined the 
coherence factors 
above and below $E_{F}$ as a function of momentum from the spectral intensity as well as from the energy dispersion 
based on BCS-Bogoliubov theory. 
The good quantitative agreement between the experiment and the theoretical prediction suggests the basic validity 
of BCS-Bogoliubov formalism in describing the superconducting state of cuprates.\\
J. van Wezel and J. van den Brink~\cite{wez08}
studied spontaneous symmetry breaking and decoherence in superconductors.
They shown that superconductors have a thin spectrum associated with spontaneous symmetry breaking similar
to that of antiferromagnets, while still being in full agreement with Elitzur's theorem, which forbids the
spontaneous breaking of local (gauge)  symmetries. This thin spectrum in the superconductors consists of
in-gap states that are associated with the spontaneous breaking of a global phase symmetry. In qubits based on
mesoscopic superconducting devices, the presence of the thin spectrum implies a maximum coherence time
which is proportional to the number of Cooper pairs in the device. Authors  presented the detailed calculations
leading up to these results and discussed the relation between spontaneous symmetry breaking in superconductors
and the Meissner effect, the Anderson-Higgs mechanism, and the Josephson effect. Whereas for the Meissner
effect a symmetry breaking of the phase of the superconductor is not required, it is essential for the Josephson
effect.\\
It is of interest to note that the authors of the recent review on the high-temperature superconductivity~\cite{saw08}
pointed out that
one of the keys to the high-temperature superconductivity puzzle is the \emph{identification of the energy scales} 
associated with the emergence of a coherent condensate of superconducting electron pairs. These might provide a measure 
of the pairing strength and of the coherence of the superfluid, and ultimately reveal the nature of the elusive pairing 
mechanism in the superconducting cuprates. To this end, a great deal of effort has been devoted to investigating the 
connection 
between the superconducting transition temperature $T_{c}$ and the normal-state pseudogap crossover 
temperature $T^{*}$. Authors analyzed  a large body of experimental data which suggests a coexisting two-gap scenario, 
i.e. superconducting gap and pseudogap, over the whole superconducting dome. They focused on spectroscopic data from 
cuprate systems characterized 
by $T_{\rm c}^{\rm max} \sim 95\,{\rm K}$ , such as $Bi_{2}Sr_{2}CaCu_{2}O_{8+\delta}$, $YBa_{2}Cu_{3}O_{7-\delta}$, 
$Tl_{2}Ba_{2}CuO_{6+\delta}$ and $HgBa_{2}CuO_{4+\delta}$, with 
particular emphasis on the $Bi$-compound which has been the most extensively studied with single-particle spectroscopies.
Their analysis have something in common with the concept of the quantum protectorate which emphasizes the importance
of the hierarchy of the energy scales. 
%
%
%
\subsection{Antiferromagnetism} 
%
%
Superconductivity and antiferromagnetism are both the spontaneously broken symmetries~\cite{stern64}.
The idea of antiferromagnetism was first introduced by L. Neel in order to explain the temperature-independent
paramagnetic susceptibility of metals like $Mn$ and $Cr$. According to his idea these materials consisted
of two compensating sublattices undergoing negative exchange interactions.
There are two complementary physical pictures of the antiferromagnetic ordering,
operated with localized spins and itinerant electrons~\cite{kuz09}.
L. Neel   formulated  also the concept
of local mean fields~\cite{kuz09}. He assumed that the sign
of the mean-field could be both positive and negative.
Moreover, he showed that below some critical temperature
(the Neel temperature) the energetically most
favorable arrangement of atomic magnetic moments is
such, that there is an equal number of magnetic
moments aligned against each other. This novel magnetic
structure became known as the antiferromagnetism~\cite{mar55}. It was established that the antiferromagnetic
interaction tends to align neighboring spins
against each other. In the one-dimensional case this corresponds
to an alternating structure, where an "up" spin
is followed by a "down" spin, and vice versa. Later it
was conjectured that the state made up from two
inserted into each other sublattices is the ground state of
the system (in the classical sense of this term). Moreover,
the mean-field sign there alternates in the "chessboard"
(staggered) order. The question of the true antiferromagnetic
ground state is not completely clarified
up to the present time. This is related to the
fact that, in contrast to ferromagnets, which have a
unique ground state, antiferromagnets can have several
different optimal states with the lowest energy. The
Neel ground state is understood as a possible form of
the system's wave function, describing the antiferromagnetic
ordering of all spins. Strictly speaking,
the ground state is the thermodynamically equilibrium
state of the system at zero temperature. Whether the
Neel state is the ground state in this strict sense or not,
is still unknown. It is clear though, that in the general
case, the Neel state is not an eigenstate of the Heisenberg
antiferromagnet's Hamiltonian. On the contrary,
similar to any other possible quantum state, it is only
some linear combination of the Hamiltonian eigenstates.
Therefore, the main problem requiring a rigorous
investigation is the question of Neel state's 
stability. In some sense, only for infinitely large lattices,
the Neel state becomes the eigenstate of the Hamiltonian
and the ground state of the system. Nevertheless,
the sublattice structure is observed in experiments on
neutron scattering, and, despite certain objections, the actual 
existence of sublattices   is beyond doubt.
It should be noted that the spin-wave spectrum of the 
Heisenberg  antiferromagnet differs from the spectrum of the Heisenberg  ferromagnet.
This point was analyzed thoroughly by Baryakhtar and Popov~\cite{bp68}.\\
The antiferromagnetic state is
characterized by a spatially changing component of magnetization which
varies in such a way that the net magnetization of the system is zero.
The concept of antiferromagnetism of localized spins which is based on
the Heisenberg model and the two-sublattice Neel ground state is
relatively well founded contrary to the antiferromagnetism of
delocalized or itinerant electrons. The itinerant-electron picture is
the alternative conceptual picture for magnetism~\cite{her22}.
The simplified band model of an antiferromagnet  has been formulated by
Slater  within the single-particle Hartree-Fock  
approximation. In his approach he used the "exchange repulsion" to
keep electrons with parallel spins away from each other and to lower
the Coulomb interaction energy. Some authors consider it as a prototype
of the Hubbard model. However the exchange repulsion was taken
proportional to the number of electrons with the same spins only and
the energy gap between two subbands was proportional to the difference
of electrons with up and down spins. In the antiferromagnetic many-body
problem there is an additional "symmetry broken" aspect.  For an
antiferromagnet, contrary to ferromagnet, the one-electron  Hartree-Fock
potential can violate the translational crystal symmetry. The period of
the antiferromagnetic spin structure $L$ is greater than the lattice
constant $a$. To introduce the two-sublattice picture for itinerant
model one should assume that $L=2a$ and that the spins of outer
electrons on neighbouring atoms are antiparallel to each other. In
other words, the alternating Hartree-Fock potential $v_{i\sigma} = -\sigma
v \exp(iQR_{i})$ where $Q = (\pi/2,\pi/2,\pi/2)$ corresponds to a
two-sublattice antiferromagnetic structure. To justify an antiferromagnetic ordering
with alternating up and down spin structure we must admit that in
effect two different charge distributions will arise concentrated on
atoms of sublattices A and B.  This  picture  accounts well for
quasi-localized magnetic behavior.\\ The earlier theories of itinerant
antiferromagnetism were proposed by des Cloizeaux and
especially Overhauser~\cite{over62,over63}.   
Overhauser invented a concept of the static spin density wave which allow the total charge density of the
gas to remain spatially uniform.  He suggested~\cite{over62,over63}  
that the mean field ground state of a three dimensional
electron gas is not necessarily a Slater determinant of plane waves.
Alternative sets of one-particle states can lead to a lower
ground-state energy.  Among these alternatives to the plane-wave state
are the spin density wave and charge density wave ground states for which the one-electron
Hamiltonians have the form
\begin{equation} H = (p^{2}/2m) -
G(\sigma_{x} \cos Qz + \sigma_{y} \sin Qz) 
\end{equation}
(spiral spin density wave; $Q = 2k_{F}z$ )  and  
\begin{equation} H = (p^{2}/2m) - 2G\cos (Qr)
\end{equation} 
(charge density wave; $Q = 2k_{F}z$). The periodic potentials in the
above expressions lead to a corresponding variation in the electronic
spin and charge densities, accompanied by a compensating variation of
the background. The effect of Coulomb interaction on the magnetic
properties of the electron gas in Overhauser's approach renders the
paramagnetic plane-wave state of the free-electron-gas model unstable
within the mean field approximation. The long-range components of the Coulomb
interaction are most important in creating this
instability.  It was demonstrated   that a
nonuniform static spin density wave is lower in energy than the uniform (paramagnetic
state) in the Coulomb gas within the mean field approximation for certain
electron density~\cite{over62,over63,fano65,fano66,fed66,penn66,penn67,verb91,kan94}.  The mean field is 
the simplest approximation but neglects
the important dynamical part. To include the dynamics one should take
into consideration the correlation effects. The role of correlation
corrections which  tend to suppress the spin density wave state as well as the role
of shielding and screening were not fully
clarified.  In the Overhauser's approach to
itinerant antiferromagnetism the combination of the electronic states
with different spins (with pairing of the opposite spins) is used to
describe the spin density wave state with period $Q$.  The first approach is
obviously valid only in the simple commensurate two-sublattice case and
the latter is applicable to the more general case of an incommensurate
spiral spin state. The general spin density wave state has the form
\begin{equation} \Psi_{p\sigma} = \chi_{p\sigma} \cos (\theta_{p}/2) +
\chi_{p+Q-\sigma} \sin (\theta_{p}/2) \end{equation}
The average spin
for helical or spiral spin arrangement changes its direction in
the $(x-y)$ plane.  For the spiral spin density wave states a spatial variation of
magnetization corresponds to $\vec Q = (\pi/a )(1,1)$.\\ The
antiferromagnetic phase of chromium  and its
alloys has been satisfactorily explained in terms of the spin density wave within a
two-band model.  It is essential to note that chromium
becomes antiferromagnetic in a unique manner. The antiferromagnetism is
established in a more subtle way from  the spins of the
itinerant electrons than  the magnetism of collective band
electrons in  metals like iron and nickel. The essential feature of
chromium which makes possible the formation of the spin density wave is the existence
of "nested" portions of the Fermi surface.  The formation
of bound electron-hole pairs takes place between particles of opposite
spins; the condensed state exhibits the spin density wave.\\
The problem of a great importance is to understand  
how broken symmetry can be produced in antiferromagnetism? 
(see Refs.~\cite{fano65,kuz99,kapl90,kotas93,tasafm,tas94,rodr97,kivel04})
Indeed, it
was written in the paper~\cite{tas94} (see also Refs.~\cite{fano65,kuz99,kotas93,tasafm,rodr97,kivel04}):
"One should recall that there are many situations in nature where we do observe a symmetry breaking in the
absence of explicit symmetry-breaking fields. A typical example is antiferromagnetism, in which a staggered
magnetic field plays  the role of symmetry-breaking field.   No mechanism can generate
a real staggered magnetic field in antiferromagnetic material. A more drastic example is the Bose-Einstein condensation,
where the symmetry-breaking field should create and annihilate particles".
The applicability of the Overhauser's  spin density wave  concept to
highly correlated tight binding electrons on a lattice  within the Hubbard model of the correlated
lattice fermions  was analyzed in Ref.~\cite{kuz99} 
It was shown the importance of the notion of
generalized mean fields~\cite{kuz09,kuznc02} which may arise in the
system of correlated lattice fermions to justify and understand the
"nature" of the local staggered mean-fields which fix the itinerant
antiferromagnetic ordering.\\    
According to Bogoliubov ideas on quasiaverages~\cite{nnb61}, 
in each condensed phase, in addition to the normal process, there is an
anomalous process (or processes) which can take place because of
the long-range internal field, with a corresponding propagator.
Additionally, the Goldstone theorem~\cite{gold61} states that, in a
system in which a continuous symmetry is broken ({\it i.e.} a
system such that the ground state is not invariant under the
operations of a continuous unitary group whose generators commute
with the Hamiltonian), there exists a collective mode with
frequency vanishing, as the momentum goes to zero. For
many-particle systems on a lattice, this statement needs a proper
adaptation.   In the above form, the Goldstone theorem is true
only if the condensed and  normal phases have the same
translational properties. When translational symmetry is also
broken, the Goldstone mode appears at a zero frequency but at
nonzero momentum, {\it e.g.}, a crystal and a helical
spin-density-wave  ordering.
The problem of
the adequate description of  strongly correlated lattice fermions
has been studied intensively during the last decade.   The
microscopic theory of the itinerant ferromagnetism and
antiferromagnetism of strongly correlated
fermions on a lattice at finite temperatures is one of the important
issues of recent efforts in the field.
In some papers  the spin-density-wave   spectrum was only used
without careful and complete analysis of the quasiparticle spectra of
correlated lattice fermions.  It was of importance to investigate
the intrinsic nature of the "symmetry broken" (ferro- and
antiferromagnetic) solutions of the Hubbard model at finite
temperatures from the many-body point of view. 
For the itinerant antiferromagnetism
 the spin density wave spectra  were calculated~\cite{kuz99} by the irreducible Green 
functions  method~\cite{kuz09}, taking into account the damping of
quasiparticles.  This alternative derivation has a close resemblance to
that of the BCS-Bogoliubov theory of superconductivity for transition
metals~\cite{kuz09},  using the Nambu representation. This aspect of the theory is connected with the
concept of broken symmetry, which was discussed in detail for that
case. 
A unified scheme for the construction of generalized mean fields (elastic scattering corrections) 
and self-energy (inelastic scattering) in terms of the Dyson equation was  generalized in
order to include  the "source fields". The "symmetry broken"
dynamic solutions of the Hubbard model  which correspond to
various types of itinerant antiferromagnetism were  clarified.
This approach complements previous studies of microscopic theory
of the Heisenberg antiferromagnet~\cite{km90} and clarifies the
 concepts of Neel sublattices for localized and
itinerant antiferromagnetism and "spin-aligning fields" of
correlated lattice fermions~\cite{mat68}.
The advantage of the Green's function method is the
relative ease with which temperature effects may be calculated.\\  
It is necessary to emphasize
that that there is an intimate connection between the adequate
introduction of mean fields and internal symmetries of the
Hamiltonian~\cite{mat68}. 
The anomalous propagators for an interacting many-fermion system
corresponding to the ferromagnetic (FM), antiferromagnetic
(AFM),  and superconducting (SC) long-range ordering are given by
\begin{eqnarray} \label{eq.71}
FM: G_{fm} \sim \langle \langle a_{k\sigma};a^{\dagger}_{k-\sigma} \rangle \rangle\\
\nonumber AFM: G_{afm} \sim \langle \langle a_{k+Q\sigma};a^{\dagger}_{k+Q'\sigma'} \rangle \rangle\\
\nonumber SC: G_{sc} \sim \langle \langle a_{k\sigma};a_{ - k -\sigma} \rangle \rangle\\
\nonumber
\end{eqnarray} In the spin-density-wave case, a particle picks up a momentum $Q - Q'$
from scattering against the periodic structure of the spiral (nonuniform) internal field, and has its spin changed from
$\sigma$ to $\sigma'$ by the spin-aligning character of the
internal field.  The long-range-order   parameters are:
\begin{eqnarray} \label{eq.72}
FM: m = 1/N\sum_{k\sigma} \langle a^{\dagger}_{k\sigma}a_{k-\sigma} \rangle\\
\nonumber AFM:
M_{Q} = \sum_{k\sigma} \langle a^{\dagger}_{k\sigma}a_{k+Q-\sigma} \rangle \\
\nonumber SC: \Delta = \sum_{k} \langle a^{\dagger}_{
-k\downarrow}a^{\dagger}_{k \uparrow} \rangle\\ \nonumber
\end{eqnarray} It is
important to note that the long-range order parameters are
functions of the internal field, which is itself a function of the
order parameter. There is a more mathematical way of formulating
this assertion. According to the paper~\cite{nnb61}, the notion
"symmetry breaking" means that the state fails to have the
symmetry that the Hamiltonian has.\\
A true breaking of symmetry can arise only if there are
infinitesimal "source fields".   Indeed, for the rotationally and
translationally invariant Hamiltonian,  suitable source terms
should be added~\cite{mat68}:
\begin{eqnarray} \label{eq.73}
FM:  \nu \mu_{B}
H_{x}\sum_{k\sigma}a^{\dagger}_{k\sigma}a_{k-\sigma}\\ \nonumber
AFM:
\nu \mu_{B} H \sum_{kQ} a^{\dagger}_{k\sigma}a_{k+Q-\sigma}\\
\nonumber SC: \nu v \sum_{k} (a^{\dagger}_{- k \downarrow}
a^{\dagger}_{k \uparrow} + a_{k \uparrow} a_{ - k \downarrow})
\end{eqnarray} where $\nu \rightarrow 0$ is to be taken at
the end of calculations.\\ For example,  broken symmetry
solutions of the spin-density-wave type  imply that the vector $Q$ is a measure
of the inhomogeneity or breaking of translational symmetry.  The
Hubbard model  (\ref{hu1})  is a very interesting tool for  analyzing   the
symmetry broken concept. It is possible to show that
antiferromagnetic state and more complicated states ( {\it e.g.}
ferrimagnetic) can be made eigenfunctions of the self-consistent
field equations within an "extended"  mean-field approach,
assuming that the "anomalous" averages
$\langle a^{\dagger}_{i\sigma}a_{i-\sigma} \rangle$ determine the behavior of
the system on the same footing as the "normal" density of
quasi-particles $\langle a^{\dagger}_{i\sigma}a_{i\sigma} \rangle$.  It is
clear, however, that these "spin-flip" terms break the rotational
symmetry of the Hubbard Hamiltonian~\cite{kiso2}. 
Kishore and Joshi~\cite{kiso2}  discussed the metal-nonmetal transition in ferromagnetic as well as in 
antiferromagnetic systems having one electron per atom and described by the Hamiltonian which consists of one 
particle energies of electrons, intra-atomic Coulomb, and interatomic 
Coulomb and exchange interactions between electrons. It was found that the anomalous correlation functions 
corresponding to spin flip processes in the Hartree-Fock approximation give rise to the metal- nonmetal transition. 
The nature of phase transition in ferromagnetic and antiferromagnetic systems was compared and clarified in their study.\\
For the single-band Hubbard
Hamiltonian, the averages $\langle a^{\dagger}_{i-\sigma}a_{i, \sigma} \rangle
= 0$ because of the rotational symmetry of the Hubbard model.  The
inclusion of  "anomalous" averages leads to the so-called
generalized mean field approximation.  This type of
approximation was  used sometimes also for the single-band Hubbard
model for  calculating  the density of states. For this aim, the
following definition of generalized mean field approximation
\begin{equation} \label{eq.74}
n_{i-\sigma}a_{i\sigma} \approx \langle n_{i-\sigma}\rangle a_{i\sigma} -
\langle a^{\dagger}_{i-\sigma}a_{i\sigma}\rangle a_{i-\sigma}
\end{equation}
was  used. Thus, in addition to the standard mean field term, the new
 so-called ``spin-flip" terms are retained. This example
clearly shows that the structure of  mean field follows from the
specificity of the problem and should be defined in a proper
way.  So, one needs a properly defined effective Hamiltonian
$H_{\rm eff}$.  In paper~\cite{kuz99} we thoroughly analyzed
 the proper definition of the irreducible Green functions which
includes the ``spin-flip" terms for the case of itinerant
antiferromagnetism  of correlated lattice fermions. For
the single-orbital Hubbard model, the definition of the
irreducible  part should be modified in the following way:
\begin{eqnarray} \label{eq.75}
^{(ir)}\langle \langle a_{k+p\sigma}a^{\dagger}_{p+q-\sigma}a_{q-\sigma} \vert
a^{\dagger}_{k\sigma} \rangle \rangle_ {\omega} =
\langle \langle a_{k+p\sigma}a^{\dagger}_{p+q-\sigma}a_{q-\sigma}\vert
a^{\dagger}_{k\sigma} \rangle \rangle_{\omega} - \nonumber\\ \delta_{p,
0}<n_{q-\sigma}>G_{k\sigma} -
<a_{k+p\sigma}a^{\dagger}_{p+q-\sigma}> \langle \langle a_{q-\sigma} \vert
a^{\dagger}_{k\sigma} \rangle \rangle_{\omega}.
\end{eqnarray}
From this definition it follows that this way of introduction of
the irreducible Green functions  broadens the initial algebra of  operators and the
initial set of the Green functions.  This means that the ``actual" algebra of
 operators must include the spin-flip terms from the beginning,
namely:  $(a_{i\sigma}$, $a^{\dagger}_{i\sigma}$, $n_{i\sigma}$,
$a^{\dagger}_{i\sigma}a_{i-\sigma})$. The corresponding initial
Green function will be of the form
$$\begin{pmatrix}
\langle \langle a_{i\sigma}\vert a^{\dag}_{j\sigma}\rangle \rangle & \langle \langle a_{i\sigma}\vert
a^{\dag}_{j-\sigma}\rangle \rangle \cr
\langle \langle a_{i-\sigma}\vert a^{\dag}_{j\sigma}\rangle \rangle & \langle \langle a_{i-\sigma}\vert
a^{\dag}_{j-\sigma} \rangle \rangle \cr  \end{pmatrix}.$$
With this definition, one
introduces the so-called anomalous (off-diagonal) Green functions which fix
the relevant vacuum and select the proper symmetry broken
solutions. 
The theory of the itinerant antiferromagnetism~\cite{kuz99} was formulated by using  sophisticated arguments of
the irreducible Green functions method in complete analogy with our description of the
Heisenberg  antiferromagnet at finite temperatures~\cite{km90}.
For the two-sublattice antiferromagnet we  used the matrix
Green function of the form
\begin{equation} \label{e92}
\hat G(k;\omega) = 
\begin{pmatrix}
   \langle \langle S^{+}_{ka} \vert S^{-}_{-ka} \rangle \rangle &
\langle \langle S^{+}_{ka} \vert S^{-}_{-kb}\rangle \rangle \cr \langle \langle S^{+}_{kb}\vert
S^{-}_{-ka}\rangle \rangle& \langle \langle S^{+}_{kb}\vert S^{-}_{-kb}\rangle \rangle \cr
\end{pmatrix}. 
\end{equation}
Here, the Green functions on the main diagonal are the usual or
normal Green functions, while the off-diagonal Green functions describe contributions
from the so-called anomalous terms, analogous
to the anomalous terms in the BCS-Bogoliubov superconductivity
theory. The anomalous (or off-diagonal)
average values in this case select the vacuum state
of the system precisely in the form of the two-sublattice
Neel state~\cite{kuz09}. The investigation of the existence of the antiferromagnetic solutions
in the multiorbital and two-dimensional Hubbard model is an active topic of research. Some complementary to the present study
aspects of the broken symmetry solutions  of the  Hubbard model were considered in Refs.~\cite{oza1,oza2,oza3,oza4,baier,bick05,mas10}
%
%
\subsection{Bose Systems} 
%

A significant development in the past decades have been experimental and theoretical studies of the Bose systems
at low 
temperatures~\cite{pit03,peth02,joa04,katz05,apos06,pnas07,lieb07,gaul08,dalib08,pit08,fet09,wid68,ferna71,jas71,imry74,bou76,roep78,heg98,leg01,wehr06,bis08}. 
Any state of matter is classified according to its order, and the type of order that a physical system can possess is profoundly affected 
by its dimensionality. Conventional long-range order, as in a ferromagnet or a crystal, is common in three-dimensional systems at low 
temperature. However, in two-dimensional systems with a continuous symmetry, true long-range order is destroyed by thermal fluctuations 
at any finite temperature. Consequently, for the case of identical bosons, a uniform two-dimensional fluid cannot undergo 
Bose-Einstein condensation, in contrast to the three-dimensional case. However, the two-dimensional system can form a 'quasi-condensate' 
and become superfluid below a finite critical temperature.
Generally inter-particle interaction is responsible for a phase transition.
But Bose-Einstein condensation type of phase transition occurs entirely due
to the Bose-Einstein   statistics. The typical situation is a many-body system made of identical bosons, e.g. atoms carrying an integer total 
angular momentum. To proceed one must construct the ground state. The simplest possibility to do so
occurs when bosons are non-interacting. In this case, the ground state is simply obtained by putting
all bosons in the lowest energy single particle state. If the number of bosons is taken to be $N$, then
the ground state is $|N, 0, \ldots \rangle$ with energy $N \varepsilon_{0}$. This straightforward observation underlies the
phenomenon of Bose-Einstein condensation: A finite or macroscopic fraction of bosons has the
single-particle energy $\varepsilon_{0}$ below the Bose-Einstein transition temperature $T_{BE}$ in the thermodynamic
limit of infinite volume $V$ but finite particle density. From a conceptual point of view, it is more
fruitful to associate Bose-Einstein condensation with the phenomenon of spontaneous symmetry
breaking of a continuous symmetry than with macroscopic occupation of a single-particle level.
The continuous symmetry in question is the freedom in the choice of the global phase of the many particle
wave functions. This symmetry is responsible for total particle number conservation. In
mathematical terms, the vanishing commutator $[H, N_{tot}]$
between the total number operator $N_{tot}$ and the single-particle Hamiltonian $H$ implies a global $U(1)$ gauge symmetry.
Spontaneous symmetry breaking in Bose-Einstein condensates was studied in Refs.~\cite{heg98,lieb07}
The structure of the many-particle wavefunction for a pair of ideal gas Bose-Einstein
condensates $a, b$  in the number eigenstate $|N_{a}N_{b}\rangle$ was analyzed~\cite{heg98}. It was found that the most
probable many-particle position or momentum measurement outcomes break the
configurational phase symmetry of the state. Analytical expressions for the particle distribution
and current density for a single experimental run are derived and found to display
interference. Spontaneous symmetry breaking is thus predicted and explained here simply
and directly as a highly probable measurement outcome for a state with a definite number of
particles. Lieb and co-authors~\cite{lieb07} presented a general proof of spontaneous breaking of gauge symmetry as a 
consequence of Bose-Einstein condensation. The proof is based on a rigorous validation of Bogoliubov's $c$-number
substitution for the $\vec{k} = 0$ mode operator $a_{0}.$\\
It has been conclusively demonstrated that two-dimensional
systems of interacting bosons do not possesses long-range order at finite temperatures~\cite{wid68,ferna71}.
Gunther, Imry and Bergman~\cite{imry74} shown that the one- and two-dimensional ideal Bose gases undergo a phase transition if the temperature is lowered 
at constant pressure. At the pressure-dependent transition temperature $T_{c} (P)$ and in their thermodynamic limit the 
specific heat at constant pressure $ c_{p} $ and the particle density $n$ diverge, the entropy $S$ and specific heat at 
constant volume $c_{v} $ fall off sharply but continuously to zero, 
and the fraction of particles in the ground state $N_{0}/N$ jumps discontinuously from zero to one. This Bose-Einstein 
condensation provides a remarkable example of a transition which has most of the properties of a second-order phase 
transition, except that the order parameter is 
discontinuous. The nature of the condensed state is described in the large but finite $N$ regime, and the width of the 
transition region was estimated. The effects of interactions in real one- and two-dimensional Bose systems and the 
experiments on submonolayer helium films were discussed.\\
A stronger version of the Bogoliubov inequality was used by Roepstorff~\cite{roep78} to derive an upper bound for the 
anomalous average $|\langle a(x)\rangle|$ of an 
interacting nonrelativislic Bose field $a(x)$ at a finite temperature. This bound is $|  a(x)^{2} | < \rho R,$  where $R$ 
satisfies $1 - R = (RT/2T_{c})^{D/2},$ with $D$ 
the dimensionality, and $T_{c}$  the critical temperature in the absence of interactions. The formation of nonzero averages is closely related to 
the Bose-Einstein condensation and $|\langle a(x)\rangle|^{2} $ is often believed to coincide with the mean 
density $\rho_{0}$ of the condensate. Author have found nonrigorous 
arguments supporting the inequality $\rho_{0} \leq |\langle a(x)\rangle|^{2}$, which parallels the result of Griffiths 
in the case of spin systems.\\
Bose-Einstein condensation   continues to be a topic of high experimental and theoretical 
interest~\cite{pit03,peth02,joa04,katz05,apos06,pnas07,gaul08,dalib08,pit08,fet09,gir69,wehr06,hadzi}. 
The remarkable realization of Bose-Einstein condensation
of trapped alkali atoms  has created
an enormous interest in the properties of the weakly
interacting Bose gas. Although the experiments are carried
out in magnetic and optical harmonic traps, the homogeneous
Bose gas has also received renewed interest. The homogeneous Bose gas is
interesting in its own right, and it was proved useful to go
back to this somewhat simpler system to gain insight
that carries over to the trapped case. Within the last twenty
years  a lot of works were done on this topic.\\ 
The pioneering paper by Bogoliubov in 1947 was the starting point for a microscopic
theory of superfluidity~\cite{bogo47}. Bogoliubov found the non-perturbative solution for a weakly
interacting gas of bosons. The main step in the diagonalization of the Hamiltonian is the
famous Bogoliubov transformation, which expresses the elementary excitations (or quasiparticles)
with momentum $q$ in terms of the free particle states with momentum $+q$ and $-q$.
For small momenta, the quasiparticles are a superposition of $+q$ and $-q$ momentum
states of free particles.
Recently
experimental observation of the Bogoliubov transformation for a Bose-
Einstein condensed gas become possible~\cite{katz05,gaul08,pit00,vog02}.
Following the theoretical suggestion in Ref.~\cite{pit00} authors of paper~\cite{vog02} observed such superposition states
by first optically imprinting phonons with wavevector $q$ into a Bose-Einstein condensate
and probing their momentum distribution using Bragg spectroscopy with a high
momentum transfer. By combining both momentum and frequency selectivity, they were
able to "directly photograph" the Bogoliubov transformation~\cite{vog02}.\\
It is interesting to note that
Sannino   and   Tuominen~\cite{sann03} reconsidered the
spontaneous symmetry breaking in gauge theories via Bose-Einstein condensation.
They proposed a mechanism leading naturally to spontaneous symmetry breaking in a gauge theory. The Higgs field was assumed 
to have global and gauged internal symmetries. Authors associated a nonzero chemical potential with one of the globally 
conserved charges commuting with 
all the gauge transformations. This induces a negative mass squared for the Higgs field, triggering the spontaneous symmetry 
breaking of the global and local symmetries. The mechanism is general and they tested the idea for the electroweak theory in which the Higgs sector is extended to 
possess an extra global Abelian symmetry. With this symmetry they associated a nonzero chemical potential. The Bose-Einstein condensation of the 
Higgs bosons leads, at the tree level, to modified dispersion relations for the Higgs field, while the dispersion relations of the gauge bosons 
and fermions remain undisturbed. The latter were modified through higher order corrections. Authors have computed some corrections to the vacuum 
polarizations of the gauge bosons and fermions. To quantify the corrections to the gauge boson vacuum polarizations with respect to the standard 
model they considered the effects on the $T$ parameter. Sannino   and   Tuominen  derived also the one loop modified 
fermion dispersion relations.
It is worth noting that Batista  and Nussinov~\cite{bati05}
extended  Elitzur's theorem~\cite{eli75} to systems with symmetries intermediate between global and local. 
In general, their theorem formalizes the idea 
of dimensional reduction. They applied the results of this generalization to many systems that are of current interest. 
These include liquid crystalline phases of quantum Hall systems, orbital systems, geometrically frustrated spin lattices, 
Bose metals, and models of superconducting arrays. 
%
%
%
%
%
%
\section{Conclusions and Discussions}
%
In this paper, we have reviewed several fundamental concepts of the modern quantum physics
which manifest the operational ability of the notion of symmetry: broken symmetry,
quasiaverages, quantum protectorate, emergence, etc. We demonstrated their power of the unification of various 
complicated phenomena and presented  certain evidences for their utility and predictive ability. Broadly speaking, these 
concepts are unifying and profound ideas "that illuminate our understanding of nature".
In particular, the Bogoliubov's method of quasiaverages   gives the deep foundation and clarification of the
concept of broken symmetry. It makes the emphasis on the notion of a degeneracy and
plays an important role in equilibrium statistical mechanics of many-particle systems. According to
that concept, infinitely small perturbations can trigger
macroscopic responses in the system if they break some
symmetry and remove the related degeneracy (or quasidegeneracy)
of the equilibrium state. As a result, they
can produce macroscopic effects even when the perturbation
magnitude is tend to zero, provided that happens
after passing to the thermodynamic limit. This approach has penetrated, directly or indirectly, many areas of the
contemporary physics as it was shown in the paper by Y. Nambu~\cite{namb09} and in the present review. Nambu emphasized rightly the
"cross fertilization" effect of the notion of broken symmetry. The same words could be said about the notion
of quasiaverages.
\begin{figure}[bt]
\centerline{\includegraphics[width=12cm,height=4cm]{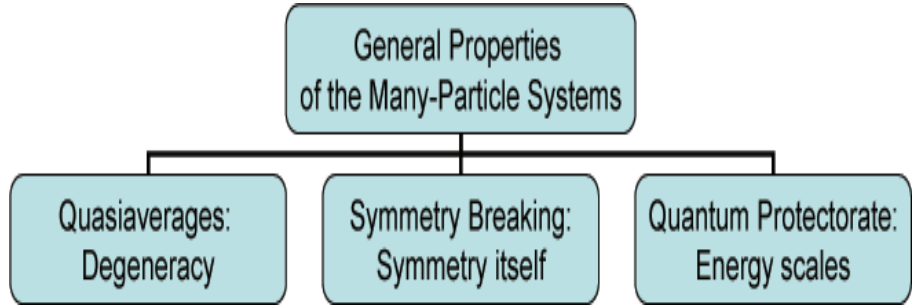}}
\vspace*{8pt}
\caption{Schematic form of the interrelation of three profound 
concepts:quasiaverages, broken symmetry and quantum protectorate} 
\label{f.1}
\end{figure}
It was shown recently  that the notion of broken symmetry can be adopted and applied to
quantum mechanical problems~\cite{zaa05,vwez08,wez08,mori06}. Thus it gives a method to approach a many body problem from an intrinsic
point of view~\cite{zur08}. From the other hand,
it is clear that only a thorough experimental and theoretical investigation of quasiparticle many-body dynamics
of the many particle systems can provide the answer on the relevant microscopic picture~\cite{kuz09}. As is well known,
Bogoliubov was first to emphasize the importance of the time scales in the many-particle systems thus
anticipating the concept of emergence macroscopic irreversible behavior starting from the reversible
dynamic equations~\cite{bog46,nnb77,bsan}.   More recently it has been possible to go step further. This step leads to a much deeper
understanding of the relations between microscopic dynamics and macroscopic behavior~\cite{pnas,sew02,adler04}.
It is worth also noticing that the notion of quantum protectorate~\cite{pnas,pines}
complements the concepts of broken symmetry and quasiaverages by making
emphasis on the hierarchy of the energy scales of  many-particle systems (see fig.~\ref{f.1}).
In an indirect way these aspects arose  already when considering the scale invariance and spontaneous 
symmetry breaking~\cite{drago87}.
D.N. Zubarev showed~\cite{zub71}  that the concepts of symmetry
breaking perturbations and quasiaverages play an
important role in the theory of irreversible processes as
well. The method of the construction of the nonequilibrium
statistical operator~\cite{zub71,kuz07,kuz09} becomes especially
deep and transparent when it is applied in the framework
of the quasiaverage concept. The main idea of
this approach was to consider infinitesimally
small sources breaking the time-reversal symmetry of
the Liouville equation~\cite{petr},
which become vanishingly small after a thermodynamic
limiting transition.\\ 
To summarize, the Bogoliubov's method of quasiaverages plays a fundamental role in equilibrium and
nonequilibrium statistical mechanics and quantum field theory and is one of the pillars of modern physics. It will serve for the future
development of physics~\cite{gross2} as invaluable tool.   
All the methods developed by N. N. Bogoliubov are and will remain the important core of a theoretician's
toolbox, and of the ideological basis behind this development.  
%
%
%
%
%
%
%
%
%
%

%
%
%
%
%
\end{document}